\pdfoutput=1
\documentclass[11pt]{article}
\usepackage{jheppub}

\usepackage{customprelude}
\usepackage{tikz}
\usepackage{tikz-cd}
\usepackage{slashed}
\usetikzlibrary{matrix,arrows}
\usepackage{multirow}

\newcommand{\eq}[1]{(\ref{#1})}
\usepackage{pifont}
\newcommand{\cmark}{\ding{51}}%
\newcommand{\xmark}{\ding{55}}%

\title{\centering Dai-Freed anomalies in particle physics}

\author[\heartsuit]{Iñaki García-Etxebarria}
\author[\spadesuit]{and Miguel Montero}

\affiliation[\heartsuit]{Department of Mathematical Sciences\\
Durham University, Durham, DH1 3LE, United Kingdom\footnote{Formerly at Max-Planck-Institut für Physik, München, Germany.}}

\affiliation[\spadesuit]{Instituut voor Theoretische Fysica, KU Leuven,\\
Celestijnenlaan 200D, B-3001 Leuven, Belgium\footnote{Formerly at ITF, Utrecht University.}}

\emailAdd{inaki.garcia-etxebarria@durham.ac.uk}
\emailAdd{miguel.montero@kuleuven.be}

\abstract{Anomalies can be elegantly analyzed by means of the
  Dai-Freed theorem. In this framework it is natural to consider a
  refinement of traditional anomaly cancellation conditions, which
  sometimes leads to nontrivial extra constraints in the fermion
  spectrum. We analyze these more refined anomaly cancellation
  conditions in a variety of theories of physical interest, including
  the Standard Model and the $SU(5)$ and $Spin(10)$ GUTs, which we
  find to be anomaly free. Turning to discrete symmetries, we find
  that baryon triality has a $\mathbb{Z}_9$ anomaly that only cancels
  if the number of generations is a multiple of 3. Assuming the
  existence of certain anomaly-free $\mathbb{Z}_4$ symmetry we relate
  the fact that there are 16 fermions per generation of the Standard
  model --- including right-handed neutrinos --- to anomalies under
  time-reversal of boundary states in four-dimensional topological
  superconductors. A similar relation exists for the MSSM, only this
  time involving the number of gauginos and Higgsinos, and it is
  non-trivially, and remarkably, satisfied for the
  $SU(3)\times SU(2) \times U(1)$ gauge group with two Higgs
  doublets. We relate the constraints we find to the well-known
  Ibañez-Ross ones, and discuss the dependence on UV data of the
  construction.  Finally, we comment on the (non-)existence of
  K-theoretic $\theta$ angles in four dimensions.}

\begin{document}

\maketitle

\newpage

\section{Introduction}

Anomalies are one of the most powerful tools that we have to analyze
quantum field theories: the anomaly for any symmetry we would like to
gauge needs to cancel, which is a constraint on the allowed spectrum.
When the symmetry is global, we have anomaly matching conditions
\cite{tHooft:1980xss} that give us very valuable information about
strong coupling dynamics.

\medskip

In the traditional viewpoint, an anomaly is a lack of invariance under
a certain gauge transformation/diffeomorphism. Local anomalies come
from transformations which are continuosly connected to the identity;
global anomalies (such as e.g. the $SU(2)$ anomaly in
\cite{Witten:1982fp}) are related to transformations that cannot be
deformed to the identity.

\medskip

However, it is becoming increasingly clear that this is not the end of
the story
\cite{Kapustin:2014dxa,Hsieh:2015xaa,Witten:2015aba}. Roughly
speaking, it also makes sense to require that the theory gives an
unambiguous prescription for the phase of the partition function when
put on an arbitrary manifold $X$, with an arbitrary gauge bundle.  We will
explain the rationale for this prescription in
section~\ref{sec:review}.

There does not seem to be a universal name for this requirement in the
literature; because the main tool to study this is the so-called
Dai-Freed theorem, we will refer to it as Dai-Freed anomaly
cancellation. Our interest stems from the fact that they result in
additional constraints on quantum field theories. The paradigmatic
example is the topological superconductor, where freedom from
gravitational anomalies on the torus requires the number of fermions
to be a multiple of 8 \cite{Hsieh:2015xaa}, and a more careful
analysis on arbitrary manifolds requires this number to be a multiple
of 16 \cite{Witten:2015aba}. As we will see, the fact that the number
of fermions in the SM, including right handed Majorana neutrinos, is a
multiple of 16 follows from Dai-Free anomaly-freedom of certain
$\bZ_4$ discrete symmetry, and it can in fact be related to the modulo
16 Dai-Freed anomaly in the topological superconductor.

\medskip

The aim of this paper is to substantiate this observation, and more
generally explore Dai-Freed anomalies in theories of interest to high
energy particle physics. We will have a look to Dai-Freed anomalies of
semisimple Lie groups, with an emphasis on GUTs and the Standard
Model, as well as discrete symmetries. To study these anomalies in
general we will compute the bordism groups of the classifying spaces
of the relevant gauge groups. We will find that both the $SU(5)$ and
$Spin(10)$ GUT's, as well as the Standard Model itself, are free from
Dai-Freed anomalies.\footnote{This result was previously obtained for
  $\Spin$ manifolds in \cite{Freed:2006mx} using different
  techniques. We rederive it, and extend it to other interesting
  classes of manifolds.} In the case of discrete symmetries, we will
find nontrivial constraints in symmetries of phenomenological
interest, such as proton triality. This symmetry has a modulo 9
Dai-Freed anomaly, which cancels only for a number of generations
which is a multiple of 3.

\medskip

This paper is organized as follows. In section~\ref{sec:review}, we
will review some useful facts about anomalies and algebraic topology
that we will use. In particular section~\ref{sec:anomalies-review} we
give a quick review of anomalies, both from the familiar viewpoint and
the more modern one based on the Dai-Freed theorem. We also explain
the connection to bordism groups. In section~\ref{sec:mtools} we then
introduce the mathematical tools that we will use to compute these
bordism groups, with particular emphasis on the Atiyah-Hirzebruch
spectral sequence. Section~\ref{sec:lie} is devoted to the computation
of the bordism groups of classifying spaces of various Lie groups. An
easy corollary of the results in this section is the absence of
Dai-Freed anomalies in the Standard Model and GUT models (including in
the case of allowing for non-orientable spacetimes). In
section~\ref{sec:discrete} we turn to the analysis of discrete
symmetries, where we will find new Dai-Freed anomalies, also in some
discrete symmetries of phenomenological interest such as proton
triality. We also identify a $\mathbb{Z}_4$ symmetry, related to
$U(1)_{B-L}$ and hypercharge, which is anomaly-free if the number of
fermions in a SM generation is a multiple of 16. In
section~\ref{sec:KOthetaangle}, which is a more theoretical aside, we
briefly review how to extend the Dai-Freed prescription to manifolds
which are not boundaries and the relationship to $\theta$ angles. We
also discuss the possibility of purely K-theoretic $\theta$
angles. Finally, section~\ref{sec:conclus} contains a brief summary of
our findings and conclusions.

\medskip

While finishing our manuscript we became aware of
\cite{Hsieh:2018ifc}, which also discusses Dai-Freed anomalies for
discrete symmetries and the connection to Ibañez-Ross constraints.

\subsection{A reading guide for the phenomenologist}

One of the main points of our paper is that a recent formal
development --- the discovery of new anomalies beyond traditional
local and global ones --- is very relevant to phenomenology, since
potentially any gauge symmetry, even the SM gauge group, could in
principle turn out to be anomalous under these more stringent
constraints. Or, from a slightly different point of view, these
developments also answer the question of whether the existence of
certain gauge symmetries imposes any constraints on spacetime
topology.

A large part of the analysis is necessarily technical, devoted to the
details of the computation of bordism groups and $\eta$ invariants. We
do encourage the reader only interested in the resulting
phenomenological constraints to skip sections \ref{sec:review} and
\ref{sec:lie}, with the exception of subsection \ref{sec:sm}, where
Dai-Freed anomalies of the SM are analyzed. Sections \ref{sec:spinzn},
\ref{sec:bt} and \ref{sec:ftop} are also of phenomenological interest
and give new constraints on gauging discrete symmetries. They contain,
in particular, explicit formulas for Dai-Freed anomaly cancellation of
discrete symmetries in $\Spin^c$ and $\Spin$ spacetimes.

\section{Review}\label{sec:review}

In this section we will briefly review the necessary background that
we will use later on. Excellent introductory references are
\cite{Witten:1985xe,AlvarezGaume:1984nf,Bilal:2008qx} for traditional
anomalies and \cite{Witten:2015aba} for the new ones. We also
recommend \cite{Hatcher:478079} for an introduction to some of the
notions in algebraic topology that will enter our analysis.

\subsection{Anomalies}
\label{sec:anomalies-review}

Suppose one has a quantum theory on which some symmetry group $G$
acts. $G$ can be a combination of internal and spacetime
symmetries. We may consider coupling the theory to a nontrivial
$G$-bundle, i.e. to a nontrivial background field. When the symmetry
group $G$ is discrete, the notion of coupling the theory to a
nontrivial $G$-bundle still makes sense (for instance, one may twist
boundary conditions along nontrivial cycles).

It can happen that physical predictions change as we act with $G$ on
the background field. More specifically, we will focus on the
partition function $Z[A]$, as a function of the background connection
$A$ for $G$. In this context, an anomaly means that $Z[A]\neq Z[A^g]$
for some gauge transform $A^g$ of $A$. Equivalently, the partition
function is not a well-defined \emph{function} of the background
connection (modulo gauge transformations), but rather a \emph{section}
of a non-trivial bundle over this space.\footnote{See for instance
  \cite{Freed:1986zx} for a more detailed discussion of this viewpoint
  aimed at physicists.}

An anomaly in a global symmetry is not an inconsistency; it just means
that we cannot gauge $G$. If we want to do this, we need to modify the
parent theory somehow. Sometimes very mild modifications suffice: in
some cases, such as in the Green-Schwarz mechanism, it is possible to
do this by introducing new non-invariant terms in the
Lagrangian. Alternatively, as discussed in
\cite{Garcia-Etxebarria:2017crf}, coupling to a topological field
theory (which introduces no new local degrees of freedom) can
sometimes be enough to cure the sickness.

\medskip

This characterization of anomalies does not require the existence of a
Lagrangian. In this paper, however, it will be sufficient for us to
restrict to Lagrangian theories, for which one can give a more
concrete description. Lagrangian theories have a path integral
formulation in terms of some elementary fields $\Phi_i$ and a
Lagrangian $\mathcal{L}(\Phi_i)$,
\begin{equation}
  Z[J_i; A]=\int [D\Phi_i]\, \exp(-S),\quad S=\int_{X}
  d^dx\, \mathcal{L}(\Phi_i, A) + J_i\mathcal{O}^i\, ,
\end{equation}
as a function of the sources $J_i$ and the background $G$-fields $A$.

Furthermore, we will further restrict to theories with some corner in
their parameter space such that the action splits as
\begin{equation}S=S_{\text{fermion}}+S_{\text{other fields}},\quad
  S_{\text{fermion}}=\frac{i}{2}\int d^dx\, \bar{\psi}
  \slashed{D}\psi. \label{fermion-action}
\end{equation} i.e. as a
fermion plus terms for the other fields, which we will take to be
non-anomalous. The fermion $\psi$ transforms on some representation
$\mathbf{R}$ of the symmetry group $G$, and (if $G$ is continuous)
couples to the background gauge field via the covariant derivative
\begin{equation}\slashed{D}=i\gamma^\mu(\partial_\mu-i A_\mu).\end{equation}
Other than that, our discussion will be completely general, applying
to real or complex fermions in an arbitrary number of dimensions. So
we will study anomalies of the theory whose partition function is
given by
\begin{equation}Z[A]=\int [D\psi]
  e^{-S_{\text{fermion}}(\psi,A)}.\label{fermion-pi}\end{equation}
This can be evaluated explicitly, since the path integral is quadratic. If $i\slashed{D}$ is self-adjoint, we can diagonalize it, and the partition function becomes
\begin{equation}
  Z[A]=\det(i\slashed{D})\, .
  \label{pfaffian-dirac}
\end{equation}
However, for anomalies we are often interested in the case where
$i\slashed{D}$ is not an endomorphism, but rather a map from one
fermion space to another. This happens when the fermions transform in
different representations (for instance, the partition function for a
Weyl fermion maps one chirality to another). In this case cases the
definition of the determinant is more subtle, but \eq{pfaffian-dirac}
still holds in an appropriate sense
\cite{AlvarezGaume:1983cs,Yonekura:2016wuc}. (Perhaps the most
conceptually clear definition is the one due to Dai and Freed,
described below.)

The above discussion holds for complex fermions. This covers most of
the cases we consider in this paper, but for completeness, we also
comment on the real case $\bar\psi=\psi$, following
\cite{Witten:2015aba}. In this case, since fermion fields
anti-commute, we can view $i\slashed{D}$ as an antisymmetric operator.

An antisymmetric operator can always be recast in block-diagonal form
\begin{equation}i\slashed{D}
  =\left(\begin{array}{ccc}\begin{array}{cc}
        0&\lambda_1\\-\lambda_1&0\end{array}& &\\ & \begin{array}{cc}
        0&\lambda_2\\-\lambda_2&0\end{array}&\\&&\ldots\end{array}\right)\end{equation}
and the quadratic path integral over $\psi$ results in
\begin{equation}
  Z[A]=\lambda_1\lambda_2\ldots=\Pf(\mathsf{D}).\label{pfaffian-dirac2}\end{equation}

An important technical point is that \eq{fermion-pi} and
\eq{pfaffian-dirac} require regularization as usual in quantum field
theory. If a regularization preserving the symmetry $G$ for an
arbitrary background gauge field can be found, then \eq{fermion-pi} is
not anomalous. In particular, this always happens whenever there is a
$G$-invariant mass term
\begin{equation}m\int d^d x\, \bar{\psi}\psi\end{equation}
for the fermions. In this case, Pauli-Villars regularization is available \cite{Witten:2015aba}, which is manifestly gauge invariant. 

\subsubsection{The traditional anomaly}
The traditional discussion of anomalies divides them in two broad classes: \begin{itemize}

\item \textbf{Local anomalies} describe a failure of \eq{fermion-pi}
  to be gauge-invariant even in a gauge transformation infinitesimally
  close to the identity. This was the first anomaly to be identified;
  it can be analyzed via the famous triangle (or more generally,
  n-gon) diagram, or more efficiently via the Wess-Zumino descent
  procedure, which relates the anomalous variation of the action
  $\delta_g S$ to a $(d+2)$ dimensional anomaly polynomial,
  \begin{equation} d(\delta_g S)=\delta_gI_{d+1},\quad
    dI_{d+1}=I_{d+2}=\left[\hat{A}(R)
      \ch(F)\right]_{d+2}.\label{wesszumino}
  \end{equation}
  The anomaly polynomial is precisely the index density in the
  Atiyah-Singer index theorem \cite{Atiyah:1968mp,Bilal:2008qx}. (A
  beautiful explanation of this fact is given by the Dai-Freed theorem
  \cite{Dai:1994kq} to be described in section~\ref{sec:daifreed}
  below.) It follows that, for the local anomaly to cancel, the
  anomaly polynomial of the theory must vanish. Because any symmetry
  transformation continuously connected to the identity can be related
  to an infinitesimal one via exponentiation, vanishing of the anomaly
  polynomial guarantees that any symmetry which can be deformed to the
  identity is anomaly free.

\item Even if a theory is free of local anomalies, it can still have a
  \textbf{global anomaly}, an anomaly in a transformation $g$ not
  continuously connected to the identity. If we are considering the
  theory on some particular manifold $X$, the relevant transformations
  are given by maps $X\rightarrow G$ to the symmetry group $G$. This
  is commonly denoted $[X,G]$. There can only be a global anomaly if
  this is nontrivial. In the particular case where $X=S^d$ is a
  sphere, $[S^d,G]=\pi_d(G)$ is the $d$-th homotopy group of
  $G$. Because the sphere is the one point compactification of
  $\mathbb{R}^d$, global anomalies on spheres are directly relevant to
  theories in flat space (or more generally, they encode the part of
  the global anomaly which is local in spacetime). For instance,
  $\pi_4(SU(2))=\mathbb{Z}_2$, related to the $SU(2)$ global anomaly
  discussed in \cite{Witten:1982fp}.

  Global anomalies were originally studied via the so-called mapping
  torus construction
  \cite{Witten:1982fp,Witten:1985xe,Witten:2015aba}. One constructs an
  auxiliary $(d+1)$ dimensional space as the quotient
  \begin{equation} X\times [0,1] /r,\quad r: (m,0) \sim (
    g(m),1),\quad \psi(m,1)=\psi(m,0)^g.\end{equation} Here, $\psi^g$
  denotes the gauge transform of $\psi$ under the potentially
  anomalous transformation. If $t\in[0,1]$, is the coordinate on the
  interval, we also have a corresponding gauge field
  \begin{equation} A_t=(1-t)A_0+tA_0^g.\label{connmapp}\end{equation}
  The mapping torus construction can be applied to study anomalies of
  any transformation, whether or not we are gauging it.  However, when
  the symmetry is gauged, so that $A_0$ and $A_0^g$ are physically
  equivalent, the mapping torus describes a non-contractible closed
  path on the space of connections on the theory on $X$; the gauge
  profile \eq{connmapp} precisely follows this non-contractible path.

  The $d$-dimensional theory will only be anomaly free if a certain
  topological invariant constructed out of a particular
  $(d+1)$-dimensional Dirac operator coupled to \eq{connmapp} actually
  vanishes. For anomalies of fermions in real representations of $G$
  in $d=4k$ dimensions (such as a 4d euclidean Weyl fermion in the
  fundamental of $SU(2)$ \cite{Witten:1982fp} --- recall that $G$
  includes the Lorentz part too), this topological invariant is the
  mod 2 index \cite{Witten:2015aba}. This is defined as the number of
  zero modes of the Dirac operator on the mapping torus, modulo
  two. For complex fermions, the anomaly is computed in terms of the
  APS $\eta$ invariant of the Dirac operator on the mapping torus
  \cite{Witten:1985xe}. We will discuss this invariant momentarily.
\end{itemize} 

\subsubsection{The Dai-Freed viewpoint on anomalies}\label{sec:daifreed}
\begin{figure}[!htb]
\begin{center}
\includegraphics[width=5cm]{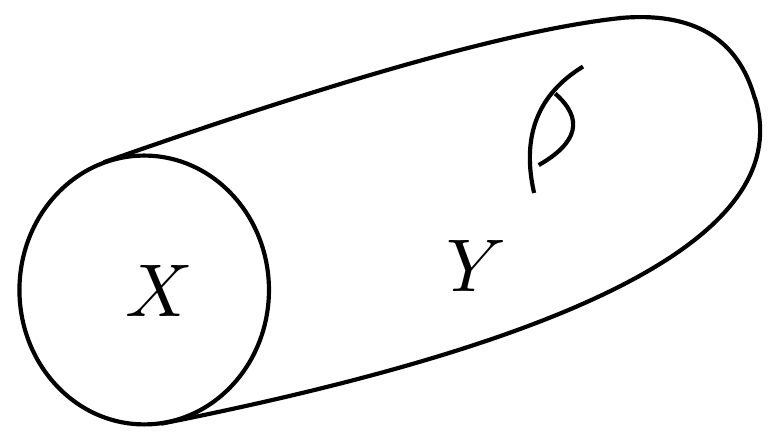}
\end{center}
\caption{The Dai-Freed construction computes the phase of the fermion path integral on a manifold $X$ via an auxiliary manifold $Y$ such that $\partial Y=X$.}
\label{f1}
\end{figure}

The more modern viewpoint on anomalies encompasses the above
discussion via what has been called the Dai-Freed theorem
\cite{Dai:1994kq,Witten:2015aba,Yonekura:2016wuc}, which for our
purposes here we can state as follows. Suppose we are interested in
studying anomalies on a fermion theory defined on some manifold $X$,
and $X$ can be written as the boundary of some other manifold $Y$,
such that both the spin/pin structure and the gauge bundle on $X$ can
be extended to $Y$, see figure~\ref{f1}.\footnote{Such a $Y$ may not
  exist. We will comment more on this situation in
  section~\ref{sec:KOthetaangle}. For now, we assume the existence of
  $Y$.}

Then, out of the Dirac operator on $X$ showing up in the fermion lagrangian \eq{fermion-action}, we construct a Dirac operator in $Y$ by the prescription that near the boundary $X \times (-\tau_0,0]$ of $Y$, $i\slashed{D}_Y$ takes the form
\begin{align}
  i\slashed{D}_Y&=i\gamma^\tau\left(\frac{\partial}{\partial \tau}+
    i\slashed{D}_X'\right),\quad \gamma^{\tau}=\text{diag}\,(
  \mathbf{I}_d,-\mathbf{I}_d),\nonumber\\&i\slashed{D}_X'=\left(\begin{array}{cc}
      0& i \slashed{D}_X\\-i\slashed{D}^\dagger_X &
      0\end{array}\right).\label{dyfdx}\end{align} As mentioned before, there is an anomaly
whenever the partition function \eq{fermion-pi} is not a well-defined
function of the connection/metric. We can rephrase this by saying that
the partition function is in general not a function on the space
$\mathcal{M}$ of connections/metrics modulo gauge
transformations/diffeomorphisms, but rather, a section of a nontrivial
complex line bundle over $\mathcal{M}$, the so-called determinant
line bundle over $\mathcal{M}$ \cite{Dai:1994kq} (or Pfaffian line bundle in the general case).

The Dai-Freed theorem tells us that there is a quantity, computed solely in terms of $\slashed{D}_Y$,
\begin{equation} \exp\left(2\pi i\eta_Y\right)\label{daifreedphase}\end{equation}
that is also a section of the \emph{same} principal $U(1)$ bundle. As a result, we can use \eq{daifreedphase} instead of working with the determinant \eq{pfaffian-dirac} directly to study anomalies. 
$\eta_Y$ is the Atiyah-Patodi-Singer (APS) $\eta$-invariant \cite{Atiyah:1975jf,Atiyah:1976jg,APS-III,Witten:1985xe,Witten:2015aba}, defined as follows. First, we pick a class of boundary conditions (called APS boundary conditions \cite{Atiyah:1975jf,Atiyah:1976jg,APS-III}, see \cite{Yonekura:2016wuc} for a nice detailed discussion) such that $i\slashed{D}_Y$ on $Y$ becomes self-adjoint. Then, $\eta_Y$ is a regularized sum of eigenvalues
\begin{equation}\eta_Y=\frac12\left(\sum_{\lambda\neq0}\text{sign}(\lambda)+\dim\ker(i\mathcal{D}_Y)\right)_{\text{reg.}}.\label{etadef}\end{equation} 
The sum is infinite and requires regularization; $\zeta$-function regularization is commonly employed in the mathematical literature. The $\eta$ invariant jumps by $\pm1$ whenever an eigenvalue crosses zero; however, $\exp(2\pi i \eta_Y)$ is a continuous function of the gauge fields and the metric.

\begin{figure}[!htb]
\begin{center}
\includegraphics[height=3.5cm]{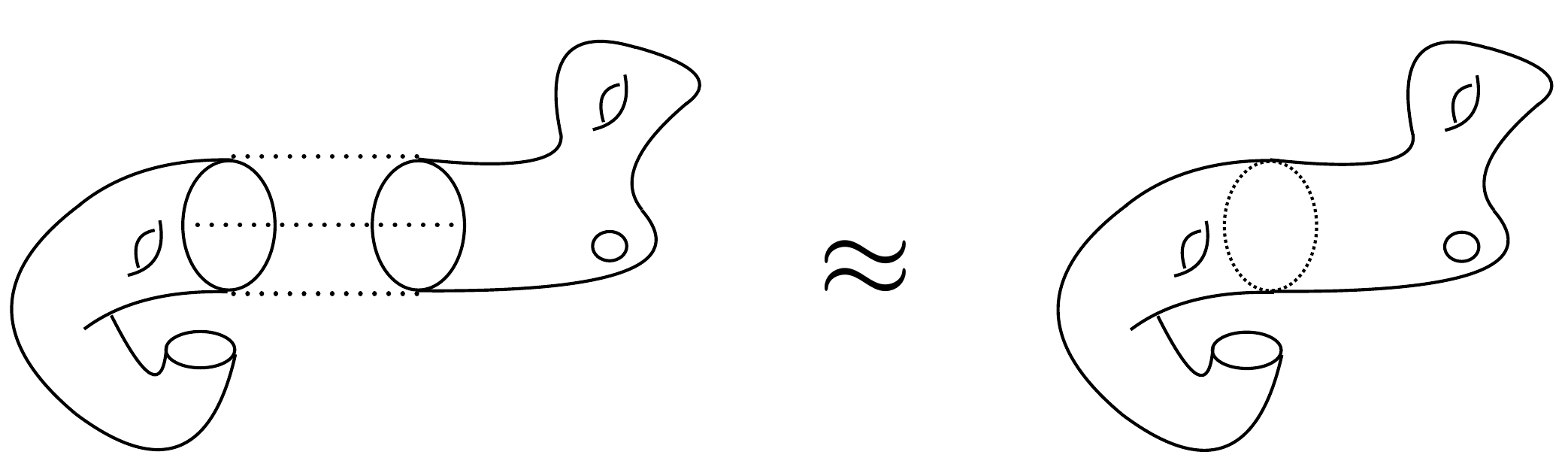}
\end{center}
\caption{The $\eta$ invariant behaves nicely under gluing as illustrated in the picture.}
\label{f2}
\end{figure}

\begin{figure}[!htb]
\begin{center}
\includegraphics[height=3cm]{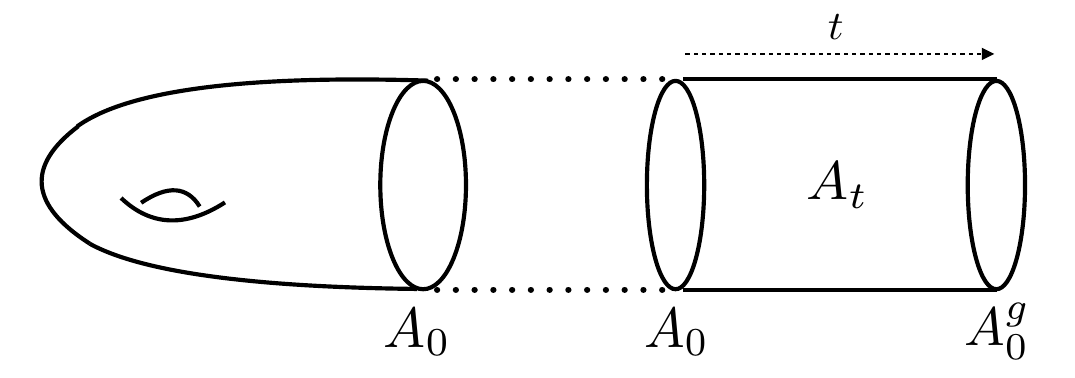}
\end{center}
\caption{To obtain the phase for a configuration $A_0^g$ starting from $A_0$, we may just attach $X\times [0,1]$ as shown in the picture. The additional contribution to the phase is identical to $\eta$ evaluated on the mapping tours obtained by gluing the two sides of $X\times [0,1]$.}
\label{f3}
\end{figure}

The advantages of this approach are that we now do not have to deal
with regularizations, etc. directly, and that we can use several
properties of the $\eta$ invariant to our advantage.  For instance,
$\eta$ behaves ``nicely'' under gluing \cite{Yonekura:2016wuc}: if we
have two manifolds $Y_1, Y_2$ glued along a common boundary as in
figure~\ref{f2}, giving the manifold $Y_1\sqcup Y_2$, we have
\begin{equation}
  \exp(2\pi i\, \eta_{Y_1\sqcup Y_2})=\exp(2\pi i\, \eta_{Y_1}) \exp(2\pi i\,
  \eta_{Y_2})\, .
\end{equation}
This means that, as discussed in \cite{Monnier:2014txa} for instance,
if we want to compute the change of the phase of the partition
function $Z[A]$, going from some configuration $A_0$ to some other
$A_0^g$ (where $g$ may or may not be continuously connected to the
identity) along a path $A_t$, we just need to compute the $\eta$
invariant on a manifold $X\times [0,1]$, since we can then glue it to
the $Y_0$ which gives the phase on $A_0$ (see figure
\ref{f3}). Because the gauge configuration at the endpoints of the
interval are gauge transformations of one another, we can glue the
sides to obtain the $\eta$ invariant in the same mapping torus that
was discussed above for global anomalies \cite{Yonekura:2016wuc}.

In this way, absence of traditional anomalies (local or global)
becomes the requirement $\exp(2\pi i\, \eta_Y)=1$ for $Y$ any mapping
torus. We indeed recover the local and global anomaly cancellation
conditions discussed above, as in \cite{Monnier:2014txa}:
\begin{itemize}
\item For $g$ continuously connected
  to the identity, one can write $Y=\partial Z$, where $Z= X\times D$
  is a $(d+2)$-dimensional manifold, since the gauge bundle can be
  extended to $Z$ without problem. In this case, we can use the APS
  index theorem for manifolds with boundary \cite{Atiyah:1975jf},
  which relates
  \begin{equation} \text{Ind} (\slashed{D}_Z)=\eta_Y +\int_Z
    \hat{A}(R) \ch(F).\label{APSindex}\end{equation} The left hand
  side is the index of a Dirac operator on $Z$, which is always an
  integer. Exponentiating, we get
  \begin{equation}
    \label{eq:exponentiated-APS}
    \exp(2\pi i\, \eta_Y)=\exp\left(2\pi i \int_Z \hat{A}(R) \ch(F)\right)=\exp\left(2\pi i \int_Z I_{d+2}\right)\, .
  \end{equation}
The only way the anomaly vanishes is if the anomaly polynomial
vanishes identically. We thus recover the traditional local anomaly
cancellation condition.
\item Global anomalies of complex fermions were already discussed in
  terms of the $\eta$ invariant. This covers almost all the cases we
  will discuss in this paper. We refer the reader to
  \cite{Witten:2015aba} for a discussion of the (pseudo-)real case.

\end{itemize}

In the present formalism, a natural question is whether the
requirement $\exp(2\pi i\, \eta_Y)=1$ should be generalized to closed
$(d+1)$ manifolds $Y$ which are not mapping tori. These conditions do
not correspond to anomalies in the traditional sense; yet demanding
their vanishing can impose nontrivial constraints on the allowed
theories. We will call them, for lack of a better term, ``Dai-Freed
anomalies'' (even though also the traditional anomalies can also be
nicely understood from the Dai-Freed point of view, as we have just
seen). The goal of this paper is the exploration of these constraints
in some interesting gauge theories. But before we start computing
$\eta$ invariants, let us review some of the reasons why it seems
plausible to us that these anomalies should cancel.

Suppose as before that we want to study the theory on some
$X=\partial Y_1=\partial Y_2$. Then, we can glue $Y_1$ and $Y_2$ with
opposite orientation along their common boundary, and we can compute
$\exp(2\pi i\, \eta_{Y_1\sqcup\bar{Y}_2})$. If this is different from
one, it means that the Dai-Freed prescription does not give a unique
answer for the phase of the path integral. Faced with this issue we
could somehow try to restrict the allowed set of $Y$'s to be used in
the Dai-Freed prescription, so that e.g. $Y_1$ is allowed but $Y_2$ is
not. However, this cannot be done arbitrarily; it has to be done in a
consistent way with cutting and pasting relations
\cite{Witten:2015aba}. Reflection positivity/unitarity also provide
further constraints. It seems more economical to impose
$\exp(2\pi i \eta_Y)=1$ for all closed $Y$ instead.

\begin{figure}
\begin{center}
\includegraphics[height=3.5cm]{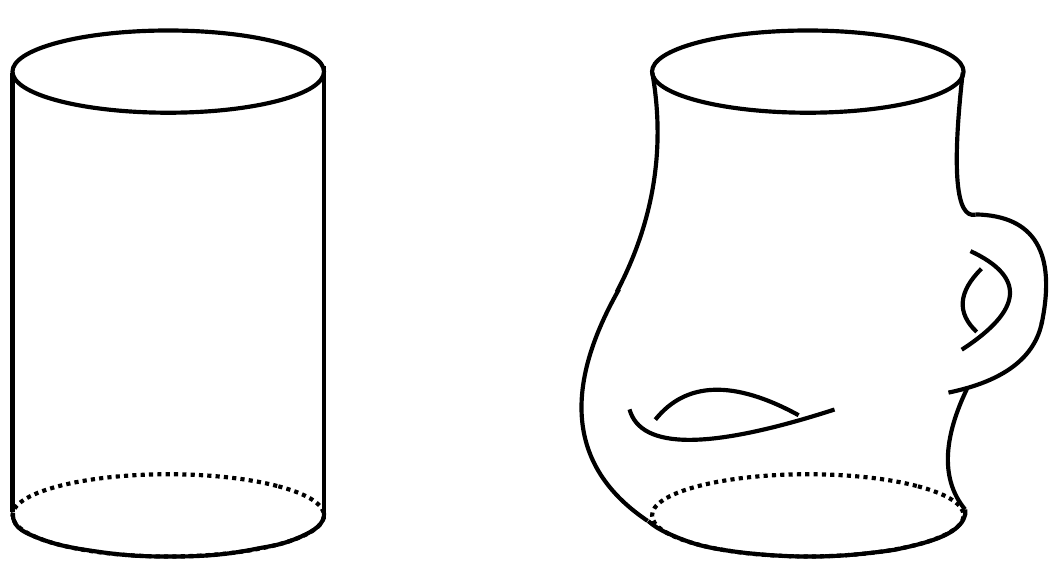}
\end{center}
\caption{Traditional global anomalies are studied via the $\eta$
  invariant on mapping tori (left figure). The general Dai-Freed
  anomaly can be regarded as a generalization in which we allow the
  ``mapping torus'' to have holes or other nontrivial topologies. In
  the same way that the traditional mapping torus follows a nontrivial
  loop in configuration space, the new anomaly can be regarded as
  coming from new nontrivial loops that arise once topology change is
  allowed, as one might expect to happen in quantum gravity.}
\label{f4}
\end{figure}

In systems coupled to dynamical gravity, there is another way to
motivate imposing these constraints. Recall that a mapping torus for a
global anomaly is just describing a non-contractible loop in the gauge
field/metric configuration space. We get one mapping torus for each
non-contractible loop. In quantum gravity, however, we generically
expect topology change (there are a myriad examples of such behaviour
understood by now in string theory, see
e.g. \cite{Giddings:1987cg,Witten:1998zw} for two examples which are
particularly close to what we are discussing here). Morally, this
enlarges the configuration space, and one can now consider closed
paths along which the topology changes. These will look like a
``mapping torus with holes'' such as the one in figure \ref{f4}, and
some of them might be non-contractible. From this point of view, the
Dai-Freed anomalies are not different from the traditional ones, at
least in a theory in which topology change is allowed.

While it is not obvious that any manifold $Y$ can be regarded as a
``generalized mapping torus'' as in figure~\ref{f4}, there is always a
perhaps different manifold $Y'_X$ with $\eta(Y)=\eta(Y'_X)$ and which
has a mapping torus interpretation over a base manifold $X$ (so that
it describes an anomaly for the theory on $X$). One can construct
$Y'_X$ by starting with a trivial mapping torus $X\times S^1$, for
which the anomaly theory is trivial since it is a boundary, and then
taking $Y'_X$ to be the connected sum $(X\times S^1)\# Y$. To display
$Y_X'$ as a generalized mapping torus, cut it open along the $S_1$,
and embed the resulting $(X\times[0,1])\# Y$ into $\mathbb{R}^K$ (such
an embedding is always possible for high enough $K$, as proven by
Whitney). Slicing with hyperplanes parallel to the $[0,1]$ factor, one
recovers the picture in figure~\ref{f4}.

\medskip

For completeness, let us mention that the rephrasing of the anomaly
for fermions in $X$ in terms of $\exp(2\pi i\, \eta_Y)$ is a specific
example of a more general construction, where one associates a
$(d+1)$-dimensional \emph{anomaly theory} $\cA[\cT]$ to any anomalous
$d$-dimensional theory $\cT$, such that the anomalous behaviour of the
partition function of $\cT$ on some manifold $X_d$ is encoded (in the
same manner as above) in the behavior of $\cA[\cT]$ on $Y_{d+1}$, with
$X_d=\partial Y_{d+1}$. In our case we have $d=4$, $\cT$ is the theory
of a Weyl fermion charged under some global symmetry $G$, and
$\cA[\cT]$ is $\exp(2\pi i\, \eta_Y)$. Other important cases for which
one can proceed analogously, and construct appropriate anomaly
theories, are theories with self-dual fields in $d=4k+2$ dimensions,
theories with Rarita-Schwinger fields, and theories where the
Green-Schwarz anomaly cancellation mechanism operates. We refer the
reader to \cite{Freed:2014iua,Monnier:2019ytc} for a systematic
discussion of such generalizations, and further references to the
literature.

\medskip

Finally, it should be pointed out that there is the possibility of
anomaly cancellation mechanisms which in some cases might weaken the
requirement of having $\exp(2\pi i \, \eta_Y)=1$ for every $Y$. The
ordinary Green-Schwarz mechanism is one example, where the anomaly can
sometimes be cancelled by adding suitable non-invariant terms to the
Lagrangian. Relatedly, as discussed in
\cite{Garcia-Etxebarria:2017crf}, anomalies which only appear for
spacetimes with specific topological properties may sometimes be
cancelled by coupling to a topological QFT with the same anomaly. When
such a possibility exists, it is perfectly fine to have a Dai-Freed
anomalous sector, as long as we ``cure'' the anomaly by coupling to
the right TQFT.  This means that any claim we make below of a theory
having a Dai-Freed anomaly should be understood to mean that the
theory is inconsistent if not coupled to any TQFT, and \emph{may} in
some cases become consistent by such a coupling, but the criterion for
which cases are fixable is currently unknown. We will present explicit
examples in section~\ref{sec:tgsd} where such a possibility plays a
very important role in connecting with known results. See
\cite{Benini:2018reh} for more examples of TQFTs with the same anomaly
as local quantum field theories of interest, also applying to
generalized global symmetries.

Luckily, the claim of consistency is not subject to such
uncertainties: for the cases for which we prove absence of Dai-Freed
anomalies one can state with certainty that anomalies are absent. It
is still interesting to couple the theory to non-trivial TQFTs, and
perhaps some of these introduce anomalies, but it is not something one
needs to do.

\subsection{Mathematical tools}\label{sec:mtools}

The rest of the paper is devoted to analyzing Dai-Freed anomalies in
theories of interest. To do this, we need a number of mathematical
tools that we review in this section.

\subsubsection[The general strategy: $\eta$ and bordism]{The general
  strategy: $\eta$ and bordism\footnote{Somewhat confusingly, the
    notion reviewed here is called both \emph{bordism} and
    \emph{cobordism} in the literature. As generalized (co)homology
    theories, what we discuss is a generalized \emph{homology}
    theory. Although it will not enter our discussion, there is an
    associated generalized \emph{cohomology} theory. It seems natural
    to call the former \emph{bordism}, and the later
    \emph{cobordism}.}}\label{sec:strategy}

In the rest of the paper, we will only consider theories in which the
local anomalies cancel. This has the very convenient consequence that
$\eta$ becomes a topological invariant, and in fact it has the
stronger property of being a \emph{bordism invariant}.

\begin{figure}
\begin{center}
\includegraphics[height=3cm]{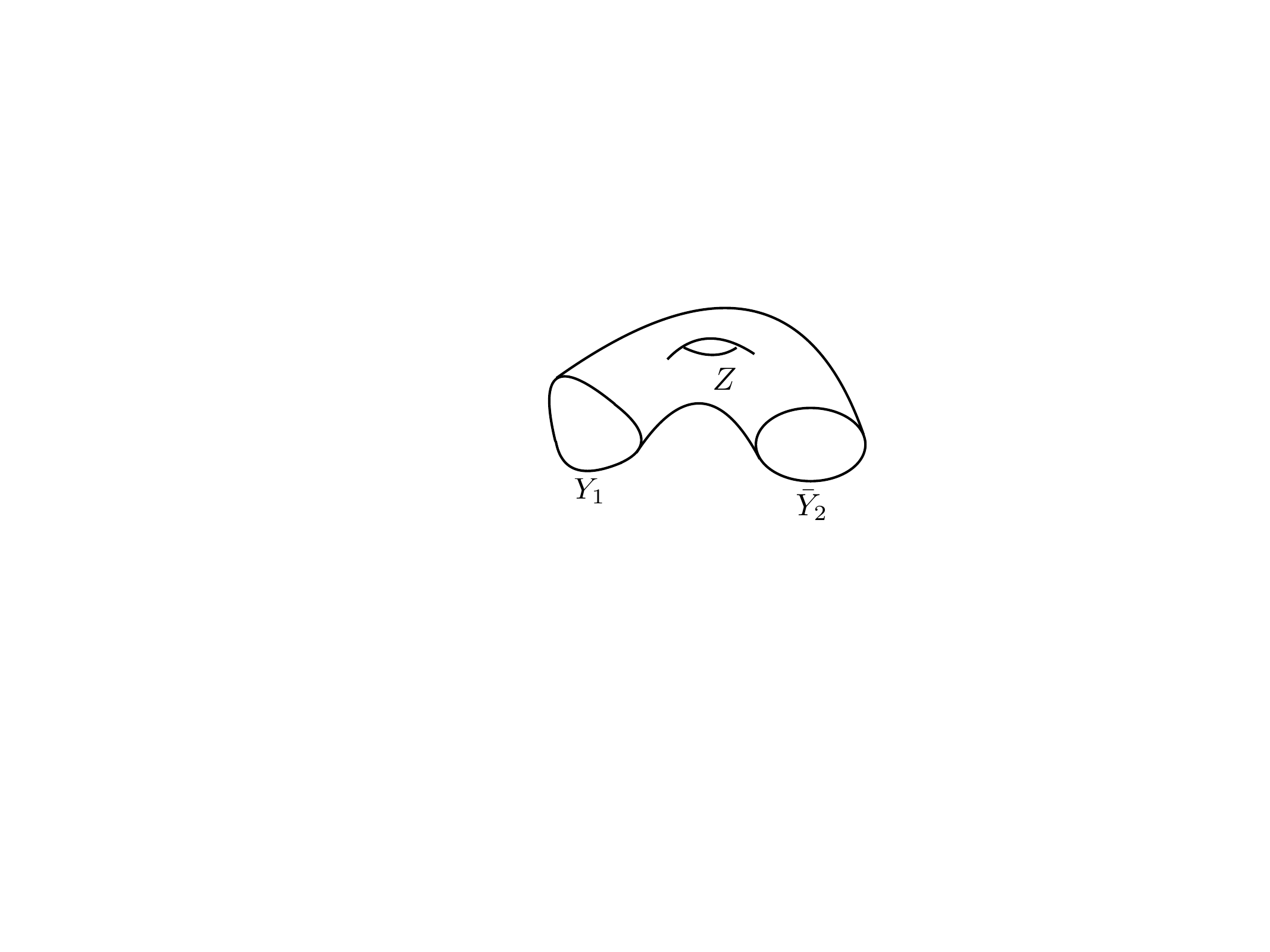}
\end{center}
\caption{The two manifolds $Y_1$ and $Y_2$ are bordant if $Y_1\sqcup \bar{Y_2}$ is boundary of another manifold $Z$.}
\label{f5}
\end{figure}

Bordism is an equivalence relation between manifolds (possibly
equipped with extra structure): $Y_1$ and $Y_2$ are bordant if their
disjoint union with a change of orientation for $Y_2$, which we denote
as $Y_1\sqcup \bar{Y}_2$, is the boundary of another manifold $Z$, as
illustrated in figure \ref{f5}. If this is the case, we write
$Y_1\sim Y_2$, which is clearly an equivalence relation. In case the
$Y_i$ carry extra structure, such as a spin structure or a gauge
bundle, we demand that this can be extended to $Z$ as well.

Bordism invariance of $\exp(2\pi i\, \eta_Y)$ is a simple consequence
of the APS index theorem \eq{APSindex} and the fact that local
anomalies cancel, so the last term in \eq{APSindex} is absent. To see
this, we use the fact that under change of orientation
\begin{equation}
  \exp(2\pi i\, \eta_{\ov Y}) = \exp(-2\pi i\, \eta_Y)
\end{equation}
so that the gluing properties of $\eta$ imply
\begin{equation}
  \exp(2\pi i\, \eta_{Y_1\sqcup \ov{Y_2}})=\frac{\exp(2\pi i\, \eta_{Y_1})}{\exp(2\pi i\,
    \eta_{Y_2})}\, .
\end{equation}
If $Y_1$ and $Y_2$ are in the same bordism class then, by definition,
$Y_1\sqcup \ov{Y_2}$ is a boundary of some manifold $Z$, so
by~\eqref{eq:exponentiated-APS} we have
\begin{equation}
  \frac{\exp(2\pi i\, \eta_{Y_1})}{\exp(2\pi i\,
    \eta_{Y_2})} = \exp(2\pi i\, \eta_{Y_1\sqcup \ov{Y_2}}) =\exp(2\pi i \int_Z I_{d+2}) = 1
\end{equation}
assuming no local anomalies.

Furthermore, the set of bordism equivalence classes forms an abelian
group under union; we define $[Y_1]+[Y_2]=[Y_1\sqcup Y_2]$. This also
works when additional structures are present.

\medskip

We will be particularly interested in the bordism groups denoted
$\Omega_d^{\Spin}(W)$, whose elements are equivalence classes of
$d$-dimensional Spin manifolds equipped with a map to $W$. To study
gauge anomalies in a theory with a symmetry group $G$, we will take
$W=BG$, the classifying space of $G$. This is an infinite-dimensional
space equipped with a principal $G$-bundle with total space $EG$, with
the universal property that any principal $G$-bundle over any manifold
$X$ is the pullback $f^*EG$ via some map $f: X\rightarrow BG$. Thus,
the set of all topologically distinct principal bundles over any given
manifold $X$ is equivalent to the set $[X, BG]$ of homotopy classes of
maps from $X$ to $BG$. The classifying space is therefore a convenient
way to describe principal bundles.\footnote{Since the physical theory
  comes equipped with a connection which must extend to the auxiliary
  manifold, the more natural data for the anomaly theory is not a
  manifold with principal bundle, but a principal bundle \emph{with
    connection}. However, the space of connections over a given
  principal bundle is an affine space \cite{nicolaescu2000notes}, and
  in particular contractible. This means we can deform smoothly any
  connection to any other. Since any bundle admits at least one
  connection \cite{nicolaescu2000notes}, it follows that as long as
  the anomaly is topological (that is, if local anomalies cancel) it
  cannot depend on the connection.} See
\cite{Guo:2017xex,Wang:2018edf} for a similar discussion in the
context of $3+1$ topological insulators, where similar bordism groups
(and twisted generalizations thereof) are computed.

In a $d$-dimensional theory with spinors and symmetry group $G$, the
Dai-Freed anomaly $\exp(2\pi i \, \eta_Y)$ is a group homomorphism from
$\Omega_{d+1}^{\Spin}(BG)$ to $U(1)$. To study these anomalies we will
follow these two steps:
\begin{itemize}
\item Compute $\Omega_{d+1}^{\Spin}(BG)$. If it vanishes, there can be
  no Dai-Freed anomaly.
\item If $\Omega_{d+1}^{\Spin}(BG)\neq0$, compute
  $\exp(2\pi i \, \eta)\colon \Omega_{d+1}^{\Spin}(BG)\rightarrow
  U(1)$, typically by explicit computation on convenient generators of
  the bordism group.
\end{itemize}

For the theories of interest in this paper, the first step can be
performed fairly systematically via spectral sequences, which we will
introduce in the next subsection. The second step is more artisanal ---
we need to analyze and compute $\eta$ in a case-by-case basis. We will
give examples in section~\ref{sec:discrete}.

\subsubsection{The Atiyah-Hirzebruch spectral sequence}\label{sec:AHSS}

A nice introduction to spectral sequences is \cite{McCleary}, we will
just cover the essentials to understand how the computation works. The
Atiyah-Hirzebruch spectral sequence (AHSS) is a tool for computing the
generalized homology groups $E_*(X)$ of some space $X$. A generalized
homology theory satisfies the same axioms as ordinary homology, except
for the dimension axiom: $H_p(\pt)$ --- the homology groups of a point
--- do not necessarily vanish for $p\neq0$. It turns out that bordism
theories $\Omega^{\Spin}_*(X)$ (and similarly $\Omega_*^{\Pin^\pm}(X)$)
are generalized homology theories on $X$.

The AHSS works as follows. Suppose we have a Serre
fibration\footnote{This means that we only require that the fibers at
  different points are homotopy-equivalent to one another
  \cite{Hatcher:478079}.} $F\rightarrow X\rightarrow B$. Then the AHSS
provides a systematic way to obtain a filtration of
$\Omega^{\Spin}_n(X)$, that is a sequence of spaces
\begin{equation}0=F_{-1}\Omega^{\Spin}_n(X)\subset
  F_{0}\Omega^{\Spin}_n(X)\subset\ldots\subset
  F_{n}\Omega^{\Spin}_n(X)=\Omega^{\Spin}_n(X)\, .\label{fil}
\end{equation}
Specifically, the AHSS provides a way to compute the quotients
\begin{equation}
  E_{k,n-k}^\infty=\frac{F_k
    \Omega^{\Spin}_n(X)}{F_{k-1}\Omega^{\Spin}_n(X)}\, .
  \label{quots}
\end{equation}
Even when all these quotients are known, they do not fully determine
$\Omega^{\Spin}_*(X)$. One has to solve the successive extension
problems associated to \eq{fil} and \eq{quots}, which may require
additional information.

The quotients $E^\infty_{p,q}$ live on the ``$\infty$ page'' of the
spectral sequence, and they are computed as follows. The ``second
page'' of the AHSS is simply\footnote{There is a subtlety here: the
  coefficient ring in~\eqref{page2} should be viewed as being
  \emph{local}. This fibration of coefficients is trivial if
  $\pi_1(B)=0$ (see for example \S9.2 in \cite{DavisKirk}), which is
  the case for our examples.}
\begin{equation}
  E^2_{p,q}=H_p(B,\Omega_q^{\Spin}(F))\, .
  \label{page2}
\end{equation}
The $r$-th page comes equipped with a differential
$d_r\colon E_{p,q}^r\rightarrow E_{p-r,q+r-1}$, with $d_r^2=0$. The
next page in the spectral sequence, $E^{r+1}$, is the cohomology of
$E^{r}$ under $d_r$.

\begin{figure}
  \centering
 \includegraphics{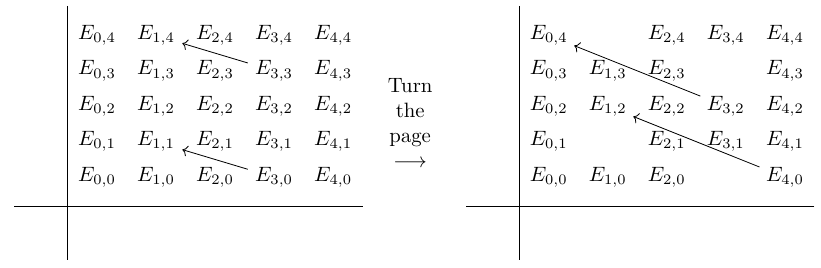}
 \caption{Generic structure of a spectral sequence.  The sequence consists of ``pages'' (in the figure we depict the second and third pages), and to turn to the next page one needs to take the cohomology with respect to the differential $d_r$. The differentials at each page are represented by arrows. Some entries might be ``killed'' by the differentials.  After we are done, at $E^\infty$,
   $\Omega^{\Spin}_n(E)$ is obtained by solving an extension problem
   involving all the entries with $p+q=n$.}
  \label{fig:SS-structure}
\end{figure}

A spectral sequence is usually presented in a diagram such as that of
figure \ref{fig:SS-structure}. The differentials are represented by
arrows. For a given entry in the spectral sequence, there are no more
differentials that can act on it after a finite number of pages; we
then say that the sequence stabilizes (for the entry of interest) and
we can read off $E^\infty_{p,q}$.

The generic strategy we will use to compute $\Omega_*^{\Spin}(BG)$ is
the AHSS associated to the fibration
$\pt \rightarrow BG\rightarrow BG$, which relates
$\Omega_*^{\Spin}(BG)$ to the groups $\Omega_{q}^{\Spin}(\pt)$,
which are given by \cite{ABP,Stong-11,Kapustin:2014dxa}\footnote{Note
  that there is a difference between \cite{Kapustin:2014dxa} and
  \cite{Stong-11} in $\Omega_{10}^\Spin(\pt)$. We have written the
  answer in \cite{Stong-11}, which agrees with the standard result
  that the free part of $\Omega_d(\pt)$ is concentrated at
  $d\in 4\bZ$ \cite{ABP}.}
\begin{equation}
  \label{eq:Omega-Spin-pt}
  \def\arraystretch{1.5}
  \arraycolsep=4pt
  \begin{array}{c|ccccccccccc}
    n & 0 & 1 & 2 & 3 & 4 & 5 & 6 & 7 & 8 & 9 & 10 \\
    \hline
    \Omega^\Spin_n(\pt) & \bZ & \bZ_2 & \bZ_2 & 0 & \bZ & 0 & 0 & 0 &
                                                                      2\bZ
                                          & 2\bZ_2 & 3\bZ_2
  \end{array}
\end{equation}
where with the notation $k\bZ$ we mean simply
$\bZ\oplus \ldots \oplus \bZ$, $k$ times.

\subsubsection{Evaluating the first nontrivial differentials: Steenrod
  squares \& their duals}
\label{sec:steenrods}

In the applications that will be discussed in section~\ref{sec:lie},
it will often be the case that the differentials in the AHSS cannot be
determined by algebraic considerations alone. In some cases, however,
we will be able to determine $d_2$ via Lemma 2.3.2 of
\cite{TeichnerPhD} (also the Lemma in pg. 751 of \cite{Teichner}),
which says that for $X$ a spectrum, the differential
$E_2^{(p,0)}\to E_2^{(p-2,1)}$, that is
\begin{equation}
  d_2\colon H_p(X,\Omega_0^{\Spin})\rightarrow
  H_{p-2}(X,\Omega_1^{\Spin})\, ,
  \label{diff2}
\end{equation}
is the composition of reduction mod 2 $\rho$ with the dual $\dSq2$
(with respect to the Kronecker pairing between homology and cohomology
\cite{adams1995stable}) of the second Stenrood square $\Sq2$. That is,
$\dsz{a}{\Sq2 b}= \dsz{\dSq2 a}{b}$ for any $a,b$, where $\dsz{}{}$ is
the Kronecker pairing between $H_n(X,\bZ_2)$ and $H^n(X,\bZ_2)$, which
is simply the evaluation map.

Note the fact that $H^n(X, \bZ_2)=H_n(X, \bZ_2)$. This follows from
the universal coefficient theorem with coefficients in an arbitrary
ring (Theorem 3.2 of \cite{Hatcher:478079})
\begin{equation}
  \label{eq:UCF-arbitrary-coeffs}
  0 \to \Ext_R^1(H_{i-1}(X, R), G) \to H^i(X, G) \to \Hom_R(H_i(X, R),
  G) \to 0
\end{equation}
and $\Ext_{\bZ_2}^1(H_{i-1}(X, \bZ_2), \bZ_2)=0$ since $\bZ_2$ is
injective as a module over itself. We thus have that
$H^i(X, \bZ_2) \cong \Hom_{\bZ_2}(H_i(X, \bZ_2), \bZ_2)$, with the
isomorphism induced by the Kronecker pairing above.

Similarly,
\begin{equation} d_2: H_p(X,\Omega_1^{\Spin})\rightarrow
  H_{p-2}(X,\Omega_2^{\Spin})
  \label{diff22}
\end{equation}
is simply the dual Steenrod square.

Steenrod squares $\Sq{i}$ are certain cohomology operations which we
can compute explicitly in the examples of interest, using the
following properties. (Here $u_i\in H^i(X,\bZ_2)$.)
\begin{subequations}
  \label{eq:Steenrod-properties}
  \begin{align}
   \Sq0 u_i & = u_i\, ,\\
    \Sq{i} u_i & = u_i^2\, ,\\
    \Sq{j} u_i & = 0 \qquad \text{for } j > i\ ,\\
    \label{eq:Cartan-formula}
    \Sq{n}(a\smile b) & = \sum_{i+j=n} (\Sq{i}a) \smile (\Sq{j}b) \, .
  \end{align}
\end{subequations}
The last equation is known as \emph{Cartan's formula}. We refer
interested readers to
\cite{WuSteenrod,Borel1953,milnor1974characteristic,Fung,Hatcher:478079}
for further details.

Reduction modulo 2 above refers to the map $\rho$ in the exact
sequence
\begin{equation}
  \label{eq:rho-long}
  \ldots \to H_i(X, \bZ) \to H_i(X, \bZ) \xrightarrow{\,\rho\,} H_i(X,
  \bZ_2) \to H_{i-1}(X, \bZ) \ldots
\end{equation}
associated to the short exact sequence $0\to \bZ \to \bZ \to \bZ_2 \to
0$.

Finally, the homology groups of a spectrum $\{X_n,s_n\}$ are defined
as
\begin{equation}H_k(X)=\text{colim}_n H_{k+n} (X_n)\end{equation}
If $X$ is the suspension spectrum of $X_0$, defined by $X_n=\Sigma^n X_0$ and $s_n$ the identity, we can use the result \cite{Hatcher:478079}
\begin{equation}H_{k+n}(\Sigma^n X_0)= H_{k} (X_0)\end{equation}
to obtain that~\eqref{diff2} and~\eqref{diff22}  also apply to an ordinary CW complex, such as the classifying spaces we will be interested in.

\medskip

We are now in position to follow the strategy outlined in
section~\ref{sec:strategy} in a number of interesting cases, which we
discuss in the following sections.

\section{Dai-Freed anomalies of some simple Lie groups}\label{sec:lie}

\subsection{$SU(2)$}\label{sec:su2}
As a warm-up, we will start with $SU(2)$. To get the AHSS to work, we
need the homology of its classifying space $BSU(2)$. This is known to
be $BSU(2)=\bH\bP^\infty$, the infinite-dimensional quaternionic
projective space (see e.g. \cite{Marathe:2010ncz}, section 5.2),
obtained as the limit of the natural inclusions
$\bH\bP^n\to\bH\bP^{n+1}$. The homology groups of this space are very
simple to obtain, we have
\begin{equation}
  H_n(\bH\bP^\infty,\bZ) = \begin{cases}
    \bZ\quad & \text{when } n \equiv 0 \text{ mod } 4\, ,\\
    0 \quad & \text{otherwise}\, .
  \end{cases}
\end{equation}

We also need a way of computing $H_p(\bH\bP^\infty,\Omega_q^\Spin)$
out of knowledge of $H_n(\bH\bP^\infty,\bZ)$ and
$\Omega_q^\Spin$. This is a task for the universal coefficient
theorem, which in its homological version implies (see
theorem~3A.3 in \cite{Hatcher:478079}) that there is a short exact
sequence
\begin{equation}
  \label{eq:UCF-SU(2)}
  0\to H_n(\bH\bR^\infty,\bZ)\otimes \Omega^{\Spin}_q \to
  H_n(\bH\bP^\infty, \Omega^\Spin_q) \to \Tor(H_{n-1}(\bH\bP^\infty,
  \bZ), \Omega_q^\Spin) \to 0\, .
\end{equation}
Since $H_n(\bH\bP^\infty,\bZ)$ is free, we have that
$\Tor(H_{n-1}(\bH\bP^\infty, \bZ), \Omega_q^\Spin))=0$, and thus
\begin{equation}
  H_n(\bH\bP^\infty, \Omega^\Spin_q) \cong
  H_n(\bH\bR^\infty,\bZ)\otimes \Omega^{\Spin}_q =
  \begin{cases}
    \Omega^{\Spin}_q\quad & \text{when } n \equiv 0 \text{ mod } 4\, ,\\
    0 \quad & \text{otherwise}\, .
  \end{cases}
\end{equation}

\begin{figure}
  \centering
  \includegraphics{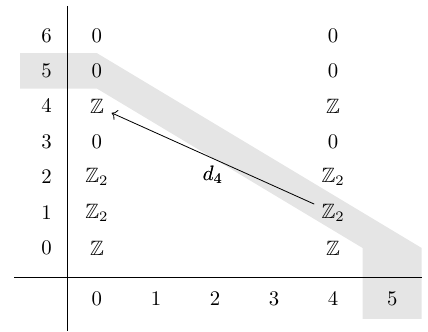}
  \caption{$E_4$ page of the AHSS for $\Omega^\Spin_*(BSU(2))$. We
    have shaded the entries of total degree 5, and indicated
    explicitly the only potentially non-vanishing differential acting
    on the shaded region.}
  \label{fig:AHSS-SU(2)}
\end{figure}

We have now the necessary information for constructing the AHSS.  It
is clear from the fact that the differential $d_r$ has bi-degree
$(-r,r-1)$, that $E_4=E_3=E_2$. More generally, it is only
differentials of the form $d_{4k}$ that can vanish.

We show this fourth page in figure~\ref{fig:AHSS-SU(2)}. There is a
priori a nonvanishing differential $d_4\colon \bZ_2\to \bZ$, but since
it is a homomorphism we necessarily have $d_4=0$. This shows that
$E_{4,1}^{\infty}=E_{4,1}^2=\bZ_2$. Since all the other elements with total degree 5
vanish already in $E_2$, we conclude that
\begin{equation}
  \label{eq:Omega_5^Spin(BSU(2))}
  \Omega_5^\Spin(BSU(2)) = \bZ_2\, .
\end{equation}
A bordism invariant that we can construct in this case, since $SU(2)$
has no local anomalies, is the $\eta$ invariant, or equivalently (in
this case) the mod-2 index. A simple example with non-vanishing mod-2
index was constructed in \cite{Witten:1982fp}. While $S^5$ itself is
trivial in $\Omega^\Spin_5$ (necessarily so, since
$\Omega^\Spin_5=0$), there is a bundle over it such that the total
space is no longer null-bordant in
$\Omega^\Spin_5(BSU(2))=\bZ_2$. What~\eqref{eq:Omega_5^Spin(BSU(2))} shows
is that the four dimensional theory of a Weyl fermion on the
fundamental of $SU(2)$ has no further gauge anomalies on any $\Spin$
manifold (the calculation in \cite{Witten:1982fp} shows absence of
anomalies in $S^4$). This was to be expected: since a Weyl fermion in
the fundamental of $SU(2)$ is in a real representation of the full
(Lorentz plus gauge) symmetry group, it has at most a $\mathbb{Z}_2$
anomaly.\footnote{One could argue similarly for some of the cases
  discussed below. For instance, some of the groups we analyze only
  have real representations, so no anomaly can arise from four
  dimensional fermions even if the bordism group happened to be
  non-vanishing.}

It is trivial to repeat the argument for other (low enough)
dimensions,\footnote{The reason we stop at degree 8 is that in page 8
  we encounter a new, potentially non-vanishing differential
  $d_8\colon E_{8,2}^8\to E_{0,9}^8$. This needs to be determined by
  other methods, since $E_{8,2}^8=\bZ_2$ and
  $E_{0,9}^8=\bZ_2\oplus\bZ_2$, so the differential is not necessarily
  vanishing a priori. This affects the computation of
  $\Omega_9^\Spin(BSU(2))$ and $\Omega_{10}^\Spin(BSU(2))$. One way of
  dealing with this differential is to use the Atiyah-Hirzebruch
  spectral sequence for reduced bordism (see
  appendix~\ref{app:reduced-bordism} and remark 2 in pg.~351 of
  \cite{Switzer}) which for our case reads
  \begin{equation}
    E_{p,q}^2 = \widetilde{H}_p(X, \Omega_q(\pt)) \Rightarrow \widetilde{\Omega}_{p+q}(X)\, .
  \end{equation}
  So in particular $E_{0,q}^2 = E_{0,q}^\infty=0$, and we learn that
  that potentially problematic differential $d_8$ vanishes.} we find
\begin{equation}
  \def\arraystretch{1.5}
  \arraycolsep=4pt
  \begin{array}{c|ccccccccc}
    n & 0 & 1 & 2 & 3 & 4 & 5 & 6 & 7 & 8 \\
    \hline
    \Omega^\Spin_n(BSU(2)) & \bZ & \bZ_2 & \bZ_2 & 0 & 2\bZ & \bZ_2
                              & \bZ_2 & 0 & 4\bZ
  \end{array}
\end{equation}
The only non-trivial case here is that of $\Omega_4^\Spin(BSU(2))$
(and $\Omega_8^\Spin(BSU(2))$, which works similarly). This has two
contributions, coming from $E_{4,0}^\infty=\bZ$ and
$E_{0,4}^\infty=\bZ$.

One point that we have neglected so far is that the spectral sequence
does not give us $\Omega^\Spin_k(BSU(2))$ directly, but rather an
associated graded module $\Gr_{p,q}$ \cite{McCleary}, which depends, as discussed in Subsection \ref{sec:AHSS}
in addition to the bordism group itself, on a suitable
filtration by graded submodules $F_p$. 
Spectral sequences compute $\Gr_{p,q}=E^\infty_{p,q}$. Tracing the
definitions, we find that
\begin{equation}
  F_3\Omega_4=F_2\Omega_4=F_1\Omega_4=F_0\Omega_4 = \Gr_{0,4} =
  E^\infty_{0,4} = \bZ\, .
\end{equation}
On the other hand, we have
$E^\infty_{4,0}=\Gr_{4,0}=F_4\Omega_4/F_3\Omega_4$. We are interested
in solving for $\Omega_4=F_4\Omega_4$. We can do this, formally, by
fitting the above into a short exact sequence
\begin{equation}
  0 \to \underbrace{F_3\Omega_4}_{\bZ} \to F_4\Omega_4 \to
  \underbrace{\Gr_{4,0}}_{\bZ} \to 0\, .
\end{equation}
Since $\Ext(\bZ,\bZ)=0$ \cite{Hatcher:478079}, the exact sequence necessarily
splits, and we have $\Omega_4=F_4\Omega_4=\bZ\oplus \bZ$.

\subsubsection{Physical interpretation}

Obstructions to a manifold being trivial in its Spin bordism class can
be detected by computation of certain suitable KO-theory classes
\cite{ABP}. This is a fancy way of saying that there is some (perhaps
mod-2) index that can detect the non-triviality of the manifold. For
instance, on an $S^1$, with the periodic structure, the mod-2 index is
non-vanishing, and similarly for the $T^2$ with completely periodic
structure (see pg.~45 of \cite{Witten:2015aba}). In these low
dimensions there is no topologically nontrivial $SU(2)$ bundle, so
what we are seeing is the fact that
$\Omega_1^\Spin(BSU(2))=\Omega_1^\Spin(\pt)$. (More formally, this
comes from the fact that every $p$-cycle is contractible in
$BSU(2)=\bH\bP^\infty$ for $p<4$.)

The $\bZ_2$ values in 5 and 6 dimensions encode global anomalies in
$SU(2)$ theories in 4d with a Weyl fermion and 5d with a symplectic
Majorana fermion \cite{Intriligator:1997pq}.

In four dimensions we get an extra factor of $\bZ$ with respect to
$\Omega_4^\Spin(\pt)$. This anomaly can be associated to the global
parity anomaly of Redlich \cite{Redlich:1983kn,Redlich:1983dv}, for a
Dirac fermion in the fundamental of $SU(2)$. To see this, we need need
to construct the right bordism invariants that detect both $\bZ$
factors. We know that the invariant that detects the class in
$\Omega^\Spin_4(\pt)$ is simply the Pontryagin number. The class
detecting the extra information in $\Omega^\Spin_4(BSU(2))$ is the
index of a Weyl fermion on the manifold, which is indeed related to
the parity anomaly in 3d.

The 8d case is related to parity anomalies in 7d. The relevant
Chern-Simons terms are those associated with $p_1(T)^2$, $p_2(T)$,
$p_1(F)^2$ and $p_1(T)p_1(F)$, with $p_i$ the Pontryagin classes of
the tangent bundle $T$ and the gauge bundle $F$.

\subsubsection{Simply connected semi-simple groups up to five dimensions}
\label{sec:Hurewicz}
The structure we have just discussed for $SU(2)$ is actually very general in low enough dimension and applies to the simply connected forms of all semisimple Lie groups, as we now explain.  First, notice that, for any such $G$, $\pi_1(G)=\pi_2(G)=0$, $\pi_3(G)=\mathbb{Z}$. We can now use the result that (see \S8.6.4 of \cite{husemoller})
\begin{equation}
  \pi_{i+1}(BG) = \pi_i(G)
\end{equation}
for any group $G$ and $i\geq 0$, to compute that
\begin{equation}
  \pi_i(BG) = \{0,0,0,0,\bZ,\pi_4(G),\ldots\}
\label{pibg}\end{equation}
 Note in particular that $BG$ is
3-connected. Applying the Hurewicz theorem \cite{Hatcher:478079} we find that
\begin{equation}
  \label{eq:BSU(n)-Hurewicz}
  H_i(BSU(n),\bZ) = \{\bZ,0,0,0,\bZ, s(\pi_4(G)), \ldots\} \, .
\end{equation}
where $s(\pi_4(G))$ denotes some subgroup of $\pi_4(G)$ to be
determined.  A couple of points require explanation. First, note that
the Hurewicz isomorphism only holds for $i>0$. We used the input
\eq{pibg} to set $H_i(BG,\bZ)=\bZ$, in contrast to $\pi_0(BG)=0$. The
standard statement for the Hurewicz isomorphism in our case is that
$\pi_i(BG)=H_i(BG,\bZ)$ up to $i=4$, see for example theorem 4.37 in
\cite{Hatcher:478079}. To set $H_5(BG,\bZ)$ we have used that the
Hurewicz homomorphism is surjective for $i=5$ in a 3-connected space,
see exercise~23 in \S4.2 of \cite{Hatcher:478079}. Whenever
$\pi_4(G)=0$, as is the case for $SU(n)$, $Spin(n)$, and the
exceptional groups, we have that $H_5(BG,\bZ)=0$.

The information in~\eqref{eq:BSU(n)-Hurewicz} is enough to compute
$\Omega_k^\Spin(BG)$ up to $k=4$ via the AHSS, with results identical
to the $SU(2)$ case. For the case $\pi_4(G)=0$, we also find that the
bordism group $\Omega_5^{\Spin}(BG)$ is given by
\begin{equation}
  \Omega_5^\Spin(BG) = \coker(d_2\colon E^{(6,0)}_2 \to
  E_2^{(4,1)})\, .
\end{equation}
Luckily, this is a differential for which we have an explicit
expression, as reviewed in section~\ref{sec:steenrods}. Part of the
rest of this section will be about the explicit computation of this
differential in various interesting examples.

Finally, we should remark that the construction of the AHSS (see
e.g. \cite{Diaconescu:2000wy}) also provides a natural candidate for
the representative of $E_2^{(4,1)}=H_4(BG,\Omega_1^{\Spin})$. We need
a manifold with a $S^1$ with a spin structure that does not bound, and
with a $G$-bundle with nontrivial second Chern class, since this is
measured by $H_4(BG)$. The natural candidate is $S^4\times S^1$, with
periodic boundary conditions along the $S^1$, and a gauge instanton on
$S^4$. The question is whether or not this is trivial in spin bordism,
which we now address in a number of examples.

If all one is interested in is the anomaly on four dimensional $\Spin$
manifolds there is a shortcut based on the previous observation: one
can detect the anomaly in the original four dimensional theory by
reducing along an $S^4$ with an instanton bundle, and seeing whether
the effective zero-dimensional theory is anomalous, as done for
instance in \cite{Garcia-Etxebarria:2017crf}.\footnote{In terms of the
  Dai-Freed viewpoint, in using compactification to detect the anomaly
  we are using the fact that
  $\eta(S^4\times S^1)=\ind(S^4)\cdot \eta(S^1)$, see Lemma 2.2 of
  \cite{bahri1987}.}

A second shortcut exists for simply connected groups in five
dimensions: say that we have a group $G$ with subgroup $H$, and we
want to understand whether we can deform any $G$ bundle over a base
$X$ to a $H$ bundle over $X$. If we can, and assuming that the $G$
theory is free of local anomalies, then we can compute the $\eta$
invariant from knowledge of the $\eta$ invariant of the $H$ theory. As
reviewed in \cite{Witten:1985bt,Diaconescu:2000wy}, for instance, the
reduction is in fact possible if $\pi_i(G/H)=0$ for all
$i<\dim(X)$. Take $H=SU(2)$, where we have already understood what
happens. One has $SU(n+1)/SU(n)=S^{2n+1}$, and in particular
$SU(3)/SU(2)=S^5$. This implies that in five dimensions any $SU(3)$
bundle can be reduced to an $SU(2)$ bundle, since $\pi_i(S^5)=0$ for
$i<5$. Similarly, by studying higher values of $n$, one can show that
every $SU(n)$ bundle can be reduced to an $SU(2)$ bundle. It is not
difficult to extend this result to the other simply connected Lie
groups, which effectively reduces the problem of computing anomalies
in these cases to a group theory analysis.

While these techniques (and related ones) often lead to an economic
derivation in specific cases, we have opted to proceed by computing of
the bordism groups using the Atiyah-Hirzebruch spectral sequence,
since it is a viewpoint that straightforwardly applies to other
situations of interest that do not admit the shortcuts above.

\subsection{$\Sp(2k)$}

\begin{figure}
  \centering
  \includegraphics{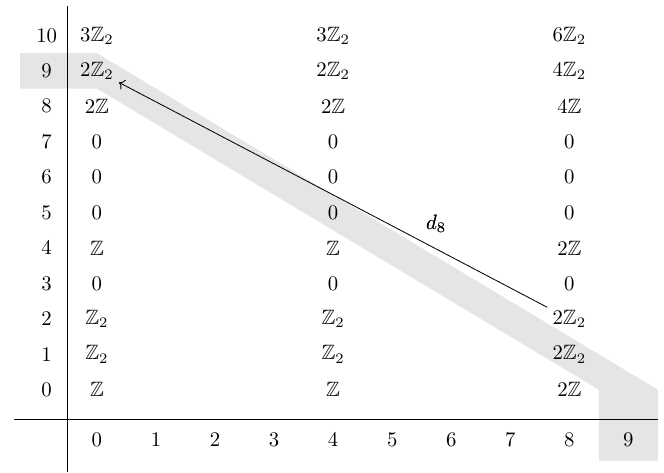}
  \caption{$E_8$ page of the AHSS for $\Omega^\Spin_*(B\Sp(2k))$ with
    $k>1$. We have shaded the entries of total degree 9, and indicated
    explicitly the only potentially non-vanishing differential acting
    on the shaded region.}
  \label{fig:AHSS-Sp(2k)}
\end{figure}

The $\Sp(2k)$ case is very similar to $\Sp(2)=SU(2)$, so we will be
brief. The classifying space $B\Sp(2k)$ is given by the infinite
quaternionic Grassmanian, we refer the reader to \cite{FuchsViro} for
details of the homology of this space. The relevant AHSS is shown in
figure~\ref{fig:AHSS-Sp(2k)}, where we have shown specifically the
$\Sp(2k)$ case with $k>1$.

From figure \ref{fig:AHSS-Sp(2k)}, it is straightforward to see that
$\Omega_5^{\Spin}(\Sp(2k))=\mathbb{Z}_2$, just like in the $SU(2)$
case. Indeed, this $\mathbb{Z}_2$ is related to a global anomaly in
four dimensions, coming from the fact that $\pi_4(\Sp(2k))\neq0$ as in
the ordinary Witten anomaly. Just as in this case, the anomaly can be
probed by a mod 2 index.

The first difference between $SU(2)$ and $\Sp(2k)$ with $k>1$ appears
in eight dimensions, and it is due to the fact that while $SU(2)$
bundles are classified by $p_1^2(F)$, $\Sp(2k)$ bundles with $k>1$ are
classified by two independent quantities: $p_1^2(F)$ and
$p_2(F)$. More formally
\begin{equation}
  H_8(B\Sp(2k),\bZ) = H^8(B\Sp(2k),\bZ) = \begin{cases} \bZ & \quad
    \text{if } k=1\, ,\\
    \bZ\oplus\bZ & \quad \text{if } k>1\, .
    \end{cases}
\end{equation}
This leads to a qualitative difference between the $k=1$ and $k>1$
cases when it comes to eight-dimensional anomalies. Consider for
example a fermion in the adjoint representation. It was shown in
\cite{Garcia-Etxebarria:2017crf} that $k>1$ had an anomaly on
spacetimes of non-trivial topology (the example analyzed there was
that of spacetimes with a $S^4$ factor, and a unit of instanton flux
on this factor, but the conclusion is clearly more general), while
$k=1$ did not have this anomaly.

\subsection{$U(1)$}

\label{sec:BU(1)-Spin}

\begin{figure}
  \centering
\includegraphics{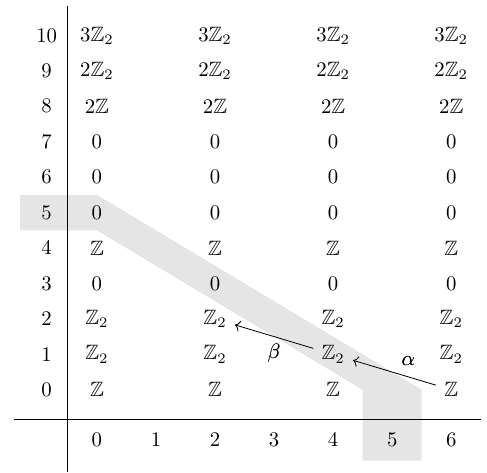}
  \caption{$E_2$ page of the AHSS for $\Omega^\Spin_*(BU(1))$. We have
    shaded the entries of total degree 5, and indicated explicitly the
    only potentially non-vanishing differential acting on the shaded
    region.}
  \label{fig:AHSS-U(1)}
\end{figure}

Let us consider now the computation of $\Omega_*^\Spin(BU(1))$. This
is the first case in which we will encounter non-vanishing
differentials in the spectral sequence for the entries of
interest. Recall that $BU(1)=K(Z,2)=\bC\bP^\infty$, so the relevant
homology groups are well known:
\begin{equation}
  H_i(BU(1), \bZ) = \begin{cases}
    \bZ \quad \text{if } i\in2\bZ\, ,\\
    0 \quad \text{otherwise.}
  \end{cases}
\end{equation}
From here, we obtain the AHSS shown in figure~\ref{fig:AHSS-U(1)}.

We see that there are two potentially non-vanishing differentials,
both on the second page, $\alpha\colon E^2_{(6,0)}\to E^2_{(4,1)}$ and
$\beta\colon E^2_{(4,1)}\to E^2_{(2,2)}$.

Let us start with $\alpha$. As reviewed in
section~\ref{sec:steenrods}, from~\cite{TeichnerPhD,Teichner} we have
that this differential is given by the composition of reduction modulo
two and the dual of the Steenrod square
\begin{equation}
  \Sq2 \colon H^4(BU(1), \bZ_2) \to H^6(BU(1), \bZ_2)\, .
\end{equation}
Recall that $H^i(BU(1),\bZ)=\bZ[x]$, with $x$ of degree two, so
analogously (by the universal coefficient theorem in cohomology)
$H^i(BU(1),\bZ_2)=\bZ_2[x]$. Now, since $x$ is of degree 2, we have
\begin{equation}
  \label{eq:U(1)-Sq2(x)}
  \Sq2(x) = x^2
\end{equation}
and for degree reasons $\Sq1(x)=0$. From here, using Cartan's formula,
we find that
\begin{equation}
  \Sq2(x^2) = \Sq0(x) \smile \Sq2(x) + \Sq2(x) \smile \Sq0(x) = 2x^2 = 0\, .
\end{equation}
This implies that the dual Steenrod square also vanishes, and we
conclude that
\begin{equation}
  \alpha = \dSq2 \circ r_2 = 0\, .
\end{equation}

We can deal with the $\beta$ differential similarly. According to
\cite{TeichnerPhD,Teichner} we have
$\beta=\dSq2$. Using~\eqref{eq:U(1)-Sq2(x)} we immediately see that
$\dSq2$ maps the generator of $H_4(BU(1),\bZ_2)$ to the generator of
$H_2(BU(1),\bZ_2)$, so we immediately conclude that
\begin{equation}
  \Omega_5^\Spin(BU(1)) = 0\, .
\end{equation}

Similar arguments can be repeated for lower degrees, with the result
\begin{equation}
  \def\arraystretch{1.5}
  \arraycolsep=4pt
  \begin{array}{c|cccccc}
    n & 0 & 1 & 2 & 3 & 4 & 5\\
    \hline
    \Omega_n^\Spin(BU(1)) & \bZ & \bZ_2 & \bZ_2 \oplus \bZ & 0 & \bZ\oplus\bZ & 0
  \end{array}
\end{equation}
The obvious interpretation of these results is that the $U(1)$ flux
adds the natural obstruction, on top of that coming from
$\Omega_*^\Spin(\pt)$.

\subsection{$SU(n)$ and implications for the Standard Model}\label{sec:sm}

Let us now compute $\Omega^\Spin_*(BSU(n))$. The classifying space of
$SU(n)$ is well known to be the infinite Grassmanian of $n$-planes in
$\bC^\infty$. The integer cohomology ring of this space is very well
known \cite{borel1955,Fung} to be the polynomial ring
\begin{equation} H^*(BSU(n),\bZ)=\mathbb{Z}[c_2,c_3\ldots c_n].\end{equation}
The generators are the Chern classes; indeed, for a $SU(n)$-bundle over a space $X$ defined by a map $f\colon X\rightarrow BG$, the Chern classes of the bundle are the pullbacks $f^*(c_i)$. 

The universal coefficient theorem for cohomology \cite{Hatcher:478079} provides a short exact sequence relating the homology groups $H_i(X,\mathbb{Z})$ with the cohomology groups $H^i(X,\mathbb{Z})$:
\begin{equation}
  \begin{tikzcd}0\arrow{r}&\text{Ext}^1(H_{i-1}(X,\mathbb{Z}),\mathbb{Z})\arrow{r}&H^i(X,\mathbb{Z})\arrow{r}&\text{Hom}(H_i(X,\mathbb{Z}),\mathbb{Z}))\arrow{r}&0.\end{tikzcd}\label{uct}\end{equation}
If the homology groups are finitely generated, the $\Ext$ term is just the torsion part of $H_{i-1}(X,\bZ)$, and the Hom is the free part of $H_{i}(X,\bZ)$. 

\begin{figure}
  \centering
\includegraphics{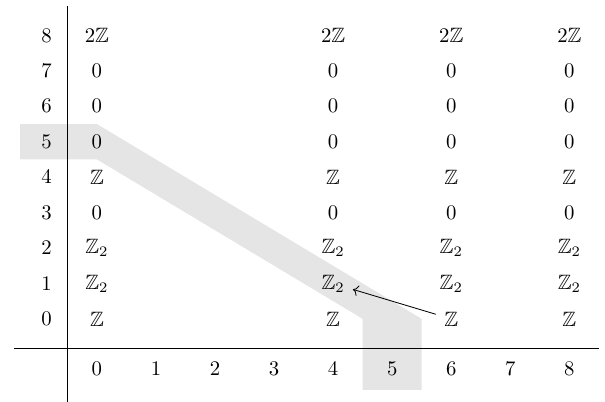}
  \caption{$E_2$ page of the AHSS for $\Omega^{\Spin}_*(BSU(n))$. We have shaded the entries contributing to the computation of $\Omega_5^{\Spin}(BSU(n))$, and indicated the only relevant differential.}
  \label{fig:AHSS-SU(n)}
\end{figure}

If $H^i(X,\bZ)=0$ for $i$ odd and there is no torsion in cohomology, such as for $BSU(n)$, we get $H_i(X,\bZ)=H^i(X,\bZ)$, with the resulting AHSS shown in figure \ref{fig:AHSS-SU(n)}.

We are now in a position to compute the differential $d_2$ in figure \ref{fig:AHSS-SU(n)}. As discussed in section~\ref{sec:steenrods}, we need to reduce modulo 2 and compose with the dual of the Steenrod square. Reduction mod 2 is the induced map in homology $H_6(BSU(n),\mathbb{Z})=\mathbb{Z}\rightarrow H_6(BSU(n),\mathbb{Z})=\mathbb{Z}_2$ from the short exact sequence
\begin{equation}
  \begin{tikzcd}0\arrow{r}&\mathbb{Z}\arrow{r}&
    \mathbb{Z}\arrow{r}&\mathbb{Z}_2\arrow{r}&0\end{tikzcd}\, .
\end{equation}
Since there is no torsion in $H_i(BSU(n),\mathbb{Z})$, the map is an
isomorphism. Since $H_6(X,\mathbb{Z}_2)$, $H_4(X,\mathbb{Z}_2)$,
$H^6(X,\mathbb{Z}_2)$ and $H^4(X,\mathbb{Z}_2)$ are all
$\mathbb{Z}_2$, $\dSq2$ will be nontrivial if and only if $\Sq2$
is. The Stenrood square operations for $BU(n)$ are computed in
\cite{10.2307/2372495}; from the remark at the start of \S12 of that
paper, together with the relationship $P_2^k=\Sq{2k}$, we obtain
\begin{equation}
  \Sq2(c_2)=c_1\smile c_2+c_3\, ,
\end{equation}
where $c_1,c_2$ are the degree two and four generators of the
cohomology ring $H_*(BU(n),\mathbb{Z}_2)$ (given by the mod 2
reduction of the generators of $H_*(BU(n),\mathbb{Z})$, the Chern
classes). The projection $BSU(n)\rightarrow BU(n)$ gives a pullback
map from $H_*(BU(n),\mathbb{Z}_2)$ to $H_*(BSU(n),\mathbb{Z}_2)$ which
sends $c_1$ to 0 and $c_2$ to the degree four generator.

As a result, $\Sq2(c_2)=c_3$, the mod 2 reduction of the third Chern
class. For $n=2$, $c_3$ vanishes identically, so the differential
vanishes in accordance with previous results. On the other hand, for
$n>2$, the map sends the generator of $H^4(BSU(n),\mathbb{Z}_2)$ to
the generator of $H^6(BSU(n),\mathbb{Z}_2)$. This means that
$\dSq2$ is the identity, so the differential kills the
$\mathbb{Z}_2$ factor. As a result,
\begin{equation}
  \Omega_5^{\Spin}(BSU(n))=0,\quad\text{for}\quad n>2\, .
  \label{boom}
\end{equation}
The result \eq{boom} is of great physical relevance. It means that the
$SU(5)$ GUT is free of Dai-Freed anomalies and therefore defines a
consistent quantum theory in any background, of any topology. But it
also implies that \emph{the Standard Model is also free of Dai-Freed
  anomalies, whatever the global form of the gauge group may be}.

To see this, recall that experiments have only probed the Lie algebra of the SM so far; there are various possibilities for the global structure. For a nice recent discussion, see \cite{Tong:2017oea}. In short, the SM gauge group is
\begin{equation}G_{SM}=\frac{SU(3)\times SU(2)\times
    U(1)}{\Gamma},\quad
  \Gamma\in\{1,\mathbb{Z}_2,\mathbb{Z}_3,\mathbb{Z}_6\}.\end{equation}
Different choices of $\Gamma$ affect quantization of monopole charges,
and also the allowed bundles when considering the theory on an
arbitrary (spin) 4-manifold. It is then conceivable that some choices
of $\Gamma$ are free of global anomalies and others are
not.\footnote{See \cite{Garcia-Etxebarria:2017crf,Monnier:2017oqd} for
  recent examples of theories that are anomalous only for specific
  choices of the global form of the gauge group.} If
$\Gamma_1\subset \Gamma_2$, all bundles for $\Gamma=\Gamma_1$ are also
bundles for $\Gamma=\Gamma_2$. In particular, the choice
$\Gamma=\Gamma_6$, is the ``potentially most anomalous'' of all.

However, this choice is also the one that embeds as a subgroup of
$SU(5)$. The SM fermions can be arranged into a representation of
$SU(5)$ which is free from local anomalies, so the Dai-Freed anomalies
of the SM can be studied just by considering Dai-Freed anomalies in a
$(SU(3)\times SU(2)\times U(1))/\mathbb{Z}_6\subset SU(5)$. But
\eq{boom} says there can be no such anomaly; hence we get the
advertised result. This was already advanced in \cite{Freed:2006mx}.

\medskip

We have shown that the $SU(5)$ GUT and the SM are anomaly free,
assuming the existence of a $\Spin$ structure. This is the simplest
possibility allowing for the existence of fermions, but it is not the
most general.  The SM breaks both $P$ and $CP$, but the $CP$ breaking
happens purely at the level of the Lagrangian -- the spectrum is
invariant under the action of $CP$ (but not $P$). One could entertain
the possibility that the $CP$ breaking in the SM is actually
spontaneous (see for example \cite{Choi:1992xp} for some early work
studying the phenomenological implications of possibility). This
theory would then make sense in unorientable spacetimes, as long as
these admit fermions. Unorientable spacetimes that admit fermions are
said to have a $\Pin$ structure (see
e.g. \cite{GilkeyBook,Witten:2015aba}). There are two possibilities,
$\Pin^+$ and $\Pin^-$.\footnote{See \cite{2001RvMaP..13..953B} for a
  discussion of potential observable differences between both
  possibilities.} We can compute the groups $\Omega^{\Pin^\pm}_5(BG)$
again via the AHSS, since we know $\Omega^{\Pin^\pm}(\text{pt})$ (see
appendix \ref{app:bgrs}). We find $\Omega^{\Pin^\pm}_5(BSU(n))=0$; we
will not reproduce the computation since the AHSS is trivial in the
$\Pin^+$ case, and very similar to the $\Spin$ case in the $\Pin^-$
case.

Another interesting question is whether the SM makes sense in
$\Spin^c$ manifolds (see e.g. \cite{GilkeyBook}). $\Spin^c$ is a
refinement of a $\Spin$ structure in which the transition functions
for the spin bundle live in $(\Spin\times U(1))/\mathbb{Z}_2$, where
the $\mathbb{Z}_2$ identifies the $\mathbb{Z}_2$ subgroup of the
$U(1)$ with the $\mathbb{Z}_2$ subgroup of $\Spin$. Every $\Spin$
manifold is $\Spin^c$, but the converse is not true; therefore, the SM
on a $\Spin^c$ manifold might in principle be anomalous.  However, we
cannot put the SM as-is in a $\Spin^c$ manifold. To have a $\Spin^c$
structure, we need to have an additional, non-anomalous $U(1)$ under
which all the fermions have odd charges. No such $U(1)$ exists in the
SM. However $U(1)_{B-L}$ satisfies these properties and, if we assume
it to be gauged, can be used to put the theory in a $\Spin^c$
manifold. We find again $\Omega_5^{\Spin^c}(BSU(5))=0$ (the relevant
AHSS entries vanish trivially; the groups $\Omega^{\Spin^c}(\pt)$ can
be found in appendix \ref{app:bgrs}).

One could consider both of the above possibilities at once, and put
the SM (plus right-handed neutrinos) on a $\Pin^c$
manifold (see appendix~\ref{app:bgrs} for the point bordism groups),
which is the refinement of $\Spin^c$ to non-orientable
spacetimes. Again $\Omega_5^{\Pin^c}(BSU(5))=0$, excluding new
anomalies in the SM.

These are more possibilities we could consider. In the presence of
certain $\bZ_4$ symmetry to be discussed in
section~\ref{sec:SMsuperconductor}, one can consider spacetimes with
$\Spin^{\bZ_4}$ structure \cite{Tachikawa:2018njr}, which do lead to a
non-trivial constraint on the spectrum of the standard model.

We have not attempted to perform a full classification of all such
possible ``twisted'' (s)pinor structures on spacetime, but it would be
clearly interesting to do so, and see if any further
phenomenologically interesting consequences can be obtained in this
way.

\subsection{$PSU(n)$}\label{sec:psun}

\begin{figure}[!ht]
  \centering
\includegraphics{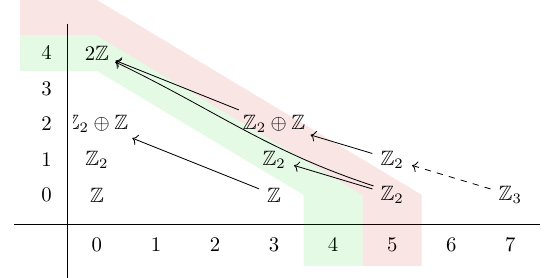}
  \caption{$E_2$ page of the AHSS for $\Omega^{\Spin}_*(PSU(n))$ associated to the fibration \eq{wfib}. We have shaded the entries with total degree four (green) and five (red).}
  \label{fig:AHSS-PSU-f1}
\end{figure}

We will now compute the bordism groups of
$PSU(n)\equiv SU(n)/\mathbb{Z}_n$. In general, we will denote by $PG$
the quotient of $G$ by its center. A direct attempt using the AHSS
associated to the fibration
$\text{pt}\rightarrow PSU(n)\rightarrow PSU(n)$ is not promising,
since there are many differentials. Instead, we will pursue an
alternate strategy, similar to the one in \cite{2016arXiv161200506G}
(the cohomology of $PSU(n)$ up to degree 10 can also be found in that
reference). Note that $PSU(n)\equiv PU(n)$, and consider the fibration
\begin{equation}
\begin{tikzcd}U(1)\arrow{r}&U(n)\arrow{r}&PSU(n).\end{tikzcd}\end{equation}
As usual, this induces a fibration of classifying spaces,
\begin{equation}
\begin{tikzcd}BU(1)\arrow{r}&BU(n)\arrow{r}&BPSU(n).\end{tikzcd}\end{equation}
We can use now the Puppe sequence \cite{2016arXiv161200506G,rotman1998introduction}, which for a fibration $F\rightarrow Y\rightarrow X$ reads
\begin{equation}
\begin{tikzcd}\ldots \arrow{r}&\Omega Y \arrow{r}& \Omega X\arrow{r}&F\arrow{r}&Y \arrow{r}& X,\end{tikzcd}\end{equation}
where $\Omega$ is a loop functor. We can act with the classifying functor $B$ and use $B\Omega X=X$ to shift the fibration to 
\begin{equation}
\begin{tikzcd}\ldots \arrow{r}& Y \arrow{r}& X\arrow{r}&BF\arrow{r}&\ldots,\end{tikzcd}\end{equation}
Since $BU(1)=K(\mathbb{Z},2)$ is an Eilenberg-MacLane space, we obtain a fibration 
\begin{equation}
\begin{tikzcd}&BU(n)\arrow{r}&BPSU(n)\arrow{r}& K(\mathbb{Z},3).\end{tikzcd}\label{wfib}\end{equation}
We will use the AHSS associated to this fibration. The homology of $K(\mathbb{Z},3)$ is computed in \cite{2013arXiv1312.5676B} to be 

\begin{equation}
  \def\arraystretch{1.5}
  \arraycolsep=4pt
  \begin{array}{c|ccccccccc}
    i & 0 & 1 & 2 & 3 & 4 & 5 & 6 & 7 & 8  \\
    \hline
    H_i(K(\mathbb{Z},3),\mathbb{Z}) & \mathbb{Z} & 0 & 0 & \mathbb{Z} & 0 & \mathbb{Z}_2  &0  & \mathbb{Z}_3  & \mathbb{Z}_2  \\
     H_i(K(\mathbb{Z},3),\mathbb{Z}_2) & \mathbb{Z}_2 & 0 & 0 & \mathbb{Z}_2 & 0 & \mathbb{Z}_2  &\mathbb{Z}_2  & 0  & \mathbb{Z}_2  
  \end{array}
\end{equation}
From this, we can construct the spectral sequence depicted in figure~\ref{fig:AHSS-PSU-f1}.

We can repeat the above procedure with the fibration
\begin{equation}
\begin{tikzcd}\mathbb{Z}_n\arrow{r}&SU(n)\arrow{r}&PSU(n).\end{tikzcd}\end{equation}
 Since $B\mathbb{Z}_n=K(\mathbb{Z}_n,1)$, proceeding as above we obtain a fibration 
\begin{equation}
\begin{tikzcd}&BSU(n)\arrow{r}&BPSU(n)\arrow{r}& K(\mathbb{Z}_n,2).\end{tikzcd}\label{wfib2}\end{equation}
Computing the homology of $K(\mathbb{Z}_n,2)$ is more laborious. Although a general algorithm to compute these in principle can be found in \cite{MR0087935}, we will only discuss the cases $n=p^k$, for $p$ prime. The main tool we will use is the following theorem\footnote{We thank Alain Cl\'{e}ment Pavon for pointing out this result to us.} (see \cite{POINTETTISCHLER19971113,TischlerPhD}) that gives $H_i(K(\mathbb{Z}_{p^k},2),\mathbb{Z})$ as follows:
\begin{align}H_i(K(\mathbb{Z}_{p^k},2),\mathbb{Z})&=M_1\oplus M_2, \quad\text{where}\nonumber\\
M_1&=\begin{cases}0&\text{if} \quad i \in 2\mathbb{Z}+1,\\\mathbb{Z}_{p^{f+s}}&\text{if}\quad i \in 2\mathbb{Z}\quad\text{and}\quad\frac{i}{2}=rp^s,\end{cases}\end{align}
where $p$ does not divide $r$. $M_2$ is a finite group whose exponent is bounded above by $S(i)$, where
\begin{align}S(i)&=\prod_{q\in\mathcal{P}(i)} q^{\varphi(q,i)}, \quad \mathcal{P}(i)=\left\{q\quad\text{prime s.t.}\quad q\leq\frac{i}{2}\right\},\nonumber\\ \varphi(q,i)&=\text{max}\left\{1,\left\lfloor\log_q\frac{i}{2q}\right\rfloor+1\right\}.
\end{align}
Using these results, we can compute the homology groups $H_i(K(\mathbb{Z}_{p^k},2),\mathbb{Z})$:
\begin{equation}
  \def\arraystretch{1.5}
  \arraycolsep=4pt
  \begin{array}{c|c|c|c|c|c|c|c|c}
    i & 0 & 1 & 2 & 3 & 4 & 5 & 6 & 7 \\
    \hline
    H_i & \mathbb{Z} & 0 & \mathbb{Z}_{2^k} & 0 & A_{p^k}\oplus\begin{cases}\mathbb{Z}_{2^{k+1}}&p=2,\\\mathbb{Z}_{p^k}&p\neq2\end{cases} & B_{p^k}  &C_{p^k}\oplus\begin{cases}\mathbb{Z}_{3^{k+1}}&p=3,\\\mathbb{Z}_{p^k}&p\neq2\end{cases}  & D_{p^k}  
  \end{array}
\end{equation}
Here, $A$ and $B$ are groups of exponent $\leq2$; this means that they are of the form $h\mathbb{Z}_2$, for some integer $h$, and $C$ and $D$ have exponent $\leq6$, meaning that all the elements have degree $\leq6$. 

We will now discuss the case at prime 2 and higher primes separately:

\subsubsection{$p=2$}
In this case, we can  use the computer program described in \cite{ClementPhD}\footnote{An updated version can be found in \url{https://github.com/aclemen1/EMM}.} to compute $A,B,C,D$ explictly. To get the homology with $\mathbb{Z}_2$ coefficients, we use the universal coefficient theorem. This produces some extensions of the form $e(\mathbb{Z}_2,\mathbb{Z}_2)$, which we know to be trivial since homology groups with coefficients in a ring $R$ must be $R$-modules (and $\mathbb{Z}_4$ is not a $\mathbb{Z}_2$-module). We obtain

\begin{equation}
  \def\arraystretch{1.5}
  \arraycolsep=4pt
  \begin{array}{c|ccccccccc}
    i & 0 & 1 & 2 & 3 & 4 & 5 & 6 & 7 & 8  \\
    \hline
    H_i(K(\mathbb{Z}_{2^k},2),\mathbb{Z}) & \mathbb{Z} & 0 & \mathbb{Z}_{2^k} & 0 & \mathbb{Z}_{2^{k+1}} & \mathbb{Z}_2  &\mathbb{Z}_{2^k}  & \mathbb{Z}_2  & \mathbb{Z}_2\oplus\mathbb{Z}_{2^{k+2}}  \\
     H_i(K(\mathbb{Z}_{2^k},2),\mathbb{Z}_2) & \mathbb{Z}_2 & 0 & \mathbb{Z}_2 & \mathbb{Z}_2 & \mathbb{Z}_2 & 2\mathbb{Z}_2  &2\mathbb{Z}_2  & 2\mathbb{Z}_2  & 3\mathbb{Z}_2  
  \end{array}
\end{equation}

\begin{figure}[!ht]
  \centering
\includegraphics{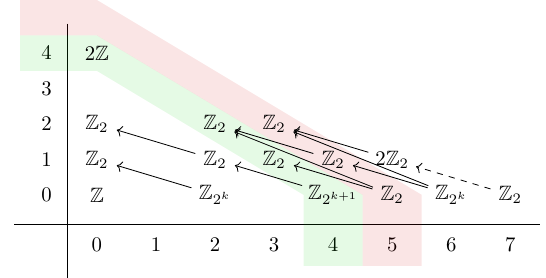}
  \caption{$E_2$ page of the AHSS for $\Omega^{\Spin}_*(PSU(n))$ associated to the fibration \eq{wfib2}, where $n=2^k$.}
  \label{fig:AHSS-PSU-f2}
\end{figure}

From this, we can construct the spectral sequence depicted in figure \ref{fig:AHSS-PSU-f2}.

Requiring the results of the two spectral sequences in figures \ref{fig:AHSS-PSU-f1} and \ref{fig:AHSS-PSU-f2} to be compatible, we can compute the relevant bordism groups up to third degree:
\begin{equation}
  \def\arraystretch{1.5}
  \arraycolsep=4pt
  \begin{array}{c|cccc}
    i & 0 & 1 & 2 & 3   \\
    \hline
    \Omega^{\Spin}_i(PSU(2^k)) & \mathbb{Z} & \mathbb{Z}_2 & \mathbb{Z}_2\oplus \mathbb{Z}_{2^k} & 0
  \end{array}
\end{equation}
Unknown differentials prevent us from proceeding any further. Note that in this case, we cannot use the result described around \eq{diff2}, since we are using the AHSS for a nontrivial fibration.

\subsubsection{$p\neq2$}
\begin{figure}[!ht]
  \centering
\includegraphics{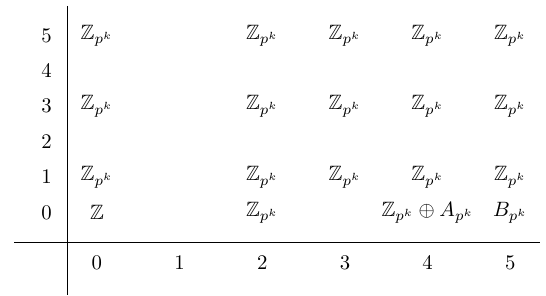}
  \caption{$E_2$ page of the Serre spectral sequence associated to the fibration \eq{fibcon}. }
  \label{fig:fibcon}
\end{figure}

In this case, we can also determine the groups $A,B,C,D$, using Serre's spectral sequence for the fibration \cite{Hatcher:478079}
\begin{equation}\begin{tikzcd}K(G,1)\arrow{r}&*\arrow{r}&K(G,2),\end{tikzcd}\label{fibcon}\end{equation}
where $*$ is a contractible space. As we know (see appendix \ref{sec:AHSSZn}), the reduced integer homology of $K(\mathbb{Z}_n,1)=B\mathbb{Z}_n$ localizes at odd degree, where it is $\mathbb{Z}_n$. In fact, direct application of the universal coefficient theorem tells us that, in the range $i\leq5$ and for odd $p$, $H_i(K(\mathbb{Z}_{p^k},2),\mathbb{Z}_n)=\mathbb{Z}_n$ with the sole exception of $i=1$, which vanishes. As a result, in the AHSS associated to the fibration \eq{fibcon}, depicted in figure \ref{fig:fibcon}, there can be no nonvanishing differentials acting on $A,B$ for $p\neq2$, and the same holds for $C,D$ for $p\neq 2,3$. Since the resulting space is contractible, we can conclude that $A=B=0$ for $p\neq2$ and $C=D=0$ for $p\neq 2,3$.

We can now compute the homology with mod 2 coefficients, which turns out to be extremely simple:

\begin{equation}
  \def\arraystretch{1.5}
  \arraycolsep=4pt
  \begin{array}{c|cccccc}
    i & 0 & 1 & 2 & 3 & 4 & 5    \\
    \hline
    H_i(K(\mathbb{Z}_{p^k},2),\mathbb{Z}) & \mathbb{Z} & 0 & \mathbb{Z}_{p^k} & 0 & \mathbb{Z}_{p^{k}} & 0 \\
      H_i(K(\mathbb{Z}_{p^k},2),\mathbb{Z}_2) & \mathbb{Z} & 0 &0 & 0 & 0 & 0   
  \end{array}
\end{equation}

\begin{figure}[!ht]
  \centering
 \includegraphics{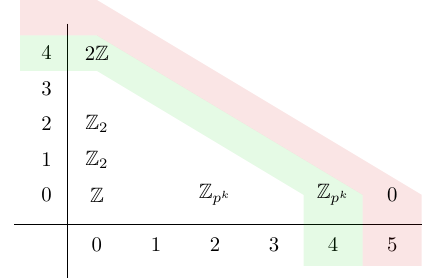}
  \caption{$E_2$ page of the AHSS for $\Omega^{\Spin}_*(PSU(n))$ associated to the fibration \eq{wfib2}, for $n=p^k$ where $p$ is an odd prime.}
  \label{fig:AHSS-PSU-f3}
\end{figure}

The AHSS associated to \eq{wfib2} is depicted in figure \ref{fig:AHSS-PSU-f3}.

Comparison with \eq{wfib} allows us to compute the bordism groups up to degree five in this case:

\begin{equation}
  \def\arraystretch{1.5}
  \arraycolsep=4pt
  \begin{array}{c|cccccc}
    i & 0 & 1 & 2 & 3 &4 & 5    \\
    \hline
    \Omega^{\Spin}_i(PSU(p^k)) & \mathbb{Z} & \mathbb{Z}_2 & \mathbb{Z}_2\oplus \mathbb{Z}_{2^k} & 0 & 2\mathbb{Z} & 0 
  \end{array}
\end{equation}

We see that there are no new anomalies in four dimensions.

\subsection{Orthogonal groups}

\subsubsection{$SO(3)$}

We now discuss $SO(n)$ groups, starting with the case $n=3$. While
$SO(3)\equiv PSU(2)$, and thus it is already covered by our discussion
in section~\ref{sec:psun} above, we will analyze it again using
different techniques as a warm-up exercise towards the case of general
$n$.

Using the results in \cite{Feshbach} for $H^*(BSO(n),\bZ)$, together
with the universal coefficient theorem, we find
\begin{equation}
  \def\arraystretch{1.5}
  \arraycolsep=4pt
  \begin{array}{c|ccccccccc}
    n & 0 & 1 & 2 & 3 & 4 & 5 & 6 & 7 & 8\\
    \hline
    H^n(BSO(3), \bZ) & \bZ & 0 & 0 & \bZ_2 & \bZ & 0 & \bZ_2 & \bZ_2 &
                                                                       \bZ\\
    H_n(BSO(3), \bZ) & \bZ & 0 & \bZ_2 & 0 & \bZ & \bZ_2 & \bZ_2 & 0 &
                                                                       \bZ\oplus\bZ_2\\
    H^n(BSO(3),\bZ_2) & \bZ_2 & 0 & \bZ_2 & \bZ_2 & \bZ_2 & \bZ_2 &
                                                                    2\bZ_2
                                  & \bZ_2 & 2\bZ_2\\
    H_n(BSO(3),\bZ_2) & \bZ_2 & 0 & \bZ_2 & \bZ_2 & \bZ_2 & \bZ_2 &
                                                                    2\bZ_2
                                  & \bZ_2 & 2\bZ_2
  \end{array}
\end{equation}

\begin{figure}
  \centering
  \includegraphics{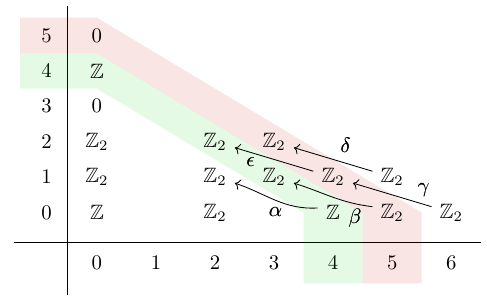}
  \caption{$E_2$ page of the AHSS for $\Omega^\Spin_*(BSO(3))$. We
    have omitted some terms which are not relevant for the computation
    of $E_\infty$ up to total degree 5, we have shaded the entries of
    total degree 4 and 5, and indicated the potentially
    non-vanishing differentials of degree 2.}
  \label{fig:AHSS-SO(3)}
\end{figure}

From here it is straightforward to write the Atiyah-Hirzebruch
spectral sequence, the result is shown in
figure~\ref{fig:AHSS-SO(3)}. We will compute the
bordism groups $\Omega_4^\Spin(BSO(3))$ and
$\Omega_5^\Spin(BSO(3))$. We see that in this range there are a number
of potentially non-vanishing differentials, so we will need extra
information to proceed. First, from \cite{TeichnerPhD,Teichner}, we
have that
\begin{align}
  d_2^{(r,0)}\colon E_2^{(r,0)} \to E_2^{(r-2,1)} & = \Sq2_*\circ
                                                    \rho_2\\
  d_2^{(r,1)}\colon E_2^{(r,1)}\to E_2^{(r-2,2)} & = \Sq2_*
\end{align}
where $\dSq2$ is the dual of $\Sq2$, and
$\rho_2\colon H_i(M,\bZ)\to H_i(M,\bZ_2)$ is reduction of coefficients
modulo 2. More precisely, it is the map induced in homology from the
exact coefficient sequence $0\to \bZ\to\bZ\to \bZ_2\to 0$. This
induces the long exact sequence
\begin{equation}
  \label{eq:homology-LES}
  \ldots \to H_i(M,\bZ) \xrightarrow{\,\cdot 2\,} H_i(M,\bZ)
  \xrightarrow{\,\rho_2\,} H_i(M,\bZ_2) \to H_{i-1}(M,\bZ) \to \ldots
\end{equation}
For our purposes we are interested in the action of $\rho_2$ on
$H_i(BSO(3), \bZ)$ with $i\in\{4,5,6\}$. These are all generated by a
single generator $e_i$. Exactness of~\eqref{eq:homology-LES} then
immediately implies $\rho_2(e_4) = m_4$ and $\rho_2(e_5)=m_5$, where
we have denoted by $m_i$ the generators of $H_i(BSO(3),\bZ_2)$. The
last remaining case, $\rho_2(e_6)$ is more subtle, since
$H_6(BSO(3),\bZ_2)=\bZ_2\oplus\bZ_2$. All we know from exactness
of~\eqref{eq:homology-LES} is that $\rho_2$ is injective when acting
on $H_6(BSO(3),\bZ)$, but not which combination of generators it maps
to.

We now pass to the evaluation of the dual Steenrod squares
\begin{equation}\dSq2\colon H_i(M,\bZ_2)\to H_{i-2}(M,\bZ_2).\end{equation}
 Recall that these are
defined by
\begin{equation}
  \dsz{\Sq2 a}{b} = \dsz{a}{\dSq2 b}
\end{equation}
for all $a\in H^i(M,\bZ_2)$ and $b\in H_{i+2}(M,\bZ_2)$, and the
pairing $\dsz{-}{-}$ is the Kronecker pairing. Notice that this
definition makes sense since
$H^i(M,\bZ_2) = \Hom_{\bZ_2}(H_i(M,\bZ_2))$, as remarked above, so
there is a natural non-degenerate pairing.

In order to proceed, we need to know the action of the Steenrod
squares on the cohomology of $BSO(3)$. This is a classic result,
originally due to Wu \cite{WuSteenrod} (see also \S8 of \cite{Borel1953}). The
$\bZ_2$-valued cohomology of $BSO(n)$ is the finitely generated ring
on $n-1$ variables
\begin{equation}
  H^*(BSO(n), \bZ_2) = \bZ_2[w_2, \ldots, w_n]\, .
\end{equation}
The Steenrod squares act on the generators of this ring as
\begin{equation}
  \Sq{i} w_j = \sum_{t=0}^i \begin{pmatrix} j - i + t -1 \\
    t \end{pmatrix} w_{i-t} w_{j+t}
\label{wu}\end{equation}
for $i\leq j$, and 0 otherwise. For the cases at hand, this implies
\begin{equation}
  \label{eq:SO(3)-basic-squares}
  \Sq1 w_2 = w_3 \quad , \quad \Sq2 w_2 = w_2^2 \quad , \quad \Sq1 w_3 =
  0 \quad \text{and} \quad \Sq2 w_3 = w_2\smile w_3\, .
\end{equation}
Steenrod squares of products of $w_i$ can then be derived via the
Cartan formula~\eqref{eq:Cartan-formula}.

Let us now finally determine the relevant differentials. We start with
$\alpha$. The relevant Steenrod square in cohomology is
\begin{equation}
  \alpha_* \colon H^2(BSO(3), \bZ_2) \to H^4(BSO(3), \bZ_2)
\end{equation}
and since $w_2$ generates $H^2(BSO(3), \bZ_2)$ this gives
$\alpha_*(w_2)=\Sq2(w_2)= w_2^2$. Since $w_2^2$ generates
$H^4(BSO(3), \bZ_2)$ we conclude that the dual map
\begin{equation}
  \dSq2 \colon H_4(BSO(3), \bZ_2) \to H_2(BSO(3), \bZ_2)
\end{equation}
is the nontrivial one, sending the generator to the generator. As
argued above, $\rho_2$ acts non-trivially on $H_4(BSO(3), \bZ)$, so we
find that $\alpha$ itself is non-trivial. A very similar argument
gives that $\beta$ is non-trivial, since $\Sq2$ maps the generator
$w_3$ of $H^3(BSO(3), \bZ_2)$ to the generator $w_2w_3$ of
$H^5(BSO(3), \bZ_2)$, and $\rho_2$ acts non-trivially on
$H_5(BSO(3), \bZ_2)$.

We now proceed to the differential $\epsilon$. The structure is very
analogous to $\alpha$, except for the fact that we do not need to
reduce coefficients. We conclude that it is non-vanishing, since
$\dSq2$ acts non-trivially on the relevant homology groups. Notice
that since $\epsilon$ is injective, we find (since
$\epsilon\circ\gamma = 0$) that $\gamma$ vanishes. We can obtain in
this way some information about $\rho_2$ acting on $H_6(BSO(3),
\bZ)$. We have
\begin{equation}
  \Sq2 (w_2^2) = 2\Sq2 w_2 \smile w_2 + \Sq1 w_2 \smile \Sq1 w_2 = w_3^2
\end{equation}
which is one of the generators of $H^6(BSO(3), \bZ_2)$, the other
being $w_2^3$. From here we learn that the dual Steenrod square is
given by
\begin{equation}
  \label{eq:dSq2-H6}
  \dSq2 \omega_3^2 = \omega_2^2 \quad ; \quad \dSq2 \omega_2^3 = 0 \, ,
\end{equation}
where by $\omega_i^k$ we mean the dual in homology of
$w_i^k$\footnote{This does not mean that we have a ring structure in homology, i.e. we do not have things like $\omega_i^k\neq (\omega_j)^k$. The notation is only meant to emphasize that we have a dual basis, in the sense of linear algebra.}. Since
$\gamma=\dSq2\circ\rho_2=0$, this implies that $\rho_2(m)=\omega_2^3$
or 0, with $m$ the generator of $H_6(BSO(3), \bZ)$. We have argued
above that the map is injective, so we conclude
$\rho_2(m)=\omega_2^3$.

Finally, we need to analyze the differential
$\delta\colon H_5(BSO(3), \bZ_2)\to H_3(BSO(3), \bZ_2)$. By the same
argument as for $\beta$, we conclude that this map is an isomorphism.

The end result of this discussion is that all of the $\bZ_2$ factors
of $E_2$ of total degree 4 or 5 vanish in $E_3$, and thus
\begin{equation}
\Omega^\Spin_4(BSO(3))=\bZ\oplus \bZ \quad ; \quad
\Omega_5^\Spin(BSO(3))=0 \, .
\label{resultsSO3}\end{equation}

\begin{figure}
  \centering
  \includegraphics{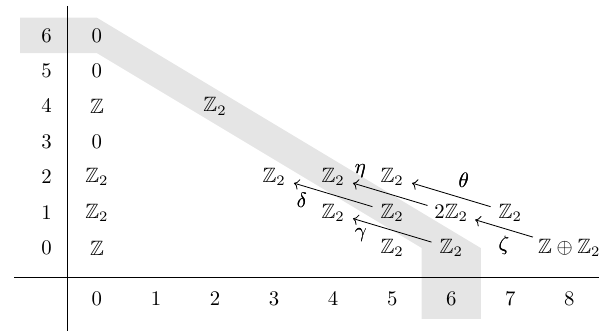}
  \caption{$E_2$ page of the AHSS for $\Omega^\Spin_*(BSO(3))$. We
    have omitted the terms which are not relevant for the computation
    of entries in $E_\infty$ of total degree 6, and we have shaded the
    entries of total degree 6.}
  \label{fig:AHSS-SO(3)-deg6}
\end{figure}

Let us also compute $\Omega_6^\Spin(BSO(3))$ via the AHSS in
figure~\ref{fig:AHSS-SO(3)-deg6}. The analysis can be performed as in
the previous case.

The $\delta$ and $\gamma$ maps have been analyzed before, with the
conclusion that $\delta$ was a bijection, and $\gamma=0$. The new maps
are $\zeta$, $\eta$ and $\theta$. Let us start with $\eta$, which is
the dual of the Steenrod square
$\Sq2\colon H^4(BSO(3), \bZ_2) \to H^6(BSO(3), \bZ_2)$. This was
computed in~\eqref{eq:dSq2-H6} above, with the result that the map is
surjective.

In order to compute $\zeta$, notice first that from the
$0\to\bZ\to\bZ\to\bZ_2\to 0$ short exact sequence, and
$H_7(BSO(3),\bZ)=0$, we obtain that
\begin{equation}
  \ldots \to H_8(BSO(3), \bZ) \xrightarrow{\,\cdot 2\,} H_8(BSO(3), \bZ)
  \xrightarrow{\,\rho_2\,} H_8(BSO(3), \bZ_2) \to 0
\end{equation}
is exact, so $\rho_2$ is surjective when acting on $H_8(BSO(3),
\bZ)$. We also need
\begin{equation}\Sq2\colon H^6(BSO(3), \bZ_2)\to H^8(BSO(3), \bZ_2).\end{equation}
The first group
is generated by $w_2^3$ and $w_3^2$, while the second is generated by
$w_2^4$ and $w_2\smile w_3^2$. Using~\eqref{eq:Cartan-formula} we find
\begin{equation}
  \begin{split}
    \Sq2 w_2^3 & = w_2\smile \Sq2 w_2^2 + \Sq1 w_2\smile \Sq1 w_2^2 + \Sq2
    w_2 \smile w_2^2 \\
    & = w_2\smile w_3^2 + w_3 \smile (2w_2 w_3) + w_2^4 \\
    & = w_2\smile w_3^2 + w_2^4\, .
  \end{split}
\end{equation}
Similarly
\begin{equation}
  \Sq2 w_3^2 = 2w_3 \Sq2 w_3 + (\Sq1 w_3)^2 = 0
\end{equation}
where we have used $\Sq1 w_3 = 0$ in $BSO(3)$. Dualizing:
\begin{equation}
  \dSq2 \omega_2^4 = \dSq2 (\omega_2\omega_3^2) = \omega_2^3
\end{equation}
using the same notation for the dual homology generators as above. As
a small check, note that $\eta\circ\zeta = 0$, as it should. (And more
precisely, $\ker \eta = \im \zeta$, so $E_3^{(6,1)}=0$.)

Finally, we need to compute $\theta\colon H_7(BSO(3), \bZ_2)\to
H_5(BSO(3), \bZ_2)$. The action of $\Sq2$ on the generator of
$H^5(BSO(3), \bZ_2)$ is easily found to be
\begin{equation}
  \begin{split}
    \Sq2 w_2w_3 & = w_2\smile \Sq2 w_3 + \Sq1 w_2 \smile \Sq1 w_3 + \Sq2
    w_2 \smile w_3 \\
    & = 2 w_2^2\smile w_3 = 0
  \end{split}
\end{equation}
using again $\Sq1 w_3 = 0$ and the basic
relations~\eqref{eq:SO(3)-basic-squares}. So the conclude $\theta=0$.

At this point we run out of technology to compute the relevant
differentials. In particular, since we find
$E_3^{(5,2)}=E_2^{(5,2)}=\bZ_2$, there is a potentially non-vanishing
differential $d_3\colon E_3^{(5,2)}\to E_3^{(2,4)}$ that we reach
before we fully stabilize. There is some discussion in
\cite{TeichnerPhD} about what these differentials are, but without
going into that, we can conclude in any case that
$\Omega_6^\Spin(BSO(3))$ is either $E_2^{(6,0)}=\bZ_2$, or (if the
differential vanishes) some extension of $\bZ_2$ by $\bZ_2$. It would
be rather interesting to characterize what this means, and whether it
signals some anomaly for the five-dimensional theory.

One observation that may be helpful here is that there is a simple
bordism invariant that characterizes $H^6(BSO(3), \bZ)=\bZ_2$. Note
that since $H^5(BSO(3), \bZ)=0$, we have an exact sequence
\begin{equation}
  0 \to H^5(BSO(3), \bZ_2) \xrightarrow{\, \beta \,} H^6(BSO(3), \bZ)
  \to \ldots
\end{equation}
We have $H^5(BSO(3), \bZ_2) = H^6(BSO(3), \bZ) = \bZ_2$, so we can
identify the generator of $H^6(BSO(3), \bZ)$ with $\beta(e)$, where
$e$ is the generator of $H^5(BSO(3), \bZ_2)$ and $\beta$ is the
Bockstein map. So a natural bordism invariant of six-manifolds is
$\dsz{\beta(e)}{M}$ where $M$ is the fundamental class of the
manifold.

\subsubsection{$SO(n)$}
\label{sec:SO(n)}

\begin{figure}[!ht]
  \centering
 \includegraphics{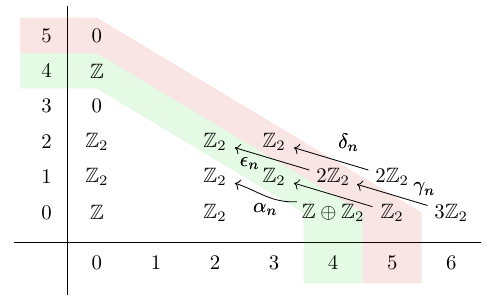}
  \caption{$E^2$ page of the AHSS for $\Omega^{\Spin}_*(BSO(n))$, for $n\geq 8$. The only differential without a label is $\beta_n$.}
  \label{fig:AHSS-SOn}
\end{figure}

We can use the above results to compute $\Omega_5^{\Spin}(BSO(n))$, for $n\geq3$ as well. The AHSS is displayed in figure \ref{fig:AHSS-SOn}, where we have also illustrated the relevant differentials.

The structure is very similar to that of figure \ref{fig:AHSS-SO(3)},
but the groups are different. $\delta_n$ and $\epsilon_n$ are again
given simply in terms of dual Steenrod squares; they are again
nonvanishing. To analyze $\alpha_n,\beta_n$ and $\gamma_n$ we need
again to know the action of $\rho$ on $H_i(BSO(n),\mathbb{Z})$ for
$i=4,5,6$. From the long exact sequence in homology, we again know
that $\rho_4$ is surjective, and sends both generators of
$\mathbb{Z}\oplus\mathbb{Z}_2$ to generators. This means that $\alpha_n$ is again nontrivial. Likewise, we know that
the image and kernel of $\rho_5$ are $\mathbb{Z}_2$, and that the
image of $\rho_6$ is $3\mathbb{Z}_2$, but we need to know the precise
action on generators. Fortunately, we can leverage our knowledge of
the $SO(3)$ case to obtain the answer for $SO(n)$ as well. To do this,
note that the inclusion $SO(3)\subset SO(n)$ induces the following
commutative diagram, where the entries are the corresponding chain
complexes,

\begin{equation}
\begin{tikzcd}0\arrow{r}&C_i(BSO(3),\mathbb{Z})\arrow{r}\arrow{d}&C_i(BSO(3),\mathbb{Z})\arrow{r}\arrow{d}&C_i(BSO(3),\mathbb{Z}_2)\arrow{d}\arrow{r}&0\\ 0\arrow{r}&C_i(BSO(n),\mathbb{Z})\arrow{r}&C_i(BSO(n),\mathbb{Z})\arrow{r}&C_i(BSO(n),\mathbb{Z}_2)\arrow{r}&0\end{tikzcd}\label{sesc}\end{equation}

which induces a commutative diagram in homology \cite{rotman1998introduction}

\begin{equation}
\begin{tikzcd}\ldots\arrow{r}&H_i(BSO(3),\mathbb{Z})\arrow{r}\arrow["\iota_*"]{d}&H_i(BSO(3),\mathbb{Z})\arrow["\rho_{SO(3)}"]{r}\arrow{d}&H_i(BSO(3),\mathbb{Z}_2)\arrow{d}\arrow["\beta'_{SO(3)}"]{r}&\ldots\\ \ldots\arrow{r}&H_i(BSO(n),\mathbb{Z})\arrow{r}&H_i(BSO(n),\mathbb{Z})\arrow["\rho_{SO(n)}"]{r}&H_i(BSO(n),\mathbb{Z}_2)\arrow["\beta'_{SO(n)}"]{r}&\ldots\end{tikzcd}\label{lesh}\end{equation}

Here, $i_*$ are the natural maps in homology induced by the inclusion. This commutative diagram in turn allows us to compute $\text{im}(\rho_{SO(n)})=\text{ker}(\beta_{SO(n)})$ by constructing $\beta'_{SO(n)}=\iota_*\circ \beta'_{SO(3)}$.

$H_5(BSO(n),\mathbb{Z}_2)$ is generated by $\xi_3\xi_2,\xi_5$, the Kronecker dual basis to $w_3w_2,w_5$, and as above $H_5(BSO(3),\mathbb{Z}_2)$ is generated by $\omega_3\omega_2$. Since in cohomology we have $\iota^*(w_3w_2)=w_3w_2$, $\iota^*(w_5)=0$ \cite{10.2307/2044298}, we obtain
\begin{equation}\iota_*(\omega_3\omega_2)=\xi_3\xi_2\label{qwqw}\end{equation}
Since in this case $\beta'_{SO(3)}=0$, the commutative diagram means that $\beta'_{SO(n)}(\xi_3\xi_2)=0$. This means that the image of the reduction modulo 2 map  is generated by $\xi_3\xi_2$, and therefore that the differential $\beta_n$ is nonvanishing.

$H_6(BSO(n),\mathbb{Z}_2)$ is generated by $\xi_2^3\xi_3^2,\xi_4\xi_2,\xi_6$, the Kronecker dual basis to the Stiefel-Whitney classes $w_2^3,w_3^2,w_2w_4,w_6$. In cohomology we have $\iota^*(w_2^3)=w_2^3$, $\iota^*(w_3^2)=w_3^2$, we have, in the same notation as above,
\begin{equation}\iota_*(\omega_2^3)=\xi_2^3,\quad \iota_*(\omega_3^2)=\xi_3^2.\end{equation}
We also have $\beta'_{SO(3)}(\omega_2^3)=0$, $\beta'_{SO(3)}(\omega_3^2)=\omega_3\omega_2$, which combined with \eq{qwqw} means that $\text{ker}(\beta_{SO(n)}')$ is generated by $\xi_2^3, \xi_6,\xi_4\xi_2$. This is also the image of the reduction modulo 2 map, so we can compute the $\gamma_n$ explicitly, to be the $\mathbb{Z}_2$ generated by $\xi_4$. 

Combining all this, we get
\begin{equation}\Omega^{\Spin}_4(BSO(n))=e(\mathbb{Z},\mathbb{Z}\oplus\mathbb{Z}_2),\quad \Omega^{\Spin}_5(BSO(n))=0.\end{equation}
Comparing with \eq{resultsSO3}, we see that we get an extra
$\mathbb{Z}_2$ factor. Presumably, this is measured by
$\int w_4$.

\subsubsection{$Spin(n)$}

We can compute the $Spin(n)$ bordism groups in the same way as above. 
First, we need the homology groups, which are (for $n\geq 8$) \cite{10.2307/2044298,10.2307/24893350,QUILLEN1971,kono1986}
\begin{equation}
  \def\arraystretch{1.5}
  \arraycolsep=4pt
  \begin{array}{c|ccccccccccc}
    n & 0 & 1 & 2 & 3 & 4 & 5 & 6 & 7 & 8 \\
    \hline
    H_i(BSpin(n),\mathbb{Z}) & \mathbb{Z} & 0 & 0 & 0 & \mathbb{Z} & 0  & \mathbb{Z}_2  & 0  &2\mathbb{Z} \\ H_i(BSpin(n),\mathbb{Z}_2) & \mathbb{Z}_2 & 0 & 0 & 0 & \mathbb{Z}_2 & 0  & \mathbb{Z}_2  & \mathbb{Z}_2  &2\mathbb{Z}_2
  \end{array}
\label{cohspinn}\end{equation}

\begin{figure}[!ht]
  \centering
  \includegraphics{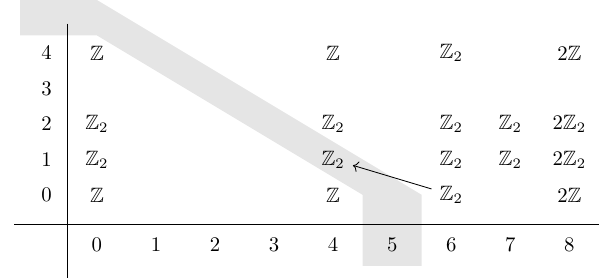}
  \caption{$E_2$ page of the AHSS for $\Omega^{\Spin}_*(BSpin(n))$, for $n\geq 8$.}
  \label{fig:AHSS-Spinn}
\end{figure}

With these we can construct the spectral sequence shown in figure
\ref{fig:AHSS-Spinn}. Since we have $H_5(BSpin(n),\mathbb{Z})=0$, the
reduction modulo 2 is an isomorphism. To compute the relevant Steenrod
square, we can use the result \cite{QUILLEN1971,2009arXiv0904.0800K}
that the cohomology with $\mathbb{Z}_2$ coefficients of $BSpin(n)$ can
be obtained from that of $BSO(n)$ via the pullback associated to the
map $f: BSpin(n)\rightarrow BSO(n)$. Now, $H_i(BSO(n),\mathbb{Z}_2)$
is a polynomial $\mathbb{Z}_2$ ring generated by the Stiefel-Whitney
classes, with $w_1=0$. The pullback map sends to zero the classes
$v_k=\Sq{2^k}\ldots \Sq1 w_2$, where $k\leq h-1$ and $h$ is the
so-called Radon-Hurwitz number, which is $\geq 9$ for $n\geq 8$.

Since $v_0=w_2$, $v_1=w_3$, the generator of
$H^4(BSpin(n),\mathbb{Z}_2)$ is just $f^*(w_4)$, and the generator of
$H^6(BSpin(n),\mathbb{Z}_2)$ is $f^*(w_6)$ . By functoriality of the
Steenrod square and Wu's formula \eq{wu},
\begin{equation}
  \Sq2(f^*w_4)=f^*(\Sq2(w_4))=f^*(w_6+w_4\smile w_2)=f^*(w_6)\, ,
\end{equation}
so the differential shown in figure \ref{fig:AHSS-Spinn} is
nontrivial. As a result, $\Omega_5^{\Spin}(BSpin(n))=0$. In
particular, this means that the $Spin(10)$ GUT is free of Dai-Freed
anomalies.

\subsection{Exceptional groups} 

We can also compute the relevant bordism groups of exceptional groups
by replacing $BG$ with a sufficiently close space which is better
understood. The familiar case is $BE_8$, which up to degree 15 has the
same homology structure as $K(\bZ,4)$
\cite{Witten:1985bt,Edwards}. Let us first review the general argument
in some detail, following \cite{Stong-11}.

Suppose we have a map $f: A\rightarrow X$. Since bordism is a generalized homology theory, we have a long exact sequence
\begin{equation}
\begin{tikzcd}\ldots\arrow{r}&\Omega_d^{\Spin}(A)\arrow{r}&\Omega_d^{\Spin}(X)\arrow{r}&\Omega_d^{\Spin}(X,A)\arrow{r}&\ldots\end{tikzcd}\label{lesgh}\end{equation}
The important point is that the relative bordism groups
$\Omega_d^{\Spin}(X,A)$ can also be computed via an AHSS with second
page $E^{2}_{p,q}=H_{p}(X,A;\Omega^{\Spin}_q(\pt))$. We will be
interested in the particular case where the induced map
$f_*:\pi_k(A)\rightarrow \pi_k(X)$ is an isomorphism for all
$k\leq n$. Then $\pi_k(X,A)=0$ for $k\leq n$, and by the relative
version of Hurewicz's theorem \cite{Hatcher:478079}, $H_k(X,A)=0$ for
$k\leq n$. The lowest corner of the AHSS is trivial, proving that
$\Omega_k^{\Spin}(X,A)$ for $k\leq n$. Then \eq{lesgh} proves that
$\Omega_d^{\Spin}(A)=\Omega_d^{\Spin}(X)$ for $d<n$, so that we may
replace $X$ by $A$ as far as low-dimensional bordisms are concerned.

Now, for any CW complex $X$, one can construct a Postnikov tower
\cite{Hatcher:478079}. This is a family of spaces $X_n$ such that
$\pi_k(X_n)=\pi_k(X)$ for $k\leq n$, $\pi_k(X_n)=0$ otherwise. There is
an inclusion $X\rightarrow X_n$ which induces a isomorphism in the
first $n$ homotopy groups. Combining with the above, we reach the
conclusion that, if we want to compute the bordism groups of some
space $X$ up to a finite degree $n$, we may replace it with the
$(n+1)$-th floor $X_{n+1}$ of the Postnikov tower.

Now consider the classifying space for $BG$, where $G$ is any exceptional group.  In fact, it is true for all exceptional groups that $\pi_4(BG)=\bZ$ and $\pi_i(BG)=0$ for $i\leq 6$. So the sixth term in the Postnikov tower for $BG$, $(BG)_6$, has homotopy groups $\pi_4((BG)_6)=\bZ$, and 0 otherwise. This means that $(BG)_6$ is by definition a presentation of the Eilenberg-MacLane space $K(\mathbb{Z},4)$\footnote{Technically, this is guaranteed by the Whitehead theorem, which \cite{Edwards,Witten:1985bt,Hatcher:478079} ensures that a continuous mapping $f: X\rightarrow Y$ between CW complexes induces isomorphisms in all homotopy groups, then $f$ is a homotopy equivalence.}.  This is turn
implies that $\Omega_i^\Spin(BG)=\Omega_i^\Spin(K(\bZ,4))$ for
$i\leq 5$. We can then immediately apply the
result in \cite{Stong-11}, and conclude

\begin{equation}
  \Omega_5^\Spin(BG) = 0
\end{equation}
for $G$ any exceptional group.

The same reasoning works for higher bordism groups whenever we have
$\pi_4(BG)=\bZ$ and $\pi_i(BG)=0$ otherwise for $i\leq d$, with $d$
large enough. For instance, for $G=E_7$ or $G=E_8$ we have
\cite{Stong-11}
\begin{equation}
  \tilde\Omega_8^\Spin(BG) =\bZ\oplus \bZ \qquad ; \qquad \tilde\Omega_9^\Spin(BG) = \bZ_2 \qquad ; \qquad \tilde\Omega_{10}^\Spin(BG) = \bZ_2\oplus \bZ_2 \, ,
\end{equation}
so for these groups we have the possibility of global anomalies in
$d=\{7,8,9\}$. (The eight dimensional case was analyzed in
\cite{Garcia-Etxebarria:2017crf}.) For $F_4$, the above results only
for $i\leq 7$, so we can only analyze anomalies up to $d=7$.

\section{Discrete symmetries and model building constraints}

\label{sec:discrete}

We now turn to anomalies of discrete symmetries. These have a long
story, see
e.g. \cite{Ibanez:1991hv,Ibanez:1992ji,Dreiner:2005rd,Mohapatra:2007vd,Lee:2011dya,Nilles:2012cy,Chen:2015aba}
among many others. Our goal will be to compute the Dai-Freed anomalies
in various cases of interest, and compare the known results. The
relevant bordism groups are nontrivial, but luckily have already been
computed in the mathematical literature in the works of Gilkey
\cite{bahri1987,GilkeyBook,Gilkey1996,Gilkey1998}, who also provides
the $\eta$ invariant for generators of the bordism groups. (Some
information about the bordism groups can also be obtained via an AHSS
sequence, as we have been doing above. However, in this case, the AHSS
is not enough to fully determine the groups, due to a nontrivial
extension problem.  Still, we have included the calculation for
$\mathbb{Z}_n$ in appendix \ref{sec:AHSSZn} for the benefit of the
curious reader.)

More concretely, we will now explore the Dai-Freed anomaly of the
so-called spherical space form groups \cite{GilkeyBook}. The main tool
we will use is the fact that there are some bordism classes for
which the $\eta$ invariants can be computed explicitly (for a
discussion, see \cite{GilkeyBook}).

A spherical space form is a generalization of a lens space, defined as
follows. Let $G$ be a finite group, and $\tau: G\rightarrow U(k)$ a
fixed-point free representation of it.\footnote{This means that the
  only matrix in the representation with unit eigenvalues is the
  identity.} Then, define
\begin{equation}
  M(\tau,G)\equiv S^{2k-1}/\tau(G).
  \label{sphericalspaceform}
\end{equation}
For $G=\mathbb{Z}_n$, this is an ordinary lens space such as the ones
employed in appendix \ref{sec:AHSSZn}.

We are naturally interested in $\Spin$ and $\Spin^c$ manifolds. For $M(\tau,G)$ to have a $\Spin$ or $\Spin^c$ structure, we just need to find a $\Spin$ or $\Spin^c$ lift of the $\tau(G)$.

For the $\Spin$ case, we have canonical spin lifts of every $\tau(G)$
up to a sign. For these to be consistent, we need that $\det(\tau)$
extends to a representation of $G$ \cite{Gilkey1996}. A particularly
simple case to ensure this is if $\tau(G)\subset SU(k)$, in which case
the determinant is 1 and $M(\tau,G)$ is always spin. As noted in
\cite{Gilkey1996}, there is no spin structure on $M(\tau,G)$ if
$\vert G\vert$ is even and $k$ odd: a finite group with even order
always has an element that squares to the identity, which in this case
has to be represented by a fixed-point-free square root of the
identity, which can only be $\text{diag}(-1,\ldots -1)$. For $k$ odd,
this has determinant $-1$.
 
The main technical result in \cite{Gilkey1996} is that $M$ represents a nontrivial class of $\Omega_{d+1}^{\Spin}(BG)$, and the $\eta$ invariant of the Dirac operator in a representation $\rho$ of $G$ is given by
\begin{equation}\eta(M(\tau,G),\rho)=\frac{1}{\vert G\vert}\sum_{\lambda \in G-\{1\}}\text{Tr}(\rho(\lambda))\frac{\sqrt{\text{det}(\tau(\lambda))} }{\text{det}(I-\tau(\lambda))}\label{etaf}.\end{equation}

For the $\Spin^c$ case, the correct expression is instead \cite{GilkeyBook,Gilkey1996}
\begin{equation}\eta(M(\tau,G),\rho)=\frac{1}{\vert G\vert}\sum_{\lambda \in G-\{1\}}\text{Tr}(\rho(\lambda))\frac{\text{det}(\tau(\lambda)) }{\text{det}(I-\tau(\lambda))}\label{etaspinc}.\end{equation}

Application of the above formulae is straightforward to a number of discrete groups of interest. 

Finally, as pointed out in section \ref{sec:daifreed}, Dai-Freed
constraints such as the ones we discuss here can sometimes be
circumvented by mild modifications, such as adding Green-Schwarz
couplings to the Lagrangian, or by coupling to a suitable topological
quantum field theory. It turns out that there are several ways of
doing this for discrete symmetries, which we discuss in subsection
\ref{sec:tgsd}.

\subsection{$\Spin-\mathbb{Z}_n$}\label{sec:spinzn}

The lens space $S^{2k-1}/\mathbb{Z}_n\equiv L^k(n)$ is not Spin for
$n$ even and $k$ odd, but it is for both $k$ and $n$ odd. As a result,
we can use \eq{etaf} to compute $\eta$ invariants corresponding
to some bordism class in $\Omega_5^{\Spin}(B\mathbb{Z}_n)$, for $n$
odd. The formula \eq{etaf} now becomes
\begin{equation}
  \eta(L^k(n),\rho_s)=\frac{1}{n}\sum_{\lambda\neq
    1}\left(\lambda^s-1\right)\left(\frac{\sqrt{\lambda}}{\lambda-1}\right)^k\,
  .\label{spinff}
\end{equation}
In this formula $s$ is the $\bZ_n$ charge of the fermion and $\lambda$
is a $n$-th root of unity. One has to be careful to define the square
root in such a way that $(\sqrt{\lambda})^n=+1$, a convenient
definition is $\sqrt{\lambda}=\lambda^{(n+1)/2}$.

As discussed in \cite{GilkeyBook}, Section 4.5.1, for odd $n$ the bordism ring $\Omega_5^{\Spin}(B\mathbb{Z}_n)$ is actually generated by only two elements, $L^3(n)$ and $\text{K3}\times L^1(n)$. This means that there are at most two independent Dai-Freed anomaly cancellation conditions. Furthermore, we have (see \cite{bahri1987}, Lemma 2.2, or \cite{APS-III,GilkeyBook} )
\begin{equation}\eta(A\times B)=\text{index}(A)\eta(B)\label{etaab}\end{equation}
which means that
\begin{equation}\eta(\text{K3}\times L^1(n))=\text{index}(\text{K3})\eta(L^1(n))=2\eta(L^1(n)).\end{equation}
So we only need to apply formula \eq{spinff}.  Using the expressions in \cite{GilkeyBook} in terms of Todd polynomials we obtain\footnote{Reference \cite{GilkeyBook} provides an alternative characterization of $\Omega_5^{\Spin}(B\mathbb{Z}_n)$, in terms of $L^3(n)$ and a generalized lens space, as well as expressions for computing their $\eta$ invariant. The computation is cumbersome, but we have checked that it agrees with \eq{eeeeee}. Details are presented in appendix \ref{sec:modular-inverse} for the benefit of the curious reader.}, after some simplifications,
 \begin{subequations}
  \begin{align}
    \sum_i \left[s_i^3 - \frac{1}{4}(n^2+3)s_i\right] & \equiv 0\mod 6n\\
    \sum_i s_i & \equiv 0 \mod n\, ,
  \label{eeeeee}\end{align}
\end{subequations}
where the $s_i$ are the $\bZ_n$ charges of the fermions in the theory.

As for the even $n$ case, \cite{Gilkey1996} provides a different family of
lens spaces which allow the computation of
$\Omega_5^{\Spin}(B\mathbb{Z}_{2^k})$. These spaces depend on two
parameters $a_1,a_2$ on top of $k$. For these, the $\eta$ invariant is
\begin{equation}\eta=2^{-k}\sum_{\lambda\neq 1}(\lambda^s-1)\frac{\lambda^{(a_1+a_2)/2}(1-\lambda^{a_1+a_2})}{(1-\lambda^{a_1})^2(1-\lambda^{a_2})^2}. \end{equation}

Since the Chinese remainder theorem means that
\begin{equation}\mathbb{Z}_{2^k m}\approx \mathbb{Z}_{2^k}\oplus \mathbb{Z}_m,\label{ancrt}\end{equation}
we can compute some $\eta$ invariants representing factors of
$\Omega_5^{\Spin}(B\mathbb{Z}_n)$ for any $n$. These are not necessarily all of the $\eta$ invariants; there might be mixed anomalies between the different factors in \eq{ancrt}.

We now apply the above anomaly cancellation conditions to some interesting cases such as $\mathbb{Z}_3$, where we obtain the constraint that the net  number of $\mathbb{Z}_3$ fermions (counted $+1$ if they have charge $1\,\text{mod}\,3$, and $-1$ if they have $2\,\text{mod}\,3$) has to vanish modulo 9,
\begin{equation}\sum_{\text{fermions}} s_i \quad\equiv\quad 0\,\text{mod}\,9,\label{mod9cons}\end{equation}
and $\mathbb{Z}_4$, where the net number of $\mathbb{Z}_4$ fermions
(counted $+1$ if they have charge $1\,\text{mod}\,4$, $-1$ if they
have $3\,\text{mod}\,4$, and $0$ otherwise) must vanish modulo 4. For
$\mathbb{Z}_5$, the net number has to vanish mod 5, where the fermions
are counted as $+1$ if they have charge 1 or 3 mod 5, -1 if they have
charge 2 or 4 mod 5, and 0 if their $\mathbb{Z}_5$ charge
vanishes. For $\mathbb{Z}_2$ the bordism group vanishes. This means,
for instance that R-parity in the MSSM is not anomalous.

On the other hand, if we have a $\mathbb{Z}_n$ bundle which can be
embedded in a $U(1)$ where local anomalies cancel, then all Dai-Freed
anomalies of the $\mathbb{Z}_n$ must vanish. This is because, as we
computed in section \ref{sec:BU(1)-Spin}, $\Omega_5^{\Spin}(BU(1))=0$,
and the $\mathbb{Z}_n$ $\eta$ invariant can also be regarded as a
$U(1)$ $\eta$ invariant, evaluated in a particular bundle whose
transition functions lie in $\mathbb{Z}_n\subset U(1)$.

\subsection{Baryon triality}\label{sec:bt}

The constraint \eq{mod9cons} has phenomenological implications, as we
will now see.  Consider the $\bZ_3$ baryon triality symmetry
\cite{Ibanez:1991pr,Dreiner:2005rd}, commonly used to ensure proton
stability in the MSSM.\footnote{Although this symmetry is typically
  introduced for phenomenological reasons in MSSM models, it can also
  be studied as a symmetry of the vanilla Standard Model.} This is a
symmetry under which the chiral superfields are charged as in
table~\ref{chbar}. The total charge mod 9, counted as above, is 3 per
generation, so we need the number of generations to be a multiple of 3
in order for baryon triality to be anomaly-free. Note that the anomaly
that we found for the $\bZ_3$ symmetry implies that baryon triality
cannot be embedded into an anomaly-free $U(1)$ as long as
generation-independent $U(1)$ charges are considered: we have just
seen that a $\bZ_3$ subgroup of the $U(1)$ is anomalous for the case
of a single generation, and introducing extra generations cannot make
an anomalous $U(1)$ anomaly-free. If we allow for generation dependent
$U(1)$ charges (but imposing that these $U(1)$ charges lead to
generation-independent $\bZ_3$ charges), then it is possible to cancel
the anomaly with three generations.\footnote{This is somewhat
  reminiscent of a similar statement in
  \cite{Gaiotto:2017yup,Gaiotto:2017tne}, which finds a mixed
  $T$-flavor anomaly when the number of flavors is a multiple of 3,
  and the gauge group is $SU(3)$. It would be interesting to see if
  the observations are related.}
\begin{table}[!hbt]
\begin{center}
\begin{tabular}{|c|c|c|c|c|c|c|c|} \hline &$Q$ & $\bar{U}$ & $\bar{D}$ & $L$ & $\bar{E}$ & $H_u$ & $H_d$\\\hline
\textbf{Triality}&0 &-1&1&-1&-1&1&1\\\hline
\textbf{Hexality}&0&-2&-5&-5&1&5&5\\\hline
\end{tabular}
\end{center}
\caption{$\mathbb{Z}_3$ and $\bZ_6$ charges of the MSSM chiral superfields under
  baryon triality and proton hexality. We use the conventions in \cite{Dreiner:2005rd}.}
\label{chbar}
\end{table}

The above analysis also extends to the proton hexality symmetry
proposed in \cite{Dreiner:2005rd}. Since
$\mathbb{Z}_6\approx \mathbb{Z}_2\oplus\mathbb{Z}_3$, and
$\Omega_5^{\Spin}(B\mathbb{Z}_2)=0$ because of a Smith homomorphism, a
$\mathbb{Z}_6$ discrete symmetry suffers from the same $\mathbb{Z}_3$
anomaly. The mod 3 reduction of the second row of table~\ref{chbar} is
minus the first row, so proton hexality suffers from the same
anomaly. Just as in the previous case, thanks to the fact that the
Standard Model has three generations, this anomaly can be fixed via
generation-dependent couplings; this indeed is what happens in section
9 of \cite{Dreiner:2005rd}.

As discussed above, all the discrete anomaly constraints that we are discussing should be automatically satisfied whenever the $\mathbb{Z}_n$ can be embedded into a non-anomalous $U(1)$. In particular, the mod 9 condition should be obtainable from local anomaly cancellation conditions. Consider a $U(1)$ with charges $q_i=(3m_i+r_i)$, where $m_i$ are integers and the $r_i$ are $-1,0,1$. As above, local anomaly imposes
\begin{equation}\sum_i 27 (m_i^3+m_i^2r_i)+9 m_i r_i^2+r_i^3=0,\quad \sum_i m_i +r_i=0.\end{equation}
Because of the definition, $r_i^3=r_i$. Taking the first equation modulo 9, we obtain
\begin{equation}\sum_i r_i\equiv 0\,\text{mod}\, 9,\end{equation}
as advertised. 

\subsection{SM fermions and the topological superconductor}\label{sec:ftop}
\label{sec:SMsuperconductor}
Here we discuss briefly one of the observations that led to this work:
that the number of fermions per generation in the SM (including right
handed neutrinos) is 16, which turns out to be the number of Majorana
zero modes of a topological superconductor that cancels the Dai-Freed
anomaly of time reversal. It turns out that the two facts can be
nicely related if we assume a certain $\mathbb{Z}_4$ subgroup of
$(B-L)$+ the SM gauge group to be gauged, as follows.\footnote{See
  \cite{vanderBij:2007fe} for a previous attempt at explaining the
  number of fermions per generation in the Standard Model using
  anomaly arguments. Reference \cite{Volovik:2016mre} also relates the Standard Model to a topological material.}

In the Standard model extended with right-handed neutrinos, there is a particular combination of hypercharge and $B-L$,\footnote{This is precisely the $X$ boson of GUT's, see e.g. \cite{Cheng:1985bj,Patrignani:2016xqp}. There are other combinations of $Y$ and $B-L$ with the same properties we use here.}
\begin{equation}X\equiv
-2Y+5(B-L),\label{qchar}
\end{equation}
such that the charges of all SM fermions under $X$ are of the form
$q_i=4k_i+1$. This means that $q^X_i\,\text{mod}\,4$ is a $\mathbb{Z}_4$
charge under which every fermion has a charge of
$1\,\text{mod}\,4$. For convenience, we have included the relevant
representations of standard model fields in table~\ref{table:tSM}.

\begin{table}[!hbt]
\begin{center}
  \setlength{\tabcolsep}{20pt}
  \begin{tabular}{|c|ccccc|}
    \hline\textbf{SM field}& $\mathbf{SU(3)}$& $\mathbf{SU(2)}$ & $\mathbf{Y}$ & $\mathbf{{B-L}}$  & $\mathbf{X}$ \\
    \hline $l^c_L$& $\mathbf{1}$& $\mathbf{2}$& $-3$ &3 & $21$\\
    \hline  $q^c_L$& ${\bar{\mathbf{3}}}$& $\mathbf{2}$& 1& $-1$ & -7 \\
    \hline $l_R$& $\mathbf{1}$& $\mathbf{1}$& 6& $-3$ & -27 \\
    \hline $u_R$& $\mathbf{3}$& $\mathbf{1}$& $-4$&1 & $13$ \\
    \hline$d_R$& $\mathbf{3}$& $\mathbf{1}$& 2&1 & 1 \\
    \hline $\nu_R$& $\mathbf{1}$& $\mathbf{1}$& 0&$-3$ & -15\\
    \hline $H$ & $\mathbf{1}$ & $\mathbf{2}$ & 3 & 0 & -6 \\
    \hline\end{tabular}
\end{center}
\caption{Charge assignments of the fields in the Standard Model. All
  fermions are right-moving chiral Weyl fermions. We have rescaled the hypercharge $Y$ and $B-L$ such that all fields have integer charges. $H$ is the Higgs
  doublet. We have included a right-handed Majorana neutrino.}
\label{table:tSM}
\end{table}

As discussed recently in \cite{Tachikawa:2018njr}, in the presence of
an extra $\mathbb{Z}_4$ symmetry, it is possible to make sense of
fermions in manifolds that are not $\Spin$. More concretely, one can
take the structure group to be
$(\Spin \times \mathbb{Z}_4)/\mathbb{Z}_2$, where the generator of the
$\mathbb{Z}_2$ subgroup of $\mathbb{Z}_4$ and $(-1)^F$ are
identified. This was called a $\Spin^{\mathbb{Z}_4}$ structure in
\cite{Tachikawa:2018njr}. Because of the above, the SM admits a
$\Spin^{\mathbb{Z}_4}$ structure.

The same reference also constructs a version of the Smith
homomorphism, along the same lines as in section~\ref{sec:toposup}
below, establishing that
\begin{equation}
  \Omega_4^{\Pin^+}\approx \Omega_{5}^{\Spin^{\mathbb{Z}_4}}\, .
\end{equation}
Physically, one can construct $\Spin^{\mathbb{Z}_4}$ bundles which contain domain walls on which 3d $\Pin^+$ fermions localize. For each 4d Weyl fermion with charge 1 modulo 4, we get one 3d $\Pin^+$ Majorana fermion. 

Using $X$ defined in \eq{qchar}, we see that we reproduce this story
once for each standard model fermion. Since the anomaly for the
topological superconductor vanishes only when the number of Majorana
fermions is a multiple of sixteen \cite{Witten:2015aba}, we learn that
the number of fermions in the standard model must be a multiple of
sixteen for the $\bZ_4$ symmetry to be anomaly-free.\footnote{We
  should note that in \cite{Chang:1994sv}, this very same condition is
  obtained from requiring that the theory makes sense in a manifold
  with a generalized spin structure.} This is precisely the number of
fermions in a generation of the standard model, once we include the
right-handed neutrino.

As discussed above, if the above $\mathbb{Z}_4$ symmetry is assumed to
embed into a $U(1)$ (in this case, the combination \eq{qchar} of
hypercharge and $B-L$), then the relevant bordism group becomes
$\Omega^{\Spin^c}_5=0$, so the constraint that the number of fermions
must be a multiple of 16 must already be implied by local anomaly
cancellation.\footnote{As discussed in section~\ref{sec:sm}, once we
  assume $U(1)_{B-L}$ we can put the standard model in a $\Spin^c$
  manifold. It is easy to see that the $\bZ_4$ subgroup of this
  $U(1)_{B-L}$ leads to a topological superconductor with 8 Majorana
  fermions of each parity under time reversal, and thus no anomaly.}
And indeed, in this case the anomaly cancellation conditions
for $U(1)$ factors (coming from $\Tr(F_{U(1)}R^2)=\Tr(F_{U(1)}^3)=0$)
\begin{equation}
  \sum_i q_i=\sum_i q_i^3=0
\end{equation}
with $q_i$ the $U(1)$ charges of the fermions, imply that the total
number of fermions $n_F$ is a multiple of 16, as follows. Define
$p_l=\sum_i k_i^l$ (recall that we defined above $q_i=4k_i+1$). The
first anomaly cancellation condition implies $n_F=-4p_1$, and the
second is
\begin{equation}
  0=n_F+12p_1+48p_2+64p_3=-2n_F+48p_2+64p_3\, ,
\end{equation}
which implies $n_F=8(3p_2+4p_3)$. This means that $n_F$ is a multiple
of 8, or equivalently that $p_1$ is an even number. But $p_1$ and
$p_2$ have the same parity, so $p_2$ is also even and $n_F$ is a
multiple of 16.

In fact, if we assume $Spin(10)$ grand unification, the $\mathbb{Z}_4$
group we are studying is just the center of $Spin(10)$, so under this
assumption we can understand the above result as coming from the fact
that $\Omega_5^{\Spin}(Spin(10))=0$.

Finally, we should also mention that at low energies there is a mass
term for $\nu_R$ that breaks $B-L$ \cite{Patrignani:2016xqp}. As a
result, the $\mathbb{Z}_4$ is broken explicitly, and there are only 15
massless fermions (before electroweak symmetry breaking, which also
breaks $\bZ_4$).

\subsubsection{Topological superconductors and the MSSM}

The above construction works straightforwardly in the
MSSM+right-handed neutrinos, since the additional fields (gauginos and
higgsinos) do not contribute to the mod 16 anomaly, given that the
$\mathbb{Z}_4$ anomaly for a charge 2 fermion vanishes. However, with
the fermion spectrum of the MSSM, there is an additional
$\mathbb{Z}_4$ whose $\mathbb{Z}_4$ anomaly cancels.
Under this symmetry, all the fermions of the MSSM transform with
charge $+1$. The bosons could have any even charge and the symmetry would remain non-anomalous, but a natural choice is to take all bosons neutral under the symmetry.\footnote{The $\mathbb{Z}_2$ subgroup of this would be
  $(-1)^F(-1)^{2s}$, where $s$ is the spin. This symmetry is related to the standard
  R-parity, which flips the sign of all the superpartners while leaving all the SM fields invariant, by a shift by ``matter parity'' $(-1)^{3(B-L)}$
  \cite{Csaki:1996ks}.} The mod 16 constraint is still satisfied
because, on top of the original 16 fermions in the SM there are 12
gauginos (one for each generator of the gauge group) and 4 higgsinos
(two for each of the Higgs doublets, since they are themselves $SU(2)$
doublets). This is only possible because of the detailed structure of
the SM - including the dimension of the gauge group and the fact that we
need two Higgses in the MSSM \cite{Csaki:1996ks}.

Again, one can find anomaly-free $U(1)$'s in which to embed this
$\mathbb{Z}_4$ symmetry, but this time there is no obvious
relationship to GUTs. A perhaps more interesting connection stems from
the observation that the symmetry we are quotienting by is
$\sqrt{(-1)^F}$, where $(-1)^F$ is fermion number - which is a
symmetry in any quantum field theory. Perhaps this symmetry is
pointing to a (possibly orientation-reversing) $\mathbb{Z}_2$
geometric symmetry in some internal space Geometric $\mathbb{Z}_2$
actions can lift to $\mathbb{Z}_4$ on the spinor bundle; this is the
case for instance for a rotation by $\pi$, or a reflection with a
$\Pin^-$ structure. A similar situation was discussed in
\cite{Tachikawa:2018njr}, where a $\Spin^{\mathbb{Z}_4}$ symmetry is
related to a $180^\circ$ rotation of the F-theory fiber.

In any case, though this anomalous $\mathbb{Z}_4$ in the MSSM may seem enticing, it is not devoid of problems. First of all, we have neglected the contribution of the gravity multiplet\footnote{We thank Luis Ibañez for bringing up this point.}. The gravitino in particular has a charge of $-1$ under the R-symmetry (in conventions where the R-charge of the graviton vanishes and that of a supercharge is $+1$), which means that it has a $\mathbb{Z}_4$ charge of $-i$. 

We therefore want to find the contribution of a gravitino with charge
$-i$ to the anomaly. As usual, the easiest way to accomplish this is
to evaluate the contribution of a vector-spinor, and then substract
another spinor with opposite chirality.

Let us recover the spinor contribution first. The generator of
$\Omega_5^{Spin^{\mathbb{Z}_4}}(\text{pt})$ is $\mathbb{RP}^5$, so we
need to evaluate the $\eta$ invariant of the Dirac operator in this
background. We will use the same trick as in \cite{Witten:2015aba} to
relate this to the index of a 6-dimensional Dirac operator on an
orbifold $T^6/\mathbb{Z}_2$. The Dirac index on this manifold is 8,
and removing the orbifold singularities we get 64 copies of
$\mathbb{RP}^5$ on the boundary. As a result,
$\eta(\mathbb{RP}^5)=1/16$, in accordance with Smith's homomorphism.

For the Rarita-Schwinger operator, the index gets multiplied by $6$ because of the extra vector index. So the Rarita-Schwinger $\eta$ invariant is $-6/16$ (taking into account the fact that the R-charge is $-1$). We need to substract the contribution of a fermion of opposite chirality (which is $1/16$), with a total result of -7/16 per gravitino. So the contribution of a gravitino is nonvanishing and spoils the agreement. One can double-check this result by using the embedding $\text{Spin}^{\mathbb{Z}_4}$ in $\text{Spin}^c$ (see Appendix \ref{sec:SpinZ4-bordism}). A fermion with $\mathbb{Z}_4$ charge of $\pm i$ embeds as a $\text{Spin}^c$ fermion of charge $q=1,3$. Since $\Omega_5^{\text{Spin}^c}=0$, the $\eta$ invariants for these two representations can be computed via the APS index theorem,
\begin{equation} \eta(q)=\frac{q^3}{6}\int_X c_1^3+\frac{q}{24}\int_X p_1c_1,\end{equation}
where $X$ is a $\text{Spin}^c$ manifold such that $\partial X=\mathbb{RP}^5$. This can then be used to compute the gravitino contribution \cite{AlvarezGaume:1983cs},
\begin{equation} \eta_\text{gravitino}=-\frac{1}{6}\int c_1^3+\frac{7}{8}\frac{1}{24}\int p_1c_1=-\frac{7}{16} \,\text{mod}\,1.\end{equation}

Even if we ignore the issues with the gravitino, there is a mixed anomaly with the non-abelian factors of the SM gauge group, since both gauginos and Higgsinos are charged under these. While a full characterization of this anomaly would involve computation of at least $\Omega^{\text{Spin}^{\mathbb{Z}_4}}(BG_{\text{SM}})$, where $G_{\text{SM}}$ is the SM gauge group, it is possible to explicitly exhibit an anomaly by looking at particular elements of this group. In particular, consider the theory on $S^1\times S^4$ with a $SU(N)$ instanton of instanton number 1 on the $S^4$, and with a nontrivial $\mathbb{Z}_4$ action on the $S^1$. Using formula \eq{etaab}, as well as $\eta(S^1)=\frac{1}{4}$ for a fermion with $\mathbb{Z}_4$ charge of 1, one obtains
\begin{equation} \eta(S^1\times S^4)=\eta(S^1)\times \text{index}(S^4)=\frac{\text{index}(S^4)}{4}.\end{equation} 
For gauginos, $\text{index}(S^4)=2N$, while for the Higgsinos in the fundamental, the index is 1. It follows that the MSSM has both mixed $SU(2)-\mathbb{Z}_4$ and $SU(3)-\mathbb{Z}_4$ anomalies, the former from the Higgsinos and the latter from the gauginos. Under these circumstances, the particular $\mathbb{Z}_4$ we discuss is clearly not as interesting as its Standard Model counterpart; at the very least one would need exotics to cancel the anomalies.

\subsection{$\Spin^c-\mathbb{Z}_n$}

From the general formula~\eqref{etaspinc}, reference \cite{bahri1987}
shows that the eta invariant for a $\Spin^c$ fermion on $L^k(n)$ on the
representation $s$ is given by
\begin{equation}\eta_{s}=\frac{1}{n}\sum_{\lambda\neq1}(\lambda^s-1)\left(\frac{\lambda}{\lambda-1}\right)^k,\label{etach0}\end{equation}
where $\lambda$ runs over all the nontrivial $n$-th roots of unity (this is a particular case of \eq{etaspinc}). This is the result for a fermion of charge $q=1$ only; in general, $\Spin^c$ fermions can have any (odd) charge under the $U(1)$. To each choice of $\Spin^c$ structure one can associate a line bundle $V$ in a canonical way, via the map
\begin{equation} (\Spin\times U(1))/\mathbb{Z}_2\rightarrow U(1)\, :\, (g,\lambda)\rightarrow \lambda^2.\end{equation}
Writing $q=2\ell+1$, a fermion of charge $q$ behaves as a fermion of charge $q=1$ coupled to an additional line bundle $\ell V$. As discussed in  \cite{bahri1987}, for the $\Spin^c$ structure such that \eq{etach0} is valid, one has $c_1(V)=k\zeta$, where $\zeta$ is the generator of $H^{2}(L^k(n),\mathbb{Z})=\mathbb{Z}_n$ (for $k>1$).

On top of this, the result \eq{etach0} is derived for a particular $\Spin^c$ structure on $L^k(n)$. $\Spin^c$ structures over a manifold are affinely parametrized by line bundles over the manifold; in the $\Spin^c$ structure corresponding to the line bundle $L$, a fermion of charge $q$ gets an additional factor of $L^q$.

Putting all of the above together, a fermion of charge $q$ in the $\Spin^c$ structure related to the one just discussed by an element $\beta\in H^{2}(L^k(n),\mathbb{Z})$ is coupled to an additional line bundle with class $q\beta + \ell k\zeta\in H^{2}(L^k(n),\mathbb{Z})$.

This means that the $\eta$ invariant in a lens space for a fermion of charge $q=2\ell+1$ and spin structure $\beta$ is obtained as
\begin{equation}\eta_{s,q,\beta}=\frac{1}{n}\sum_{\lambda\neq1}(\lambda^{s+k\ell+q\beta}-\lambda^{k\ell+q\beta})\left(\frac{\lambda}{\lambda-1}\right)^k.\label{etach01}\end{equation}

This formula, for different values of $k$ and $\beta$, is sufficient to address all possible anomalies, thanks to Theorem 0.1 of \cite{bahri1987}, which guarantees that, for $k=3$, independent $\eta$ invariants in the $\Spin^c$ case come only from four different manifolds, namely
\begin{equation}\eta(L^3(n)),\ \eta(L^2(n)\times CP^1),\ \eta(L^1(n)\times CP^1\times CP^1),\ \eta(L^1(n)\times CP^2).\label{thmgil}\end{equation}
On each of these manifolds we must in principle consider all possible $\Spin^c$ structures. We will parametrize spin structures as follows, where the $\beta_i$ are integers modulo $n$, and the $\gamma_i$ are integers:

\begin{center}
  \begin{tabular}{|c|c|c|}
    \hline\textbf{X}& $\mathbf{H^2(X)}$& \textbf{Basis coefficients}   \\ \hline 
  $L^3(n)$ & $\mathbb{Z}_n$& $\beta_3$\\\hline
  $L^2(n)\times CP^1$ & $\mathbb{Z}_n\oplus\bZ$ & $\beta_2,\gamma_1$\\\hline
   $L^1(n)\times CP^1\times CP^1$ & $2\bZ$ & $\gamma_2,\gamma_3$\\\hline
   $L^1(n)\times CP^2$ & $\bZ$ & $\gamma_4$\\\hline
   \end{tabular}
\end{center}

Using formula \eq{etaab}, we can express the last three $\eta$ invariants in \eq{thmgil} in terms of Dirac indices in projective spaces and $\eta$ invariants on lens spaces,
\begin{align}\eta(L^2(n)\times CP^1)&= q\gamma_1 \eta(L^2(n)),\quad \eta(L^1(n)\times CP^1\times CP^1)=q^2\gamma_2\gamma_3 \eta(L^1(n)),\nonumber\\\eta(L^1(n)\times CP^2)&= \left(\frac{q^2-1}{8}+q^2\frac{\gamma_4(\gamma_4+1)}{2}\right)\eta(L^1(n)).\end{align}
To evaluate the $\Spin^c$ index of $CP^2$, we use the fact that its signature is $1$ \cite{FuchsViro}, together with the index theorem for the $\Spin^c$ complex \cite{dillen1999handbook} and the fact that any complex manifold has a canonical $\Spin^c$ structure whose associated line bundle $V$ equals the determinant line bundle. 

From the above, it is clear that the anomaly cancellation conditions that we get from the above set is redundant. In particular, we can take $\gamma_1=\gamma_2=\gamma_3=1$ and $\gamma_4=0$ without loss of generality. Using the expressions around Example 1.12.1 in \cite{GilkeyBook}, we find that demanding absence of Dai-Freed anomalies on an arbitrary $\Spin^c$ manifold amounts to the constraints
\begin{align}\sum_{\text{fermions}}&\frac{s \left(2 n^2+6 n (3 q+s)+27 q^2+18 q s+4 s^2-3\right)}{24 n}\in\mathbb{Z},\quad
\sum_{\text{fermions}}&\frac{q s (n+2 q+s)}{2 n}\in\mathbb{Z},\nonumber\\ 
\sum_{\text{fermions}}&\frac{q^2 s}{n}\in\mathbb{Z},\quad\text{and}\quad
\sum_{\text{fermions}}\frac{(q^2-1)s}{8 n}\in\mathbb{Z} \label{anocanc}.\end{align}
Notice that there is no dependence in the $\beta_i$; this because all the $\beta_i$-dependent terms can be rewritten as linear combinations of the \eq{anocanc}.

\subsubsection{Connection to mapping tori anomaly and Ibañez-Ross constraints}\label{sec:IR}

Anomalies of $\mathbb{Z}_n$ discrete symmetries have a long story, starting with the work of Iba\~{n}ez and Ross \cite{Ibanez:1991hv}. This work considers $\mathbb{Z}_n$ symmetries that come from Higgsing a non-anomalous $U(1)$ in the UV. As a result, the UV fermion spectrum satisfies the corresponding (local) anomaly cancellation conditions.  Iba\~{n}ez and Ross then work out which part of these anomaly conditions still survive as constraints in the infrared theory, taking into account that some fermions can become massive as we break the $U(1)$ symmetry.  These are the well-known Iba\~{n}ez-Ross constraints. We are interested in the case where the symmetry is $U(1)^2$ in the UV and $\mathbb{Z}_n-U(1)$ in the infrared (the $U(1)$ will be our $\Spin^c$ connection). Then there are two linear Ibañez-Ross constraints (here, $(x_i,q_i)$ are the UV charges, and $x_i= k_i n + s_i$), coming from mixed and gravitational anomalies,

\begin{equation} \sum_{\text{fermions}} s_i = a\frac{n}{2}, \quad \sum_{\text{fermions}} q_i^2s_i = bn,\label{IRlin}\end{equation}
and two nonlinear, coming from mixed and cubic anomalies,
\begin{equation} \sum_{\text{fermions}} s^2_iq_i = cn, \quad \sum_{\text{fermions}} s^3_i = dn+e\frac{n^3}{8},\label{nlIR}\end{equation}
where $a,b,c,d,e$ are integers which are constructed out of the UV data. It was already pointed out in \cite{Ibanez:1991hv} that the second condition in \eq{IRlin} is not a useful constraint in the infrared, because the normalization of the $U(1)$ charges is not known. It was later pointed out in \cite{Banks:1991xj} that the nonlinear constraints are UV-sensitive, in the sense that they depend on the global structure of the UV gauge group. For instance, suppose that we don't change the fermion spectrum, but change $U(1)$ that is fixed to an $l$-fold cover of the original. Equivalently, we demand that the charge quantum is not $1$, but $1/l$ in the above units. Then, in terms of the fundamental charge, the breaking is not to $\mathbb{Z}_n$ but to $\mathbb{Z}_{nl}$. At the same time, the $s_i$ rescale as $s_i\rightarrow s_i\,\text{mod}\, nl$, so the left and right hand sides of \eq{nlIR} scale differently. The linear constraints, on the other hand, are independent of the particular normalization of $U(1)$ charges. As we will see, this distinction is also present in some of the Dai-Freed anomalies \eq{anocanc}.

The constraint \eq{anocanc} is
particularly interesting in examples where the discrete $\mathbb{Z}_n$
symmetry cannot be embedded into a continuous unbroken $U(1)$ in the
field theory regime, such as e.g. discrete symmetries coming from
discrete isometries in Calabi-Yau compactifications.\footnote{These
  particular examples can be embedded into continuous group actions in
  supercritical string theory
  \cite{Berasaluce-Gonzalez:2013sna,Garcia-Etxebarria:2015ota}, but it
  is hard to argue for standard, local anomaly cancellation in these
  exotic scenarios.} We will now see that, in this framework, the
linear Ibañez-Ross constraints can be recovered from the eta invariant on
mapping tori. Therefore, they correspond to ``traditional'' global
anomalies in the sense of section \ref{sec:review}.

As discussed in section \ref{sec:review}, restricting to mapping tori leads to an anomaly cancellation condition which is in general weaker than full Dai-Freed anomaly cancellation; for instance, as discussed in \cite{Witten:2015aba}, for the 3d topological superconductor one obtains a $\mathbb{Z}_{16}$ anomaly by demanding $\exp(\pi i\eta)=1$ for arbitrary 4-manifolds, but if we restrict to mapping tori only a $\mathbb{Z}_8$ is visible. This $\mathbb{Z}_8$ can be studied by standard anomaly techniques, such as e.g. modular anomalies in appropriate backgrounds \cite{Hsieh:2015xaa}.

The same happens with the $\mathbb{Z}_n-U(1)$ anomaly \eq{anocanc}. A
particularly interesting subset of mapping tori in this context are of
the form $X_d\times S^1$, where $X_d$ is an arbitrary $d$-dimensional
manifold, and we pick up a $\mathbb{Z}_n$ gauge transformation as we
move around the $S^1$. (A low dimensional analogue of this fibration
would be obtained by regarding $S^1$ as the lens space $L^1(n)$ in the
sequence $\mathbb{Z}_n\rightarrow S^1\rightarrow L^1(n)$.) Studying
anomalies on this background is equivalent to studying anomalies on
the zero-dimensional theory obtained from dimensional reduction on
$X_d$. Now, we have \cite{bahri1987}
\begin{equation}\eta_{s,q}(L^1(n))=-\frac{s}{n}\,\text{mod}\, 1,\end{equation}
which together with the formula \eq{etaab} implies the anomaly condition
\begin{equation}\sum_{\text{fermions}}\text{index}(X_d)s=0\,\text{mod}\, n.\label{qwer}\end{equation}
Notice that the formula \eq{etaab} agrees with the dimensional reduction picture: reducing on $X_d$ produces $\text{index}(X_d)$ zero-dimensional fermion zero modes, and we must take into account the $\eta$ invariant for each of these.
For instance, consider the case $d=4$, $X_d=S^2\times S^2$ with the canonical $U(1)$ bundle over each $S^2$, and fermions with $U(1)$ charges $q_i$. Then \eq{qwer} becomes
\begin{equation}\sum_{\text{fermions}} q^2s=0\,\text{mod}\, n,\label{irl1}\end{equation}
which is the mod $n$ reduction of the would-be mixed local anomaly cancellation condition, one of the Iba\~{n}ez-Ross constraints. If on the other hand we choose $X_d$ to be e.g. a K3, we obtain
\begin{equation}2\sum_{\text{fermions}} s=0\,\text{mod}\, n,\label{irl2}\end{equation}
another of the Iba\~{n}ez-Ross constraints. We therefore recover the linear Iba\~{n}ez-Ross constraints \eq{IRlin}, which are precisely the ones that are not UV-sensitive \cite{Banks:1991xj,Ibanez:1992ji}.

A natural question is the precise relationship between Dai-Freed
anomaly cancellation and whether the $\mathbb{Z}_n$ symmetry can be
embedded into a non-anomalous $U(1)$. If such an embedding is
possible, then all Dai-Freed anomalies must necessarily vanish, since
$\Omega_5^{Spin^c}(BU(1))=0$.\footnote{While we did not discuss this
  case explicitly in section \ref{sec:lie}, the computation via the
  AHSS is very simple, and similar to that of figure
  \ref{fig:AHSS-U(1)}. We just need to know the $\Spin^c$ bordism
  ring, which can be found in \cite{GilkeyBook} and
  appendix~\ref{app:bgrs}.}
  
Let us now discuss the converse statement. If Dai-Freed anomalies cancel, does this mean that the $\mathbb{Z}_n$ can be embedded into an anomaly-free $U(1)$? To address this point, consider a set of charges $(q_i,s_i)$ which satisfy the cubic constraint for the $U(1)$ as well as the Dai-Freed constraints \eq{anocanc} (since we are in the $\text{Spin}^c$ case, all of the $q_i$ are odd). If the $\mathbb{Z}_n$ arises from Higgsing from a $U(1)$, a fermion in a representation with charge $s_i$ comes from a representation with charges $r_i=s_i+ np_i$, $p_i\in\mathbb{Z}$. On top of this, pairs of fermions with charges $(q_j,r_j)$ and $(-q_j, r'_j)$ can acquire a mass after Higgsing, as long as $r_j+r_j'\equiv0\, \text{mod}\, n$. The UV theory has four mixed anomaly cancellation conditions, which we encode as
\begin{equation}\mathcal{A}_i= \left(q_i^3, q_i^2r_i, r_i^2q_i, r_i^3\right),\quad \sum_i \mathcal{A}_i=0.\end{equation}
For a particle of charge $r_i=s_i+np_i$,
\begin{equation}\mathcal{A}_i=\left( q_i^3, q_i^2s_i, s_i^2q_i, s_i^3\right)+ \mathcal{E}_i,\quad \mathcal{E}_i\equiv n\left( 0,p_iq_i^2,q_ip_i(2_is+p_i),3(s_ip_i^2n+s_i^2p_i)+n^2p_i^3\right),\end{equation}
while the anomaly for the pair of fermions which becomes massive after Higgsing is (writing $r_j'=-r_j+l_j n$)
\begin{equation} \mathcal{A}^{(\text{massive})}_j= n \left(0,q_j^2l_j,ql_j(2r_j+l_jn), 3(r_j^2l_j-r_jl_j^2n)+l_j^3 n^2\right).\label{vmassive}\end{equation}
Notice that $\mathcal{E}_i$ is of the same form as the $\mathcal{A}^{(\text{massive})}_j$. Embedding of the $\mathbb{Z}_n$ in an anomaly-free $U(1)$ will be possible if there is some choice of massive particles such that the anomaly can cancel. This means that we can pick any set of $(r_j,l_j)$ that will do the trick. We can always pick some of these to cancel the $\mathcal{E}_i$, so without loss of generality, embedding will be possible if and only if
\begin{equation}\sum_i \left( q_i^3, q_i^2s_i, s_i^2q_i, s_i^3\right) \quad\in\quad\mathcal{L}^{(\text{massive})},\label{acanccond}\end{equation}
where $\mathcal{L}^{(\text{massive})}$ is the lattice generated by all linear combinations of all vectors of the form \eq{vmassive}. 

We have checked the condition \eq{acanccond} numerically for values of $n$ up to $15$. For every trial spectrum we checked where Dai-Freed anomalies \eq{anocanc} are cancelled, \eq{acanccond} is satisfied as well. This suggests that Dai-Freed anomaly cancellation is sufficient to ensure embedding into an anomaly-free $U(1)$, though we have not proven this. On the other hand, there are spectra which satisfy the full set of Ibañez-Ross constraints \eq{IRlin} and \eq{nlIR}, but not \eq{acanccond} or \eq{anocanc}. One such example is $n=2$ and a spectrum with charges $(q_i,s_i)$ given by
\begin{equation} (3,0),(-5,0),(3,1),(-1,1).\label{noluisexample}\end{equation}

To sum up, the full set of Dai-Freed constraints \eq{anocanc} is stronger than the Ibañez-Ross constraints, and numerical evidence suggests that it is equivalent to anomaly cancellation in the UV. The example \eq{noluisexample} shows this is not the case for Ibañez-Ross. Both the non-abelian Ibañez-Ross and the nonlinear Dai-Freed constraints are UV sensitive.  In the Dai-Freed case, this is made manifest by the presence of a topological GS term, as we will discuss in subsection \ref{sec:tgsd}.

Finally, all these considerations apply equally well to the $\Spin$ case discussed in subsection \ref{sec:spinzn}. Here, the only linear Ibañez-Ross constraint is the mod $n$ reduction of gravitational anomaly cancellation. For instance for $n=3$, this is just the requirement that the charges vanish modulo 3; we have instead a stronger, modulo 9 constraint, \eq{mod9cons}. We have focused on the $\Spin^c$ case because of its richer structure.

\subsubsection{$n=2$ and the topological superconductor}\label{sec:toposup}

For the $n=2$ case there is a nice connection to the theory of the
boundary modes of a 4d topological superconductor. In this context,
there is a well-known $\mathbb{Z}_{16}$ constraint, obtained in the
same way as above, by requiring that the anomaly theory (recall our
discussion in \S\ref{sec:daifreed}) provided by the $\eta$ invariant
in one dimension more should be trivial.

Physically, the connection between the two comes from the fact that one can introduce a scalar which breaks the $\mathbb{Z}_2$ symmetry. The associated $\mathbb{Z}_2$ domain walls contain localized fermions, with a $\Pin^c$ structure. When the anomaly theory of the domain wall admits a $\Pin^+$ structure, one such fermion becomes equivalent to two copies of an ordinary topological superconductor.

We will now explicitly construct these $\mathbb{Z}_2$ domain walls. We consider two Euclidean fermions $\psi_1, \psi_2$, charged under a $U(1)$, as well as under an additional $\mathbb{Z}_2$ symmetry, as indicated in table \ref{tf1} (we take $q\neq 0$).
\begin{table}[!hbt]
\begin{center}
\begin{tabular}{|c|c|c|}
\hline \textbf{Fermion}& $\mathbb{Z}_2$ & $U(1)$\\\hline
$\psi_1$& $0$ & $q$\\\hline
$\psi_2$& $1$ & $-q$\\\hline
\end{tabular}
\end{center}
\caption{Two-fermion system which gives rise to a 3d $\Pin^c$ zero mode.}
\label{tf1}
\end{table}

We see that the $U(1)$ anomalies cancel, but the Dai-Freed anomaly \eq{anocanc} does not. In fact, the fermion with charge $0$ does not contribute to the anomaly, so the anomaly theory is just that of a single fermion in the sign representation of $\mathbb{Z}_2$. The kinetic term will be
\begin{equation} \frac{i}{2}\sum_{i=1,2} \psi_i^T \mathcal{C} \slashed{D} \psi_i.\end{equation}
The most general mass term is of the form\footnote{There is another allowed mass term, with an extra insertion of $\gamma_5$, but locally this can be removed by a change of basis.}
\begin{equation}M_{ij} \psi^T_i \mathcal{C} \psi_j,\end{equation}
where $M_{ij}$ is a symmetric matrix. The diagonal mass terms are forbidden by the $U(1)$ charge, and the only nondiagonal one is forbidden by the $\mathbb{Z}_2$ charge, so no mass terms are allowed. However, let us introduce a real scalar $\psi$, transforming under the sign representation of $\mathbb{Z}_2$,  coupled to the fermions via the Yukawa coupling
\begin{equation}
  g \phi\, \psi_1^T\mathcal{C}\psi_2\, .
\end{equation}
A vev for $\phi$ will completely break the $\mathbb{Z}_2$ symmetry,
and gap the fermions. On the $\phi=0$ locus there will be a localized
3d zero mode, which we now construct locally. Pick coordinates on a
neighborhood of a point on the $\phi=0$ locus such that $\phi=0$
corresponds locally to $x^3=0$.  The equations of motion are
\begin{equation} \slashed{D} \psi_1=g\phi(x) \mathcal{C} \psi_2, \quad \slashed{D} \psi_2=g\phi(x) \mathcal{C} \psi_1.\label{eomf2f}\end{equation}
We are interested in localized 3d zero modes, for which the $x^3$ part of \eq{eomf2f} vanishes identically,
\begin{equation} \gamma^3\partial_3 \psi_1=g\phi(x) \mathcal{C} \psi_2, \quad \gamma^3\partial_3 \psi_2=g\phi(x) \mathcal{C} \psi_1.\label{eomf2fo}\end{equation}
To solve these, introduce $\xi_\pm=\psi_1\pm \mathcal{C}\psi_2$. The equations become
\begin{equation} \gamma^3\partial_3 \xi_\alpha= \alpha g\phi(x) \xi_\alpha.\end{equation}
Now, we are interested in solutions of the form
\begin{equation}\xi_\alpha (x^1,x^2,x^3)=\zeta_{\alpha,\beta} (x^0, x^1, x^2) f_{\alpha,\beta}(x^3).\end{equation}
Plugging back on \eq{eomf2f}, we get
\begin{equation}\gamma^3\zeta_{\alpha,\beta}=\beta\zeta_{\alpha,\beta},\quad   \partial_3 f_{\alpha,\beta}(x^3)= \alpha\beta\, g\phi(x)\,f_{\alpha,\beta}(x^3).\end{equation}
The local profile for $f_{\alpha,\beta}$ can be found explicitly,
\begin{equation} f_{\alpha,\beta}(x^3)= f_{\alpha,\beta}(0) \exp\left(\alpha\beta g\int_0^{x^3} \phi(x) d\, x\right).\end{equation}
Two of the functions $f_{\alpha,\beta}(x^3)$ localize around $x^3=0$. For instance, if $\phi(x^3)=x^3$, then $f_{+,-}$ and $f_{-,+}$ are both Gaussians. The other two solutions are not normalizable (although in a compact space there will be a small component of these as well). A similar construction can be found in \cite{Hellerman:2004zm}

The localized modes are two 3d fermions, which we will label as $\lambda_1=\zeta_{+,-}$ and $\lambda_2=-i\gamma_5\zeta_{-+}$. Acting with a $U(1)$ gauge transformation with angle $\theta$, which acts as $\psi_1\rightarrow e^{i\theta q \gamma_5}\psi_1$, $\psi_2\rightarrow e^{-i\theta q \gamma_5}\psi_2$ in accordance with table \ref{tf1}, we get the transformation law
\begin{align}\left(\begin{array}{c}\lambda_1\\\lambda_2\end{array}\right)\rightarrow \left(\begin{array}{cc}\cos\theta&-\sin\theta\\\sin\theta&\cos\theta\end{array}\right)\left(\begin{array}{c}\lambda_1\\\lambda_2\end{array}\right),\end{align}
so we can equivalently describe the zero mode sector by a complex 3d fermion $\lambda\equiv\lambda_1+i\lambda_2$ of charge $q$. On top of this, a rotation by $180^\circ$ degrees on the $x^2-x^3$ plane, with $i=1,2,3$, acts on $\psi_1,\psi_2$ by multiplication by $\gamma^3\gamma^i$. This maps
\begin{equation}\zeta_{\alpha,\beta}\rightarrow \zeta_{-\alpha,-\beta},\end{equation}
which maps normalizable modes to normalizable modes, so it is a good
symmetry of the theory and implements a spin lift of a reflection
along the $x^2$ coordinate. As a result, the symmetry group of the 3d
fermion includes reflections. Crucially, the gauge transformations
commute with the reflections. This means that the symmetry group is
$\Pin^c$. The 3d gauge field is an axial vector. Had it anticommuted,
the symmetry group would have been that of the 3d topological
insulator (see appendix~\ref{3djs} for the details).

The domain wall construction can also be understood from a mathematical point of view. As explained in \cite{bahri1987}, there is an isomorphism $\Omega_{d-1}^{\Spin^c}(B\mathbb{Z}_2)\approx \Omega^{\Pin^c}_{d-1}$, called the Smith homomorphism. This establishes explicitly  that the anomaly of the domain wall fermions is equivalent to that of the parent 5d theory. The Smith homomorphism has been discussed in the physics context before in \cite{Kapustin:2014dxa}, where it took the form 
\begin{equation}\Omega^{\Spin}_d(B\mathbb{Z}_2)\cong\Omega^{\Pin^-}_{d-1}.\end{equation}
We just use the $\Spin^c$-$\Pin^c$ version of the homomorphism instead. This has been recently discussed in the condensed matter literature \cite{Shiozaki:2016zjg}. 

The explicit construction of the homomorphism described in \cite{Kapustin:2014dxa} also works in our case. Consider a 5d $\Spin^c$ manifold $Y$ with a $\mathbb{Z}_2$ principal bundle. The sign representation gives a $\mathbb{Z}_2$ vector bundle $V$ over $Y$, and consider the class $w_1(V)$. Let $X$ be the Poincar\'{e} dual to this class; this always can be represented by a submanifold by a theorem of Thom \cite{doi:10.1142/9789812772107_0002}. Over $X$, the $\Spin^c$ structure on $Y$ restricts to a $\Spin^c$ structure on $TY=TX\oplus NX$. $NX=V\vert_X$. We can compute
\begin{equation} 0=w_1(TY)=w_1(TX)+w_1(V),\end{equation}
and 
\begin{equation}w_2(TY)= w_2(TX)+w_1(TX) w_1(V)=w_2(TX)+w^2_1(V).\end{equation}
Since $w_2(TY)$ can be lifted to an integer class (since $Y$ is $\Spin^c$, and  $w^2_1(V)$ can always be lifted to an integer class\footnote{The complex line bundle $\mathcal{L}=\mathbb{C}\otimes V=V\oplus iV$ has $w_2(\mathcal{L})=w_1(V)^2=c_1(\mathcal{L})\, \text{mod} 2$.}), it follows that $w_2(TX)$ can also be lifted, which is precisely the condition to have a $\Pin^c$ structure on $X$ (see e.g. \cite{Gilkey1998}). 

Physically, the scalar $\phi$ of the previous subsection is a section
of $V$, which therefore vanishes on the Poincar\'{e} dual of $w_1(V)$-
in other words, on $X$ we have a $\mathbb{Z}_2$ domain wall with
$\Pin^c$ fermions on it. There is also an inverse map, given by
dimensional oxidation \cite{Kapustin:2014dxa}: Start with a 3d
$\Pin^c$ manifold $X$, and consider the real 2-dimensional bundle
$W=\epsilon_X\oplus t$, where $\epsilon_X$ is the orientation bundle
of $X$ and $t$ is a trivial real line bundle. Then $Y$ can be taken as
the total space of the circle bundle of $W$.

Finally, this system is also closely connected to the $\mathbb{Z}_{16}$ obstruction of the topological superconductor. This is obtained from the $\eta$ invariant of 4d $\Pin^+$ manifolds. Every $\Pin^+$ manifold is also $\Pin^c$, and if we forget the $U(1)$ gauge field the worldvolume theory in the domain wall is exactly two copies of the topological superconductor, so we can understand the $\mathbb{Z}_8$ as coming from  $\Omega^{\Pin^+}_4=\mathbb{Z}_{16}$ after multiplication by two. 

To sum up: A 5d fermion system with a unitary $\mathbb{Z}_2$ symmetry gives rise to domain walls with a $\Pin^c$ structure. Consequently, the bordism group classifying the anomalies $\Omega_{d-1}^{\Spin^c}(B\mathbb{Z}_2)\approx \Omega^{\Pin^c}_{d-1}$, where the isomorphism is obtained explicitly by domain wall construction.

\subsection{Quaternionic groups in six dimensions}
The quaternionic groups $Q_\nu$ are defined as follows: Consider the
sphere $S^3\approx SU(2)$ as the unit quaternions $H$. Define
$n=2^{\nu-1}$ (for $\nu \geq 3$) and $\xi=e^{2\pi i/n}$. $Q_\nu$ is
generated by the quaternions $\xi$ (viewed as the quaternion
$\cos(2\pi i/n) + \mathbf{i}\sin(2\pi i/n)$) and $\mathbf{j}$. It has
order $2^\nu$.  We will analyze the $Q_\nu$ anomaly cancellation
conditions in six dimensions (see \cite{Chen:2013dpa} for a recent
study of non-abelian discrete symmetries in four dimensions). Since
the $Q_\nu$ are subgroups of $SU(2)$, there is a nice interplay with
$SU(2)$ anomaly cancellation. Since $\Omega_7^{\Spin}(BSU(2))=0$, in
the $SU(2)$ case we need to concern ourselves only with local
anomalies.

In \cite{Gilkey1996}, the seven-dimensional bordism group $\Omega_7^{\Spin}(BQ_\nu)$ was computed explicitly, and the $\eta$ invariant of all the generators given. In this section we will look at only one of the anomaly cancellation conditions, and study its interplay with $SU(2)$ and its Green-Schwarz mechanism.

Concretely, we will look at anomalies in the spherical space form $S^7/\tau(G)$, where the action $\tau(G)$ in \eq{sphericalspaceform} is given in this particular case as follows: Pick quaternionic coordinates $(q_1,q_2)$ in $\mathbb{H}^2$, and consider the unit sphere $S^7\subset \mathbb{H}^2$. The spherical space form under consideration is obtained as the quotient of this $S^7$ by the generators
\begin{equation} \left(\begin{array}{c}q_1\\q_2\end{array}\right)\rightarrow R(g) \left(\begin{array}{c}q_1\\q_2\end{array}\right),\label{ggggggg}\end{equation}
for $g\in Q_\nu$ and representation matrices for the generators
\begin{equation}  R(\xi)=\left(\begin{array}{cc}e^{2\pi i/n}&0\\0&e^{-2\pi i/n}\end{array}\right),\quad R(\mathbf{j})=\left(\begin{array}{cc}0&1\\-1&0\end{array}\right). \label{matse}\end{equation}
With the definition in \cite{Gilkey1996}, the $Q_\nu$-bundle on $S^7/Q_\nu$ for which the $\eta$ invariant is computed is precisely the tangent bundle of $S^7/Q_\nu$, which has a natural $Q_\nu$-structure. More precisely, if $E$ is the corresponding principal $Q_\nu$-bundle, we have
\begin{equation} T(\mathbb{C}^4/Q_\nu)\vert_{S^7/Q_\nu}=T(S^7/Q_\nu)\oplus L= E_{f}\oplus E_{f},\label{prprpr}\end{equation}
where $L$ is a trivial line bundle and $E_{f}$ is the associated vector bundle in the fundamental $SU(2)$ representation. This splitting can be seen explictly by writing the biquaternion as 
\begin{equation} \left(\begin{array}{c}q_1\\q_2\end{array}\right)=  \left(\begin{array}{c}z_1\\z_2\end{array}\right)+j \left(\begin{array}{c}z_3\\z_4\end{array}\right),\end{equation}
where the $z_i$ are complex numbers, and noticing that each of these subspaces is invariant under the action of  \eq{matse}.

Anomalies can be computed using \eq{etaspinc}, after choosing a
particular representation $\rho$ of $Q_\nu$. We will consider the case
of the irreducible complex two-dimensional representation that embeds
$\xi$ and $\mathbf{j}$ into the fundamental of $SU(2)$ as in
\eq{matse}. A fermion in the fundamental of $SU(2)$ transforms under
this representation under the $Q_\nu$ subgroup.

Lemma 3.1 (b) of \cite{Gilkey1996} means that, for the representation \eq{matse}, the $\eta$ invariant \eq{etaf} takes the value
\begin{equation}\eta=\frac{a}{2^{\nu+2}},\quad\text{for some odd integer $a$}.\end{equation}
This means that a theory containing only fermions in the representation \eq{matse} must satisfy the anomaly cancellation condition that the total number of such fermions is a multiple of $2^{\nu+2}$. 

A similar calculation can be carried out for a field in the adjoint of $SU(2)$, which we then decompose in terms of $Q_\nu$ representations. The adjoint of $SU(2)$ reduces to a direct sum of a two-dimensional and a one-dimensional $Q_\nu$ representations, as can be seen explicitly from the representation matrices of the generators:
\begin{align}R(\xi)=\left(\begin{array}{ccc}\cos\left(\frac{4\pi}{n}\right)&0-\sin\left(\frac{4\pi}{n}\right)&0\\\sin\left(\frac{4\pi}{n}\right)&\cos\left(\frac{4\pi}{n}\right)&0\\0&0&1\end{array}\right),\quad
  R(\mathbf{j})=\left(\begin{array}{ccc}1&0&0\\0&-1&0\\0&0&-1\end{array}\right).\end{align}
The invariant \eq{etaf} is an odd multiple of $2^{\nu-1}$ in this
case. Put together, a theory with $a$ fundamentals and $b$ adjoints of
$SU(2)$ has an anomaly cancellation condition in the $Q_\nu$ subgroup
measured by
\begin{equation} \frac{\alpha_1}{2^{\nu+2}}a+\frac{\alpha_2}{2^{\nu-1}}b\in\mathbb{Z},\label{6deta0}\end{equation}
where $\alpha_1,\alpha_2$ are odd numbers explicitly given by the recurrence relations
\begin{equation}\alpha_1(\nu)= \alpha_1(\nu-1)+(1+2^{\nu-3})2^{\nu-1},\quad \alpha_2(\nu)=\alpha_2(\nu-1)+2^{2\nu-5}.\end{equation}
There is an interesting interplay between $SU(2)$ anomaly cancellation and \eq{6deta0}.  Consider a theory with $a$ $SU(2)$ fundamentals and $b$ adjoints. The $Q_\nu$ anomaly cancellation conditions lead to the constraints
\begin{equation}b\,\eta_{\text{Adj.}}+a\,\eta_{\text{Fund.}}\, \in \, \mathbb{Z}.\label{anoqnu7d}\end{equation}
These are only satisfied for $a=8b$. This can be understood in terms of $SU(2)$ local anomaly cancellation. The relevant anomaly polynomial is (ignoring the purely gravitational anomaly, which can always be cancelled by adding uncharged fermions)
\begin{equation}I=(a+4b)\frac{p_1c_2}{24}+(a+16b)\frac{c_2^2}{12},\end{equation}
where $c_2$ is the second Chern class of the $SU(2)$ bundle, and $p_1$ is the first Pontryagin class of the tangent bundle. The anomaly always factorizes in this case, so in principle it can be cancelled by a Green-Schwarz term
\begin{equation}- \int B_2\wedge \left[(a+4b)\frac{p_1}{24}+(a+16b)\frac{c_2}{12}\right].\label{gst}\end{equation}
and a modified Bianchi identity $dH=c_2$ for the $B_2$ field. The Green-Schwarz term amounts to an extra contribution to the seven-dimensional anomaly theory, given by
\begin{equation} -\int H \wedge
  \left[(a+4b)\frac{p_1}{24}+(a+16b)\frac{c_2}{12}\right].\label{gsan7d}\end{equation}
The anomaly theory of the fermions together with \eq{gsan7d} is
trivial. As discussed in section~\ref{sec:su2},
$\Omega_7^{\Spin}(BSU(2))=0$. Additionally, under the assumption that
the Green-Schwarz term can be extended to an 8-manifold, \eq{gsan7d}
can be rewritten, by using the modified Bianchi identity, as
\begin{equation} -\int_{8d} c_2 \wedge \left[(a+4b)\frac{p_1}{24}+(a+16b)\frac{c_2}{12}\right].\label{gsan8d}\end{equation}
where the integral is on some $8$-manifold that bounds the 7-manifold we use to study the anomaly. This is precisely minus the anomaly polynomial of the fermions, by construction.

We can now restrict the above construction to $SU(2)$ bundles that sit in $Q_\nu$. Since thanks to the GS term anomalies cancel for any $a,b$, it is clear that no anomaly cancellation such as \eq{anoqnu7d} is at play. From the point of view of the $Q_\nu$ theory, there is a topological GS term \cite{Garcia-Etxebarria:2017crf} which in practice can be computed by embedding the $Q_\nu$ gauge bundle into $SU(2)$, and then computing \eq{gst}. Stated like this the GS term is not a honest TQFT; there is some ambiguity in its definition, since the 7d theory \eq{gsan7d} is not trivial on an arbitrary 7-manifold with $Q_\nu$ bundle. Nevertheless, this ambiguity is compensated with that of the $Q_\nu$ fermions to provide a well-defined partition function.
 
Even though the theory makes sense for any $a,b$, \eq{gsan7d} can be trivial for special values of $a,b$. In these cases, the anomalies of the $Q_\nu$ fermions have to cancel by themselves - and so \eq{anoqnu7d} should be satisfied. Let us work out precisely when this happens. The anomaly cancellation condition \eq{anoqnu7d} comes from computing the $\eta$ invariant on a particular manifold obtained as the quotient of $S^7$ by some discrete group. For \eq{anoqnu7d} to be satisfied, we have to show that \eq{gsan7d} is trivial in this manifold, or equivalently, that \eq{gsan8d} is trivial on any 8-manifold $N$ which has \eq{sphericalspaceform} as its boundary.

To simplify \eq{gsan8d} in this case, notice that it actually only depends on the restriction of the bundles to the boundary, so we can use \eq{prprpr} to replace $p_1(TM)$ by $p_1(E_f\oplus E_f)=2p_1(E_f)$, where $E_f$ is now to be regarded as an $SU(2)$ bundle via the natural embedding. On the other hand, we have $p_1(E_f)=-2c_2(E_f)$, so that $p_1=-4c_2$. Plugging this into \eq{gsan8d}, we get 
\begin{equation}-\frac{1}{12}\int_{\mathbb{C}^4/\Gamma} (8b-a)c^2_2.\label{gsan8d2}\end{equation}
The integral of $c_2^2$ in the above orbifold will not vanish in general, but if $a=8b$ we recover the condition that the $Q_\nu$ anomalies of the fermions must vanish, as advertised.

\subsection{Coupling to TQFTs}\label{sec:tgsd}

So far we have explored the constraints that Dai-Freed anomaly
cancellation impose on theories of interest. These results can be
altered by adding Green-Schwarz terms to the action, or more generally
by coupling to a suitable topological field theory, without changing
the local degrees of freedom. We review some examples in this
subsection. Our present understanding of this phenomenon is rather
incomplete, so we will simply discuss some examples.

\subsubsection{Embedding $\mathbb{Z}_n$ in an anomalous $U(1)$}

As a simple example of how anomalies of discrete symmetries can be
cancelled by topological terms, let us look at standard Green-Schwarz
anomaly cancellation (see \cite{Bilal:2008qx} for a review, and
\cite{Chen:2015aba} for a discussion for discrete symmetries). In four
dimensions, an anomalous $U(1)$ can sometimes be rendered consistent
via the Green-Schwarz mechanism: one introduces a scalar $\phi$
transforming as $\phi\rightarrow \phi+q\lambda$ under a $U(1)$ gauge
transformation with parameter $\lambda$, and with a coupling of the
form $-c\int \phi p_1(R)$ into the action.\footnote{This is a GS for
  mixed gravitational-gauge anomalies, which are related to
  $\mathbb{Z}_n$ anomalies as discussed in subsection
  \ref{sec:spinzn}. Other types of GS terms can in some cases be
  introduced to cancel gravitational and pure gauge anomalies.} The
anomalous variation of this coupling is then $-cq\int p_1$. The
anomalous variation coming from the fermions is of the same form,
$S \int \lambda p_1$ where $S=\sum q_i$. It follows that if
\begin{equation}cq=S\label{gscanc}\end{equation}
then anomalies cancel. On the other hand, invariance under $\phi\sim\phi+2\pi$ implies that $c$ has to be an integer (in units where the elementary $U(1)$ charge is just 1). The same mechanism also works for e.g. mixed or cubic anomalies; the one caveat is that one should make sure that the coefficients $c_i$ in front of the topological terms are adequately quantized.

One could then imagine embedding e.g. a $\mathbb{Z}_n$ symmetry into a possibly anomalous $U(1)$, cancel any anomalies via Green-Schwarz couplings, and then higgs down to $\mathbb{Z}_n$. Since higgsing a non-anomalous theory cannot produce new anomalies, it would seem that in  this way one can evade any kind of anomaly constraint for $\mathbb{Z}_n$ symmetries. 

The catch is that, as discussed in \cite{Chen:2015aba}, once one introduces a Green-Schwarz term the $U(1)$ symmetry (and therefore a generic $\mathbb{Z}_n$ subgroup) are spontaneously broken by the vev of $\phi$. As a result, higgsing produces a non-anomalous theory, but the $\mathbb{Z}_n$ symmetry is gone.

Another way to see this is to look at the spectrum of charged $\mathbb{Z}_n$ strings. In a higgsing perspective, the $\mathbb{Z}_n$ strings are vortices of the UV $U(1)$. However, the Green-Schwarz axion $\phi$ has a Stuckelberg coupling to the $U(1)$. This implies (see e.g. \cite{Banks:2010zn,BerasaluceGonzalez:2012vb}) that $q$ $\mathbb{Z}_n$ strings can break by having a $U(1)$ monopole at the endpoint. 

In general, there there will be a honest
$\mathbb{Z}_{r}$ symmetry in the infrared, where $r=\text{gcd}(q,n)$ (in case we have several GS axions with charges $q_i$, $r=\text{gcd}(q_1,q_2,\ldots,n)$). In this case, the $\mathbb{Z}_r$ symmetry may avoid some of the Iba\~{n}ez-Ross constraints, but not all of them. For instance, \eq{gscanc} implies that $S$ vanishes modulo $r$, so the corresponding linear Ibañez-Ross constraint still holds. On the other hand, the cubic anomaly cancellation condition requires $\sum_i s_i^3$ to vanish modulo $r^3$, at least for odd $r$; in the presence of a GS term, it only has to vanish modulo $r$. 

In contrast with the Ibañez-Ross constraints, we cannot get rid of \emph{any} Dai-Freed constraints for $\mathbb{Z}_r$ in this way. Part of the reason is that, unlike the Ibañez-Ross constraints, even the cubic Dai-Freed constraints are linear in $r$. But the way to prove it in general is to show that the $U(1)$ GS terms are trivial for $\mathbb{Z}_r$ bundles embedded in $U(1)$. For a GS term of the form $c \int \phi\, W$, the contribution to the 5-dimensional anomaly theory is
\begin{equation}\mathcal{A}_{\text{GS}}=\exp\left(2\pi i\, c\, \int d\phi\wedge W\right).\end{equation}
Now, by assumption, $W$ is an integral cohomology class. On a generic 5-manifold, $W$ will be Poincar\'{e}-dual to a 1-cycle $\alpha$, and we get
\begin{equation}\mathcal{A}_{\text{GS}}=\exp\left(2\pi i\, c\, \int_\alpha d\phi\right)=\exp\left(2\pi i\, cq\, \int_\alpha A\right),\end{equation}
where we have used the modified Bianchi identity $d\phi= qA$. If we now restrict to $\mathbb{Z}_r$ bundles, the Wilson line $\int_\alpha A$  is of the form $m/r$, where $m$ is an integer. Since $r$ divides $q$, we have $\mathcal{A}_{\text{GS}}=1$, and the Dai-Freed anomalies for $\mathbb{Z}_r$ must cancel by themselves. 

To sum
up, the $U(1)$ GS term either breaks the discrete symmetry we are interested
in or does nothing useful, which is why we will not consider it any
further.

\subsubsection{Nonlinear Dai-Freed constraints}

Even if one cannot get rid of Dai-Freed constraints by embedding in an
anomalous $U(1)$, they are affected by the same pathology that affects
the nonlinear Ibañez-Ross constraints (see section \ref{sec:IR}). In
essence, what happens is that an observer with access only to
low-energy local physics cannot tell the difference between a
$\mathbb{Z}_{n l}$ theory with a spectrum with discrete charges
$s_{i,l}= l s_i$ for different values of $l$; they all provide the
same selection rules for couplings in the Lagrangian. Because the
groups $\Omega_{5}^{\Spin}(B\mathbb{Z}_{nl})$ and
$\Omega_{5}^{\Spin^c}(B\mathbb{Z}_{nl})$ are different for different
values of $l$, the Dai-Freed constraints are sensitive to $l$. The
low-energy observer is entitled to impose the ones that are
present for any value of $l$; these are precisely the Dai-Freed
constraints that are linear on the charges.

This does not mean that the $\mathbb{Z}_{n l}$ are all physically
equivalent; they differ on the set of allowed bundles, and spectrum of
stable strings. Due to the completeness principle
\cite{Polchinski:2003bq,Banks:2010zn}, when coupled to gravity they
must also necessarily differ in their charged spectrum. However, none
of these features can be detected via local experiments in the
infrared.\footnote{Naturally, the situation changes if one is has a
  specific string theory model at hand; in this case the precise gauge
  group is in principle completely specified.}

Since the fermion charges are also multiplied by $l$, the transition functions of the vector bundles in which the fermions live in are always in $\mathbb{Z}_n$; one way to understand the $l$-sensitivity of the results is that for $l\neq1$ we also require that the $\mathbb{Z}_n$ bundle admits a lift to $\mathbb{Z}_{nl}$. Since not all bundles can be lifted, we obtain a topological obstruction, which forbids some of them and their associated Dai-Freed constraints.

$\mathbb{Z}_n$ bundles over a base $X$ are classified by homotopy
classes from $X$ to the Eilenberg-MacLane space $B\bZ_n= K(\bZ_n,
1)$. Since the Eilenberg-MacLane spaces $K(G,\bullet)$ are the
spectrum that defines ordinary (co)homology with coefficients in $G$, we
have that $\bZ_n$ bundles are classified by $H^1(X, \bZ_n)$. The
$\mathbb{Z}_n$ bundle describing the fermion transition functions
embeds in $\mathbb{Z}_{nl}$ in a canonical way. In the theory with
$l\neq1$, this bundle describes fermions with charge $q_i l$, so the
associated principal $\mathbb{Z}_{nl}$-bundle is the $l$-th root of
the embedding. This root does not always exist, which is the technical
reason why we lose constraints sometimes. For instance, for $n=l=3$,
and $H^1(X,\bZ_9)=\bZ_3$ (this is the case, for instance, for the lens
space $L^3(3)$) with generator $\xi_3$, a $\mathbb{Z}_9$ bundle with class $\xi_3$
does not admit a 3rd root (which morally would have a characteristic
class of ``$\xi_3/3$'').

This obstruction can also be recast in terms of a coupling to a topological field theory that forbids some of the bundles. Let $Z(\xi)$ be the partition function in the topological sector specified by the class  $\xi\in H^1(X,\bZ_n)$. Then the total partition function of the theory is simply
\begin{equation}
  \label{eq:Zn-bundle-sum}
  \sum_{\xi\in H^{1}(X,\bZ_n)} Z(\xi)\, .
\end{equation}
The restriction that only bundles that are $l$-th roots contribute to
the partition function can be implemented at the level of the path
integral by modifying this equation to
\begin{equation}
  \sum_{\xi\in H^{1}(X,\bZ_n)} \sum_{\substack{\beta \in H^{1}(X,\bZ_n)\\  \chi\in H^{d-1}(X,\bZ_n)}}\exp\left(2\pi i \int_X (\xi-l\beta)\smile \chi\right)\, Z(\xi)\, ,\label{gzth}
\end{equation}
where the integral is just the pairing against the $\mathbb{Z}_n$
fundamental class of the manifold (which is henceforth assumed to be
$\mathbb{Z}_n$-orientable). The sum over $\chi$ runs over
$H^{d-1}(X,\bZ_n)$, and thus $\chi$ might be regarded as the
characteristic class classifying a $\mathbb{Z}_n$ $(d-1)$-gerbe over
the manifold; so \eq{gzth} means coupling to the topological field
theory which describes the gauging of a $\mathbb{Z}_n$ $(d-2)$
generalized global symmetry \cite{Gaiotto:2014kfa}.

We will now prove that \eq{gzth} implements the restriction on bundles we advertised. To do this, we just have to show that the function
\begin{equation}\delta(\alpha)=\frac{1}{N}\sum_{\chi\in H^{d-1}(X,\bZ_n)} \exp\left(2\pi i \int_X \alpha \smile \chi\right),\label{ddelta}\end{equation}
where $N$ is the order of $H^{d-1}(X,\bZ_n)$, evaluates to $1$ if $\alpha$ vanishes, and to $0$ otherwise. Since $\Ext_\bZ^1(H_0(X,\mathbb{Z}),\mathbb{Z}_n)=0$, the universal coefficient theorem for cohomology gives an isomorphism
\begin{equation}H^1(X,\mathbb{Z}_n)\approx \text{Hom}_{\mathbb{Z}_n}(H_1(X,\mathbb{Z}),\mathbb{Z}_n). \label{ucth}\end{equation}
In fact, this isomorphism is precisely (see \cite{DavisKirk}),
\begin{equation} \alpha\rightarrow \alpha(c)=\int_{\mu(c)}\alpha,\end{equation} 
 where $\alpha\in H^1(X,\mathbb{Z}_n)$, $c \in H_1(X,\mathbb{Z})$, $\mu$ is the canonical map in the universal coefficient theorem for homology sending a class in $H_1(X,\mathbb{Z})$ to one in $H_1(X,\mathbb{Z}_n)$, and $\int_c\alpha$ is the Kronecker pairing
\begin{equation} H^1(X,\mathbb{Z}_n)\times H_1(X,\mathbb{Z}_n)\rightarrow \mathbb{Z}_n.\end{equation}
In fact, since $\text{Tor}(H_0(X,\mathbb{Z}),\mathbb{Z}_n)=0$, we have
\begin{equation}\text{Hom}_{\mathbb{Z}_n}(H_1(X,\mathbb{Z}),\mathbb{Z}_n)=\text{Hom}_{\mathbb{Z}_n}(H_1(X,\mathbb{Z})\otimes\mathbb{Z}_n,\mathbb{Z}_n)= \text{Hom}_{\mathbb{Z}_n}(H_1(X,\mathbb{Z}_n),\mathbb{Z}_n)\end{equation}
which means that the map $\mu$ is an isomorphism. As a result, the Kronecker pairing is nondegenerate. We can now use Poincar\'{e} duality (assuming the manifold is $\mathbb{Z}_n$-orientable) to obtain a perfect bilinear pairing between $\mathbb{Z}_n$ modules
\begin{equation} H^1(X,\mathbb{Z}_n)\times H^{d-1}(X,\mathbb{Z}_n)\rightarrow \mathbb{Z}_n,\end{equation}
which we will denote by
\begin{equation} \int_X \alpha \smile \chi,\quad \alpha\in H^1(X,\mathbb{Z}_n),\quad \chi \in H^{d-1}(X,\mathbb{Z}_n).\end{equation}
One may then recognize \eq{ddelta} as an expression for the Dirac delta on discrete groups. In more detail, pick a generating set of  $\{\chi_i\}$ such that the order of $\chi_j$ is $N_j$, and $N=\prod_j N_j$.
\begin{equation}\quad \chi=\sum_i d_i \chi_i,\quad \int_X \alpha \smile \chi_i= c_i.\end{equation}
Then, \eq{ddelta} can be rewritten as
\begin{equation}\delta(\alpha)=\prod_{j} \left(\frac{1}{N_j}\sum_{d_j=1}^{N_j}e^{2\pi i\, d_j c_j}\right)=\prod_j \delta_{c_j,0}.\end{equation}
But if $c_j=0$ for all $j$, it must be the case that $\alpha=0$, since the pairing is nondegenerate.

\subsubsection{Green-Schwarz and the topological superconductor}
It is also possible to cancel Dai-Freed anomalies a la Green-Schwarz in the standard topological superconductor. 
Reference \cite{Tachikawa:2016cha} introduces several tQFT's which have the anomaly of $\nu$ copies of the topological superconductor, for $\nu=2,8$. 

On their own own, these theories do not yield an acceptable partition function. For instance, the $\nu=8$ theory fails to be reflection positive \cite{Witten:2015aba}. However, we can now couple this topological theory to 8 copies of the topological superconductor to obtain a Dai-Freed anomaly free theory. 

Via the Smith homomorphism, we can uplift the anomaly theory of 8 copies of the topological superconductor to the $\Spin^{\bZ_ 4}$ case. As discussed in subsection \ref{sec:ftop} and \cite{Witten:2015aba}, a $\Spin^{\bZ_ 4}$ manifold comes equipped with a $\mathbb{Z}_2$ bundle $V$, and the Smith homomorphism describes fermions living in the Poincar\'{e} dual locus to $w_1(V)$. As a result, the 4d term $\int w_4$ can be rewritten in terms of a 5d manifold $Y$ as
\begin{equation}\int_Y w_4(TY)\smile w_1(V).\label{rrrr}\end{equation}
Although we have not been able to write down a 4d topological field theory that gives rise to \eq{rrrr} as an anomaly theory, the Smith homomorphism suggests that it does exist.

\section{K-theoretic \texorpdfstring{$\theta$}{theta} angles}
\label{sec:KOthetaangle}

The Dai-Freed prescription introduced in section~\ref{sec:review}
provides a way to define the phase of the partition function for a
null-bordant manifold $X=\partial Y$. However, it is not always the
case that $Y$ exists. For instance, in four dimensions,
$\Omega_4^{\Spin}=\bZ$, generated by K3. So the Dai-Freed prescription
as we introduced it does not work for defining the phase of the
partition functions on K3. We will now review how to understand these
cases, following \cite{Witten:1996hc} (see also \cite{Freed:2004yc}).

Let us start by describing what happens when the relevant bordism
group is discrete, for instance
$\Omega_1^{\Spin^c}(B\mathbb{Z}_n)=\mathbb{Z}_n$. While the Dai-Freed
prescription does not apply to the generator $X$ of
$\Omega_1^{\Spin^c}(B\mathbb{Z}_n)$, it does apply to the manifold
obtained by taking $n$ disjoint copies of $X$. We would like to define
the phase of the partition function on $X$ by taking an $n$-th root,
but this procedure is ambiguous, so we need to specify additional data
(a choice of $n$-th root). Different choices differ from each other by
a map from $\Omega_1^{\Spin^c}(B\mathbb{Z}_n)$ to a phase. We can
think of this map as a topological field theory that we can couple to
our system, parametrized in terms of a coupling defined modulo $n$ --- a
sort of discrete $\theta$ angle.

It is easy to see that the case in which the bordism group includes
free factors can be understood in similar terms. There is an ambiguity
in the Dai-Freed procedure that we fix by specifying the phase in the
generator of the bordism group; this removes the ambiguity. Different
choices of this phase are related by coupling to a topological field
theory. For instance, the non-trivial elements in
$\Omega_4^{\Spin}=\bZ$ are measured by $\int\! p_1(TX)$, and the (now
continuous) coupling is the usual gravitational $\theta$ angle.

\medskip

There is one interesting question arising naturally from this
viewpoint, which we now briefly explore. It arises from the fact that
it is not true that every non-trivial bordism class can be detected by
integrating characteristic classes. Rather, often one must resort to
computations in K-theory
\cite{Hattori,STONG1965267,HopkinsHovey}. That is, we can detect
certain bordism classes by taking indices (perhaps mod 2) of suitable
Dirac operators. So the more general possibility is that we have
``K-theoretic $\theta$ angles'': bordism-invariant characteristic
numbers not expressible as integrals of characteristic classes. A
five-dimensional example is simply the $\eta$ invariant that appears
in Witten's $SU(2)$ anomaly \cite{Witten:1982fp}. We can view this as
a $\bZ_2$-valued TQFT, and introduce a discrete $\theta$ angle. This
angle is the usual ``discrete $\theta$ angle'' in 5d. The same happens
in 9d, see for example \cite{Bergman:2013ala}, which implies that
Sethi's string \cite{Sethi:2013hra} can also be understood in this
framework.

Can we find any example of this phenomenon for Lie groups in four
dimensions? A review of the results in previous sections does not give
rise to any example, suggesting that the answer may be negative, at
least on $\Spin$ manifolds.\footnote{Similarly to how
  $\Omega_d^\Spin(\pt)$ bordism groups themselves provide examples of
  such exotic angles in one and two dimensions,
  $\Omega^{\Pin^+}_4(\pt)=\bZ_{16}$ provides an example in four
  dimensions. (This group is generated by $\bR\bP^{4}$.) So there is a
  notion of K-theoretic $\theta$ angle in the gravitational sector
  once one allows for non-orientable $\Pin^+$ manifolds. In the text
  we are interested in ``gauge-theoretic'' angles, namely those in the
  reduced $\Spin$ bordism group (see
  appendix~\ref{app:reduced-bordism}).}  More specifically, the
argument in section~\ref{sec:Hurewicz} shows that for all simply
connected forms of semi-simple Lie groups $\Omega_4^\Spin(BG)$ only
receives contributions that can be measured via characteristic
classes. One obtains the same result for various non-simply connected
cases: $SO(n)$ in section~\ref{sec:SO(n)}, and $SU(n)/\bZ_n$ in
section \ref{sec:psun}, at least when $n$ is an odd prime power; these have not yielded any examples of
K-theoretic angles either.

Another potential candidate comes from manifolds with $\Spin^\bZ_4$
structure, discussed in appendix~\ref{sec:SpinZ4-bordism}, but we
argue there that there is no K-theoretic $\theta$ angle in this case
either.

We can in fact prove that, at least in the four-dimensional case,
there are no purely real K-theoretic $\theta$ angles. By definition, a
K-theory $\theta$ angle is a topological field theory that only
depends on a (real) K-theory class. Such a class can always be
represented by a stable real vector bundle, i.e. a $SO(n)$ vector
bundle with $n$ large enough. In \cite{10.2307/1970030}, it is proven
that such a bundle over an arbitrary four-dimensional manifold is
completely determined by its second and fourth Stiefel-Whitney classes
together with its Pontryagin class (see \cite{woodward1982} for a
partial result in dimension up to 8). This means that all K-theory
invariants can be described in terms of cohomology.  However, we
emphasize that this does not mean that all topological couplings in 4d
can be described via cohomology; this is just the case if the relevant
data can be encoded as a real K theory class. While this is often the
case e.g. for the index of a Dirac operator, there may be more general
topological theories which rely on finer topological data. We hope to
come back to this issue in future work.

\section{Conclusion and summary}\label{sec:conclus}

We have explored Dai-Freed anomalies in four-dimensional theories,
both for continuous and discrete groups, as well as a few selected
higher-dimensional examples. Morally, these anomalies can be
understood as an extension of the traditional global anomaly
computation where the mapping torus is replaced by a more general
manifold, as in figure~\ref{f4}.

Since, in the absence of local anomalies, the $\eta$ invariant used to
study the anomaly is a bordism invariant, the first step is the
computation of the relevant bordism groups. We have summarized our
results in table~\ref{tresumen}. The fact that the GUT groups $SU(5)$
and $Spin(10)$ have a vanishing group means that they are free of
Dai-Freed anomalies. We have also argued that this conclusion also
extends to the SM gauge group, whatever its global structure. Overall,
we find that for simple Lie groups there are no new anomalies, since
all the nonzero entries in table~\ref{tresumen} can be accounted for
by known global anomalies.

\begin{table}[!hbt]
\begin{center}

\begin{tabular}{|c|ccccccccc|}
\hline

\multirow{2}{*}{$\mathbf{G}$}&\multicolumn{8}{c}{\quad\quad$\mathbf{\Omega_d^{\Spin}(BG)}$}& \\\cline{2-10}
& 0&1&2&3&4&5&6&7&8\\\hline
$SU(2)$ & $\bZ$ & $\bZ_2$ & $\bZ_2$ & $0$ & $2\bZ$ & $\bZ_2$&$\bZ_2$ & $0$ & $4\bZ$\\\hline
$SU(n> 2)$ & $\bZ$ & $\bZ_2$ & $\bZ_2$ & $0$ & $2\bZ$ & $0$&-- &-- & --\\\hline
$USp(2k>2)$&$\bZ$ & $\bZ_2$ & $\bZ_2$ & $0$ & $2\bZ$ & $\bZ_2$&$\bZ_2$ & $0$ & $5\bZ$\\\hline
$U(1)$ &$\bZ$ & $\bZ_2$ & $\bZ_2 \oplus \bZ$ & $0$ & $2\bZ$ & 0&-- &-- & --\\\hline
$PSU(2^k)$ & $\mathbb{Z}$ & $\mathbb{Z}_2$ & $\mathbb{Z}_2\oplus \mathbb{Z}_{2^k}$ & $0$ & -- &--&-- &-- & --\\\hline
$PSU(p^k,p\,\, \text{odd})$ & $\mathbb{Z}$ & $\mathbb{Z}_2$ & $\mathbb{Z}_2\oplus \mathbb{Z}_{p^k}$ & $0$ & $2\mathbb{Z}$ & 0&-- &-- & --\\\hline
$Spin(n\geq 8)$ & $\bZ$ & $\bZ_2$ & $\bZ_2$ & $0$ & $2\bZ$ & $0$&-- &-- & --\\\hline
$SO(3)$ & $\bZ$ & $\bZ_2$ & $e(\bZ_2,\bZ_2)$ & $0$ & $2\bZ$ & $0$&-- &-- & --\\\hline
$SO(n> 3)$ & $\bZ$ & $\bZ_2$ & $e(\bZ_2,\bZ_2)$ & $0$ & $e(\bZ,\bZ\oplus\bZ_2)$ & $0$&-- &-- & --\\\hline
$E_6,E_7,E_8$ & $\bZ$ & $\bZ_2$ & $\bZ_2$ & $0$ & $2\bZ$ & $0$&0&0 & $2\bZ$\\\hline
$G_2$& $\bZ$ & $\bZ_2$ & $\bZ_2$ & $0$ & $2\bZ$ & $0$&-- &-- & --\\\hline
$F_4$& $\bZ$ & $\bZ_2$ & $\bZ_2$ & $0$ & $2\bZ$ & $0$&0 &0 & --\\\hline
 \end{tabular}
\end{center}
\caption{Bordism groups of semisimple Lie groups computed in the
  text. The discrete groups we use have been computed in
  \cite{GilkeyBook} (see also appendix \ref{sec:AHSSZn}).}
\label{tresumen}
\end{table}

We also studied discrete symmetries in four dimensions. In this case,
the result is different, and one gets genuinely new Dai-Freed
anomalies. The constraints we obtain are stronger than the (linear)
Iba\~{n}ez-Ross constraints. A particularly interesting case is the
$\mathbb{Z}_3$ or $\mathbb{Z}_6$ discrete symmetries that are commonly
imposed in the MSSM to guarantee proton stability. While these have
long been known to be free of Ibañez-Ross anomalies even for a single
generation, we find a nonvanishing modulo 9 Dai-Freed anomaly. The
charge of a single generation is 3 modulo 9, so while the MSSM with
one generation is anomalous, the full MSSM with three generations is
Dai-Freed anomaly free.

These particular Dai-Freed anomalies can also be cancelled by coupling
to a suitable topological quantum field theory, in a discrete version
of the Green-Schwarz mechanism. This coupling forbids the bundles
which give rise to the anomalies, thereby removing the constraints
from the spectrum. As a result, cancellation of Dai-Freed anomalies is
not necessary for consistency of the IR theory - but these anomalies
provide information about topological terms in the theory and on which
manifolds does the theory make sense. For instance, proton triality in
the MSSM with just one generation cannot be coupled to an arbitrary
$\mathbb{Z}_3$ bundle, in spite of the fact that the IR theory seems
to have a $\mathbb{Z}_3$ symmetry.

One of the first discussions of Dai-Freed anomalies was in the
condensed matter literature, where it was found that a 3d Majorana
fermion (topological superconductor) on a nonorientable manifold has a
modulo 16 anomaly, so we need 16 fermions to cancel it. Interestingly,
the Standard Model with right-handed neutrinos also has 16
(four-dimensional) fermions per generation. We were able to relate
these two 16's, if we gauge a particular $\mathbb{Z}_4$ symmetry of
the Standard Model + right-handed neutrinos to make sense of the
theory on manifolds with a $\Spin^{\mathbb{Z}_4}$ structure.

Interestingly, the same construction is possible in the MSSM --- the
theory makes sense on manifolds with a $\Spin^{\mathbb{Z}_4}$
structure. This may be either a coincidence, or a clue about the UV
completion; for instance, a geometric $\mathbb{Z}_2$ symmetry in the
internal space can give rise to a $\Spin^{\mathbb{Z}_4}$ structure.

The same theories that we use to describe anomalies in $d$ dimensions
also provide interesting topological field theories in $(d+1)$
dimensions. These can be viewed as a generalization of $\theta$
angles. Sometimes these angles are purely KO-theoretic, i.e. they
cannot be described by the integral of a cohomology class. We
discussed the situation in four dimensions in
section~\ref{sec:KOthetaangle}.

\medskip

We have only explored cancellation of Dai-Freed anomalies in a few
examples, and it is possible that we missed some phenomenologically
interesting cases. A more systematic exploration of anomaly
cancellation for discrete symmetries seems very worthwhile. And more
generally, it would also be important to determine whether examples of
mixed \mbox{discrete-$G_{SM}$} anomalies exist, where
$G_{SM}=(SU(3)\times SU(2)\times U(1))/\Gamma$ is the gauge group of
the standard model.

 Furthermore, our discussion for Lie groups in section~\ref{sec:lie}
 admits a very natural generalization. The classifying space of an
 abelian group is another abelian group, so we can view abelian
 $p$-form theories as the gauge theories for the abelian groups
 $K(\bZ, p)$. So one could try to compute the bordism groups for any of
 these theories. These will have potential anomalies (recall the
 results for $K(\bZ,4)$ \cite{Stong-11}), and it would be rather
 interesting to understand if any of these non-trivial bordism groups
 give rise to non-trivial physical anomalies. A related direction is to compute the bordism groups
 for $K(\Gamma, p)$, with $\Gamma$ some discrete group. Presumably, these would classify anomalies of discrete generalized symmetries. 

\acknowledgments

We thank Alain Cl\'{e}ment-Pavon, Markus Dierigl, Peter Gilkey, Sebastian Greiner, Ben Heidenreich, Chang-Tse Hsieh, Luis Ibañez, Angel Uranga and Gianluca Zoccarato for very useful discussions and comments on the manuscript. We especially thank Diego Regalado for initial collaboration, and
many illuminating discussions. We also thank the organizers and participants of Strings 2018 and StringPheno 2018 for many fruitful conversations. MM is supported by a postdoctoral fellowship from the Research Foundation -- Flanders.

\appendix

\addtocontents{toc}{\protect\setcounter{tocdepth}{1}}

\section{On reduced bordism groups }

\label{app:reduced-bordism}

Consider the bordism group $\Omega_d$, which we think of as the group
of $d$-dimensional manifolds (possibly with some structure, such as an
orientation, framing, $\Spin$ structure, \ldots), under the
equivalence relation $X_1=X_2$ iff there is some manifold $Y$ such
that $\partial Y = X_1-X_2$. The group operation is given by the
disjoint union of manifolds.

We can construct the group $\Omega_d(Z)$ by decorating the structure
above with maps $\mu\colon X\to Z$ and $\nu\colon Y\to Z$, compatible
in the natural way. (Clearly, $\Omega_d =\Omega_d(\pt)$.) This provides
a potential refinement of the bordism classes: a pair $(X_1,\mu_1)$
may not be equivalent to $(X_2,\mu_2)$, even if $X_1\sim X_2$ in
$\Omega_d$.

In this appendix we would like to discuss the forgetful map
\begin{equation}
  \Phi\colon \Omega_d(Z) \to \Omega_d(\pt)\cong \Omega_d
\end{equation}
defined by $\Phi([X,\mu])=[X]$, where we have picked an arbitrary
representative of a given class $\omega\in \Omega_d(Z)$. This map is
well defined: if $(X_1,\mu_1)$, $(X_2,\mu_2)$ are two distinct
representatives of $\omega$, we can choose any $(Y,\nu)$ such that
$\partial Y = X_1-X_2$ (and $\nu|_{\partial Y}=(\mu_1,\mu_2)$), and
then $Y$ gives a bordism between $X_1$ and $X_2$ in $\Omega_d$.

Furthermore, this map in surjective: every element in $\Omega_d$ can
be understood as $\Phi(\omega)$ for some (potentially many)
$\omega\in\Omega_d(Z)$. To see this, note that we can construct a
partial converse $\Psi\colon \Omega_d(\pt) \to \Omega_d(Z)$: pick an
arbitrary point ``$\pt$'' in $Z$. Choosing a representative $X$ of
$\omega$, we set $\Psi(X)=(X, \pt)$.\footnote{In all the applications
  in this paper $Z$ will be the classifying space of some group, so
  the statement that we are making in this case is that there is a
  natural notion of decorating an arbitrary manifold with a trivial
  principal bundle of the group.}  This map is well defined: given $Y$
such that $\partial Y = X_1 - X_2$, we have that $(Y,\pt)$ is a
bordism in $\Omega_d(Z)$ between $(X_1,\pt)$ and $(X_2,\pt)$. Clearly,
$\Phi\circ\Psi$ is the identity.

Since $\Phi$ is surjective, we can construct the short exact
sequence
\begin{equation}
  \label{eq:reduced-bordism}
  0 \to \ker\Phi \to \Omega_d(Z) \xrightarrow{\,\Phi\,}
  \Omega_d(\pt) \to 0\, .
\end{equation}
A convenient notation is $\tOmega_d(Z) \equiv \ker\Phi$, and
$\tOmega_d(Z)$ is usually called the ``reduced bordism group''.

It is perhaps not immediately clear whether~\eqref{eq:reduced-bordism}
splits, but the answer follows from the fact that $\Phi\circ\Psi=1$
and the splitting lemma for abelian groups \cite{Hatcher:478079}. We
have
\begin{equation}
  \label{eq:reduced-splitting}
  \Omega_d(Z) \cong \Omega_d(\pt)\oplus \tOmega_d(Z) \, .
\end{equation}

\medskip

These facts about the map $\Phi$ can in principle be useful when
computing the action of AHSS differentials: the end result should
never be ``smaller'' than the bordism class of a point, and we get
partial information about the extension problem from the splitting of
the exact sequence. They also have an interesting physical
interpretation: in some sense $\Omega_d(BG)$ encodes all anomalies of
the theory, both gravitational, gauge and mixed, while $\tOmega_d(BG)$
encodes the purely gauge and mixed gravity-gauge ones. So coupling to
a gauge bundle cannot remove gravitational anomalies, as one
intuitively expects.

\section{Tables of bordism groups of a point}\label{app:bgrs}

For reference, here we list tables of $\Omega_d(\pt)$ for different
bordism theories that appear in the text. The original reference is
\cite{ABP} for the $\Spin$ case (see \cite{Stong-11} for explicit
tables), \cite{KirbyTaylor} for $\Pin^+$, \cite{ABPPin-} for $\Pin^-$,
and \cite{bahri1987} for $\Spin^c$ and $\Pin^c$. A similar table appears in
\cite{Kapustin:2014dxa}.

\begin{equation}
  \label{eq:bordism-table}
  \def\arraystretch{1.5}
  \arraycolsep=4pt
  \begin{array}{c|c|c|c|c|c}
    d & \Omega^\Spin_d(\pt) &  \Omega^{\Pin^-}_d(\pt) & \Omega^{\Spin^c}_d(\pt) & \Omega^{\Pin^+}_d(\pt)  &  \Omega^{\Pin^c}_d(\pt)\\\hline
    0&\bZ&\bZ_2&\bZ&\bZ_2&\bZ_2\\
    1&\bZ_2&\bZ_2&0&0&0\\
    2&\bZ_2&\bZ_8&\bZ&\bZ_2&\bZ_4\\
    3&0&0&0&\bZ_2&0\\
    4&\bZ&0&2\bZ&\bZ_{16}&\bZ_2\oplus\bZ_8\\
     5&0&0&0&0&0\\
     6&0&\bZ_{16}&2\bZ&0&\bZ_4\oplus\bZ_{16}\\
     7&0&0&0&0&0\\
     8&2\bZ&2\bZ_2&4\bZ&\bZ_2 \oplus \bZ_{32}&\begin{array}{c}2\bZ_2 \oplus \bZ_{8}\\[-2.67mm]\oplus \bZ_{32}\end{array}\\
     9&2\bZ_2&2\bZ_2&0&0&0\\
     10&3\bZ_2&\begin{array}{c}\bZ_2\oplus \bZ_8\\[-2.67mm] \oplus\bZ_{128}\end{array}&4\bZ\oplus\bZ_2&3\bZ_2&\begin{array}{c}\bZ_2\oplus 2\mathbb{Z}_4\\[-2.67mm]\oplus\bZ_{16}\oplus  \bZ_{64}\end{array} \\
      11&0&\bZ_2&0&3\bZ_2&\bZ_2\\
      12&3\bZ&\bZ_2&7\bZ& \begin{array}{c}\bZ_{256}\oplus\bZ_{16}\\[-2.67mm]\oplus\bZ_{4}\end{array}&\begin{array}{c}4\bZ_2\oplus 2\bZ_8\\[-2.67mm]\oplus \bZ_{32}\oplus\bZ_{128}\end{array}
    \end{array}
\end{equation}

\section{Bordism groups for \texorpdfstring{$\bZ_k$}{Zk}}\label{sec:AHSSZn}

We want to compute various bordism groups for $B\bZ_n$, the
classifying space for $\bZ_n$. We have that $B\bZ_n=K(\bZ_n,1)$ is the
infinite dimensional lens space $L_n^\infty$ defined as follows (see
\S1.B of \cite{Hatcher:478079}). Consider the space $\bC^k$, and take the
$S^{2k-1}$ embedded in it at radius one, using the natural
metric. Consider the action given by multiplication of all the $z_i$
coordinates of $\bC^k$ by a simultaneous phase
$\omega_n \equiv e^{2\pi i/n}$
\begin{equation}
  \Lambda\colon (z_1,\ldots, z_k) \to (\omega_n z_1, \ldots, \omega_n
  z_n)\, .
\end{equation}
We denote $L_n^k=S^{2k-1}/\Lambda$. There is an obvious family of
inclusions $\iota\colon L_n^{k}\subset L_n^{k+1}$, obtained by setting $z_{k+1}=0$
in $L_n^{k+1}$. These embeddings in fact provide generators for the
(torsion) odd homology groups of $L_n^{k+1}$. The homology groups of
$L_n^k$ are (\cite{Hatcher:478079}, Example 2.43)
\begin{equation}
  H_i(L_n^k,\bZ) = \begin{cases}
    \bZ & \quad \text{when } i=0\, ,\\
    \bZ_n & \quad \text{when } 1\leq i<2n-1 \text{ and } i\in 2\bZ+1\, ,\\
    \bZ & \quad \text{when } i = 2n-1\, ,\\
    0 & \quad \text{otherwise.}
  \end{cases}
\end{equation}

We define $B\bZ_n=L_n^\infty$ to be the formal limit of the inclusions
$\iota$ when $k\to\infty$, with the homology
\begin{equation}
  H_i(L_n^\infty,\bZ) = \begin{cases}
    \bZ & \quad \text{when } i=0\, ,\\
    \bZ_n & \quad \text{when } i\in 2\bZ+1\, ,\\
    0 & \quad \text{otherwise.}
  \end{cases}
\end{equation}

As above, we are ultimately interested in the case with coefficients
in some bordism ring. We obtain these by application of the universal
coefficient theorem~\eqref{eq:UCF-SU(2)}, which in our current context
can be easily seen to imply
\begin{equation}
  \label{eq:Zn-homology}
  H_i(B\bZ_n, \Omega) = \begin{cases}
    \Omega & \quad \text{when } i = 0\, ,\\
    \Omega\otimes \bZ_n \cong \Omega/n\Omega & \quad \text{when }
    i\in 2\bZ+1\, ,\\
    \Tor(\bZ_n, \Omega) & \quad \text{otherwise.}
  \end{cases}
\end{equation}
For the cases of interest to use we will need that \cite{Hatcher:478079}
\begin{equation}
  \Tor(\bZ_n, \bZ) = 0 \qquad \text{and} \qquad \Tor(\bZ_n, \bZ_k) =
  \bZ_n\otimes \bZ_k = \bZ_{\gcd(k,n)}\, .
\end{equation}

\subsection{$\Spin^c$ bordism}

\begin{figure}
  \centering
  \includegraphics{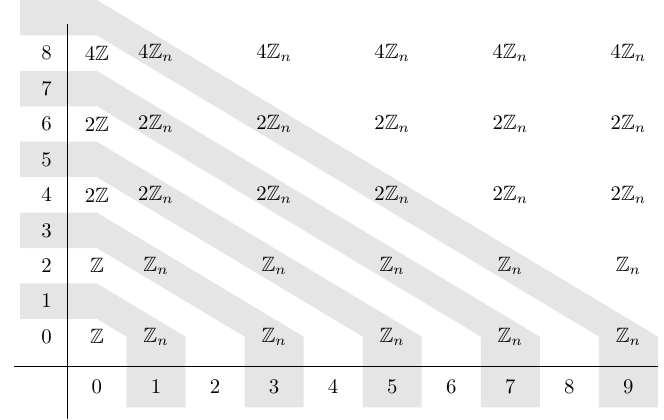}
  \caption{$E_2$ page of the AHSS for $\Omega^{\Spin^c}_*(B\bZ_n)$,
    with the odd degree entries shaded.}
  \label{fig:Spin^c-Zn-AHSS}
\end{figure}

We will start by computing the $\Spin^c$ bordism groups, in order to
compare with the results in \cite{GilkeyBook}. The basic ingredient will
be the $\Omega_k^{\Spin^c}(\pt)$ groups, given by \cite{GilkeyBook}
\begin{equation}
  \def\arraystretch{1.5}
  \arraycolsep=4pt
  \begin{array}{c|ccccccccccc}
    n & 0 & 1 & 2 & 3 & 4 & 5 & 6 & 7 & 8 & 9 & 10 \\
    \hline
    \Omega^{\Spin^c}_n(\pt) & \bZ & 0 & \bZ & 0 & 2\bZ & 0 & 2\bZ
                                  & 0 & 4\bZ & 0 & 4\bZ\oplus \bZ_2
  \end{array}
\end{equation}
It is now immediate to construct the first page of the AHSS spectral
sequence, which we show in figure~\ref{fig:Spin^c-Zn-AHSS}.

One simplifying feature of the $\Spin^c$ case is that there is no
torsion in $\Omega^{\Spin^c}_k(\pt)$ for $k<10$, so using the fact
that $d\colon \bZ_n\to \bZ$ necessarily vanishes (either for degree
reasons, or because $\bZ_n\to\bZ$ homomorphisms are always vanishing)
we see that $E_{p,q}^2=E_{p,q}^\infty$ for $p+q<10$. So we immediately
conclude that $\Omega^{\Spin^c}_k(B\bZ_n)=\Omega^{\Spin^c}_k(\pt)$ for
$k<10$, $k\in2\bZ$.

Since all torsion in $\Omega^{\Spin^c}(\pt)$ comes from $\bZ_2$ factors
\cite{GilkeyBook}, in the case that $n\in2\bZ+1$ we have that all
differentials vanish, the spectral sequence collapses at the second
page already, and in addition (looking to the degree of the
differentials) $\Omega^{\Spin^c}_k(B\bZ_n)=\Omega^{\Spin^c}_k(\pt)$
for all $k\in2\bZ$.

For $k\in \{1,3,\ldots,9\}$ we also have that the relevant
differentials all vanish, so we conclude that
\begin{equation}
  \label{eq:Omega^{Spin^c}(BZn)}
  \begin{split}
    \Omega^{\Spin^c}_1(B\bZ_n) = \bZ_n \quad ; \quad
    \Omega^{\Spin^c}_3(B\bZ_n) = e(\bZ_n, \bZ_n) \quad ; \quad
    \Omega^{\Spin^c}_5(B\bZ_n) = e(2\bZ_n, \bZ_n, \bZ_n) \\
    \Omega^{\Spin^c}_7(B\bZ_n) = e(2\bZ_n, 2\bZ_n, \bZ_n, \bZ_n) \quad
    ; \quad \Omega^{\Spin^c}_9(B\bZ_n) = e(4\bZ_n, 2\bZ_n, 2\bZ_n,
    \bZ_n, \bZ_n)\, .
  \end{split}
\end{equation}
Here we have defined $e(A,B)$ to be some (yet unknown) extension of
$B$ by $A$, i.e. some $C$ such that $0\to A \to C \to B \to 0$ is
exact. We then define
\begin{equation}
  e(A_1, A_2, \ldots, A_n) = e(e(e(\ldots e(A_1,A_2), A_3), \ldots A_n)
\end{equation}
to be the left associative generalization of $e(A,B)$.

One can easily compare these results to those listed in
\cite{GilkeyBook}. For instance, consider the case $n=4$. According to
\cite{GilkeyBook} we have $\Omega^{\Spin^c}_3(B\bZ_n) = \bZ_8\oplus
\bZ_2$. This is compatible with~\eqref{eq:Omega^{Spin^c}(BZn)} since
\begin{equation}
  0 \to \bZ_4 \xrightarrow{\,f\,} \bZ_8\oplus \bZ_2 \xrightarrow{\,g\,} \bZ_4 \to 0\, .
\end{equation}
is exact if we choose $f(1)=(2,1)$ and $g(1,0)=1$, $g(0,1)=2$.

To finish the comparison with \cite{GilkeyBook}, let us note that for even
$n$ there is a non-vanishing contribution to $E_{2,10}^2$ from
\begin{equation}
  \Tor(\bZ_n,\Omega^{\Spin^c}_{10}(\pt))=\Tor(\bZ_n, 4\bZ\oplus \bZ_2) =
  \Tor(\bZ_n, \bZ_2) = \bZ_2\, ,
\end{equation}
which explains the $\bZ_2$ contribution to
$\Omega^{\Spin^c}_{12}(B\bZ_n)$ shown in \cite{GilkeyBook}. (Note that
\cite{GilkeyBook} lists the \emph{reduced} bordism groups, the full
bordism group is $\Omega^{\Spin^c}_{12}(B\bZ_n) =
\Omega^{\Spin^c}_{12}(\pt)\oplus \bZ_2 = 7\bZ\oplus\bZ_2$.)

\subsection{$\Spin$ bordism, with $n$ odd}

\begin{figure}
  \centering
  \includegraphics{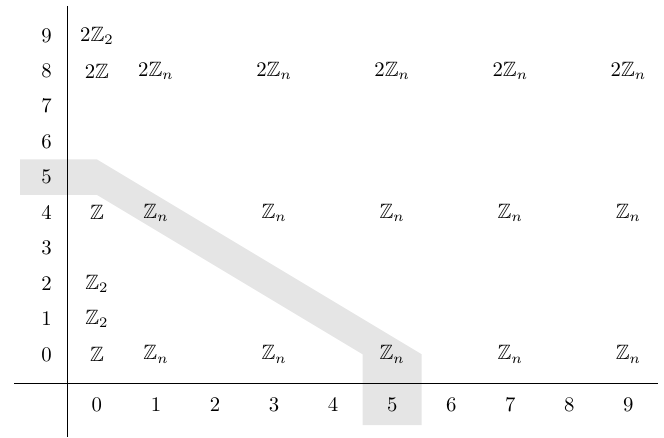}
  \caption{$E_2$ page of the AHSS for $\Omega^\Spin_*(B\bZ_n)$, for
    $n$ odd. We have shaded the contributions relevant for the
    computation of four-dimensional anomalies.}
  \label{fig:Spin-Zn-AHSS}
\end{figure}

The exercise for $\Omega^{\Spin}(B\bZ_n)$ proceeds similarly. For
simplicity we specialize to $n\in2\bZ+1$. In this case, since
$\Tor(\bZ_n,\bZ_2)=0=\bZ_n\otimes\bZ_2$, we are led to a rather
simple spectral sequence, shown in figure~\ref{fig:Spin-Zn-AHSS}.

We will restrict to $p+q<9$. Since the differentials $d_r$ have
bidegree $(-r,r-1)$ we immediately see that there is no non-vanishing
differential acting on the degrees of interest.\footnote{One can show
  that the torsion components of $\Omega^\Spin(\pt)$ are all of the
  form $\bZ_{2^m}$ \cite{Novikov,ABP}, so the result follows in
  general. We then have that $\Omega_k^{\Spin}(\pt)\otimes
  \bZ_n=\bZ_n$ if $k\in 4\bZ$, and vanishes otherwise.} We find
\begin{equation}
  \def\arraystretch{1.5}
  \arraycolsep=4pt
  \begin{array}{c|cccccc}
    d & 0 & 1 & 2 & 3 & 4 & 5 \\
    \hline
    \Omega^{\Spin}_d(B\bZ_n) & \bZ & \bZ_2\oplus\bZ_n & \bZ_2 & \bZ_n &
                                                                    \bZ & e(\bZ_n, \bZ_n)                
  \end{array}
\end{equation}
and also
\begin{equation}
  \def\arraystretch{1.5}
  \arraycolsep=4pt
  \begin{array}{c|ccccc}
    d &6 & 7 & 8 & 9 & 10\\
    \hline
    \Omega^{\Spin}_d(B\bZ_n)   & 0 & e(\bZ_n, \bZ_n) &
                                                                    2\bZ
                                          & 2\bZ_2\oplus e(2\bZ_n,
                                            \bZ_n, \bZ_n) &3\bZ_2 
  \end{array}
\end{equation}

\subsection{$\Spin$ bordism for $B\bZ_2$}

\label{sec:BZ2-Spin}

The case of even $n$ is more involved, as there are many more
non-vanishing entries. We do not attempt a general discussion here,
but rather focus on some features of the $\bZ_2$ case. As we discuss
below, there is a more efficient way of computing
$\Omega_*^{\Spin}(B\bZ_2)$ than using the Atiyah-Hirzebruch spectral
sequence, but the spectral sequence computation will come useful in
the next section. The homology groups relevant for this case can be
read off from~\eqref{eq:Zn-homology}
\begin{equation}
  H_i(B\bZ_2, \bZ) = \begin{cases}
    \bZ & \quad \text{when } i=0\, ,\\
    \bZ_2 & \quad \text{when } i\in 2\bZ+1\, ,\\
    0 & \quad \text{otherwise.}
  \end{cases}
\end{equation}
and $H_i(B\bZ_2, \bZ_2)=\bZ_2$ for all $i\geq 0$. (Alternatively,
these results follow simply from the fact that
$B\bZ_2=\bR\bP^\infty$.)

\begin{figure}
  \centering
  \includegraphics{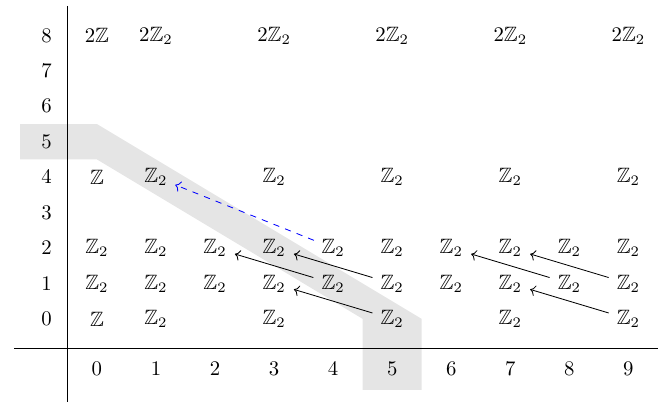}
  \caption{$E_2$ page of the AHSS for $\Omega^\Spin_*(B\bZ_2)$. We
    show the non-vanishing differentials $d_2$ in solid black, and a
    $d_3$ in dashed blue that should vanish in order to reproduce the
    results of the Smith isomorphism~\eqref{eq:Spin-Z2-Smith}.}
  \label{fig:Spin-Z2-AHSS}
\end{figure}

The second page of the AHSS resulting from this is shown in
figure~\ref{fig:Spin-Z2-AHSS}. We see that there are many potentially
differentials, and many extension problems to be solved, so we will
not solve the issue completely. Nevertheless, some useful information
can be teased out of the spectral sequence. Clearly,
$\Omega_0^\Spin(B\bZ_2)=\bZ$ and
$\Omega_1^\Spin(B\bZ_2)=\bZ_2\oplus \bZ_2$, simply because there are
no differentials that could enter act on the corresponding entries of
the spectral sequence. (In the second identity we have used the splitting
result~\eqref{eq:reduced-splitting}.)

Going beyond this requires computing some differentials, using the
technology discussed in section~\ref{sec:steenrods}. We have that,
as a ring, $H^*(B\bZ_2,\bZ_2)$ is freely generated by $w_1$, the
generator of $H^1(B\bZ_2, \bZ_2)$:
\begin{equation}
  H^*(B\bZ_2,\bZ_2) = H^*(\bR\bP^\infty, \bZ_2) = \bZ_2[w_1]\, .
\end{equation}
Using the properties~\eqref{eq:Steenrod-properties} it is then simple
to show the relations
\begin{equation}
  \Sq1 (w^n) = w\Sq1(w^{n-1}) + w^{n+1}\quad ; \quad \Sq2(w^n) =
  w^2\Sq1(w^{n-1}) + w\Sq2(w^{n-1})\, .
\end{equation}
Using $\Sq1(w) = w^2$ and $\Sq2(w)=0$, these are solved by
\begin{equation}
  \Sq1 (w^n)=n w^{n+1}\quad \text{and} \quad \Sq2(w^n)= \frac{n(n-1)}{2}w^{n+2}\, ,
\end{equation}
with coefficients understood modulo 2. The result is that the
differentials which are non-vanishing on the second page are those
shown in figure~\ref{fig:Spin-Z2-AHSS}, where we have used in addition
that the reduction modulo two map
$\rho\colon H_{2k+1}(B\bZ_2, \bZ)\to H_{2k+1}(B\bZ_2, \bZ_2)$ is
surjective, which follows easily from $H_{2k}(B\bZ_2, \bZ)=0$ and
exactness of~\eqref{eq:rho-long}.

It is not straightforward to make much further progress using the
Atiyah-Hirzebruch spectral sequence, but luckily there is a Smith
isomorphism that comes to the rescue here \cite{Kapustin:2014dxa}:
\begin{equation}
  \label{eq:Spin-Z2-Smith}
  \Omega^\Spin_d(B\bZ_2) \cong \Omega_{d-1}^{\Pin^-}(\pt)\oplus
  \Omega^\Spin_d(\pt)\, .
\end{equation}
Using this isomorphism one finds
\begin{equation}
  \def\arraystretch{1.5}
  \arraycolsep=4pt
  \begin{array}{c|ccccccccc}
    d & 0 & 1 & 2 & 3 & 4 & 5 & 6 & 7 & 8 \\
    \hline
    \Omega^{\Spin}_d(B\bZ_2) & \bZ & 2\bZ_2 & 2\bZ_2 & \bZ_8 & \bZ & 0
                              & 0 & \bZ_{16} & 2\bZ
  \end{array}
\end{equation}
which can be easily checked to be compatible with the structure of the
exact sequence above.

\subsection{$\Spin^{\bZ_4}$ bordism in four dimensions}
\label{sec:SpinZ4-bordism}

As discussed in \cite{TeichnerPhD,Teichner}, the Atiyah-Hirzebruch
spectral sequence for $\Omega^{\Spin^{\bZ_4}}$ agrees on the first
page with the one for $\Omega_*^\Spin(B\bZ_2)$ we have just computed,
but the differentials are different, being twisted.

\begin{figure}
  \centering
  \includegraphics{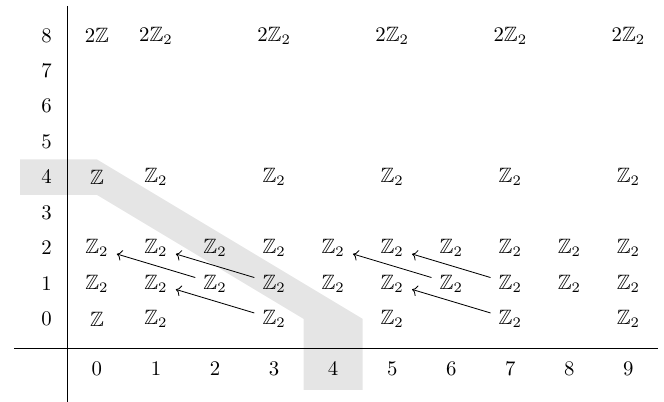}
  \caption{$E_2$ page of the AHSS for
    $\Omega^{\Spin^{\bZ_4}}(\pt)$. We have shaded the entries relevant
    for the computation of four dimensional $\theta$ angles.}
  \label{fig:Spin-Z4-AHSS}
\end{figure}

Clearly the $E_2^{(0,4)}=\bZ$ entry survives to $E_\infty$. We would
now like to argue that $E_2^{(2,2)}$ does too. To see this, notice
that it can only be killed either by being the target of a
differential coming from a term of total degree 5. But no such
differential can exist, since otherwise
$|\Omega_5^{\Spin^{\bZ_4}}|<16$, and this was proven not to be the
case in \cite{Tachikawa:2018njr}. On the other hand, the differential
$d_2^w\colon E_2^{(3,1)}\to E_2^{(1,2)}$ is non-vanishing. This is
because $d_2^w$ is the dual of $\Sq2_w$ \cite{TeichnerPhD,Teichner},
defined as $\Sq2_w(x)=\Sq2(x)+w^2\smile x$, with $w^2$ the generator
of $H^2(B\bZ_2,\bZ_2)$. We $\Sq2 w =0$, so $\Sq2_w(w)=w^3\neq 0$.

We then find that
\begin{equation}
  \label{eq:Z4-extension}
  0 \to \bZ \to \Omega^{\Spin^{\bZ_4}}_4(\pt) \to \bZ_2 \to 0
\end{equation}
is exact. The physical interpretation of this computation depends on
whether this extension is trivial or not. If it is trivial, and
$\Omega^{\Spin^{\bZ_4}}_4(\pt)=\bZ\oplus \bZ_2$, this would give a
candidate for the K-theoretical $\theta$ angles discussed in
\ref{sec:KOthetaangle}. If the extension is non-trivial, so that
$\Omega^{\Spin^{\bZ_4}}_4(\pt)=\bZ$, we would instead have that there
are some $\Spin^{\bZ_4}$ manifolds which have
$\int\! \hat{A}=\frac{1}{24}p_1=1$. (Recall that for four dimensional
$\Spin$ manifolds one has $\int\!\hat{A}\in2\bZ$.)

Either way, an example of a four-dimensional space that is not $\Spin$
but it is $\Spin^{\bZ_4}$ is given by the Enriques surface
$\cE=\text{K3}/\sigma$ (see \cite{barth2015compact} for a review),
where $\sigma$ is a fixed-point-free $\bZ_2$ action on K3. This
surface is not $\Spin$: its signature is 8, while Rochlin's theorem
states that the signature is always a multiple of 16 on
four-dimensional $\Spin$ manifolds. Nevertheless, it admits a
$\Spin^{\bZ_4}$ structure: consider the Voisin-Borcea (Calabi-Yau, and
thus $\Spin$) manifold $X=(\text{K3}\times T^2)/\hat{\sigma}$, where
$\hat\sigma$ acts as $\sigma$ on K3, and as reflection along both
coordinates of the $T^2$. This space can be understood as a $T^2$
fibration with base $\cE$. If we consider spinors on $X$, and reduce
along the $T^2$, we obtain a natural $\Spin^{\bZ_4}$ structure on
$\cE$ (since reflections square to $(-1)^F$, on fermions they act as a
$\mathbb{Z}_4$).

We can now discard the possibility of a trivial extension by the
following argument. Assume that the sequence~\eqref{eq:Z4-extension}
does split. We then have that K3 is a generator of
\mbox{$\Omega_d^{\Spin^{\bZ_4}}=\bZ\oplus\bZ_2$}. The other generator
is some space $X$ which is not $\Spin$, and such that $2X\sim 0$ in
$\Spin^{\bZ_4}$ bordism. Since we showed above that $\cE$ is
$\Spin^{\bZ_4}$, it should be the case that $2\cE\sim 0$ in
$\Omega_4^{\Spin^{\bZ_4}}$. But this is not the case: the embedding
$\bZ_4\to U(1)$ induces an homomorphism
$(\Spin(d)\times \bZ_4)/\bZ_2\to (\Spin(d)\times U(1))/\bZ_2$ which in
turn induces a natural homomorphism
\begin{equation}
  \sigma\colon \Omega_d^{\Spin^{\bZ_4}}\to \Omega_d^{\Spin^c}\, .
\end{equation}
So $2\cE\sim 0$ in $\Omega_4^{\Spin^{\bZ_4}}$ would induce the
relation $2\cE\sim 0$ in $\Spin^c$. A manifold is trivial in $\Spin^c$
iff all its Pontryagin and Stiefel-Whitney characteristic numbers
vanish (see theorem 3.1.1 of \cite{GilkeyBook}), but we have
$p_1(\cE)=24$, so $p_1(2\cE)=2p_1(\cE)=48$, and we arrive to a
contradiction.\footnote{In fact, one has $2\cE\sim \text{K3}$ in
  $\Omega_4^{\Spin^c}$. We can show this by comparing their
  characteristic numbers, and using theorem 3.1.1 of
  \cite{GilkeyBook}. The Stiefel-Whitney numbers of $2\cE$ vanish identically, since Stiefel-Whitney classes are additive under disjoint union. For the integral characteristic numbers, $c_1^2=0$
  in both cases, and $p_1(\cE)=\frac{1}{2}p_1(\text{K3})=-24$, which
  completes the proof.}

Finally, let us list some low degree groups that are easily computable
from the Atiyah-Hirzebruch spectral sequence:
\begin{equation}
  \def\arraystretch{1.5}
  \arraycolsep=4pt
  \begin{array}{c|ccccccc}
    d & 0 & 1 & 2 & 3 & 4 & 5 & 6 \\
    \hline
    \Omega^{\Spin^{\bZ_4}}_d(\pt) & \bZ & e(\bZ_2,\bZ_2) & 0 & 0 & \bZ
                          & \bZ_{16} & 0
  \end{array}
\end{equation}

\section{3d currents}\label{3djs}
Suppose we have two 3d fermions $\lambda_1,\lambda_2$, with Lagrangian
\begin{equation}\lambda_1^T\epsilon \slashed{\partial}\lambda_1+ \lambda_2^T\epsilon \slashed{\partial}\lambda_2.\end{equation}
This system has a $U(1)$ symmetry
\begin{align}\left(\begin{array}{c}\lambda_1\\\lambda_2\end{array}\right)\rightarrow \left(\begin{array}{cc}\cos\theta&-\sin\theta\\\sin\theta&\cos\theta\end{array}\right)\left(\begin{array}{c}\lambda_1\\\lambda_2\end{array}\right).\end{align}
which is associated via Noether's theorem to the current
\begin{equation}J^\mu=\lambda_1^T\epsilon \gamma^\mu\lambda_2.\end{equation}
Mass terms can be constructed with the invariant $\epsilon$ tensor. There is just one possibility compatible with the $U(1)$ symmetry, namely
\begin{equation}\lambda_1^T\epsilon\lambda_1+\lambda_2^T\epsilon\lambda_2.\label{mterm33}\end{equation}
We can also consider a $R$ or $CR$ discrete symmetry, which we will call $\mathcal{S}$, which acts on the fermions with a phase:
\begin{equation}\mathcal{S}_\alpha\lambda_1=\lambda_1, \quad \mathcal{S}_\alpha\lambda_2=\alpha\lambda_2.\end{equation}
The sign $\alpha$ in the second term can be mapped to whether or not $\mathcal{S}$ commutes or anticommutes with the generator of $U(1)$ rotations. If $\alpha=-1$, it anticommutes: when continuing $\mathcal{S}_\alpha$ to Minkowskian signature, it will become a $T$ transformation which commutes with the electric charge, as is usually the case. If $\alpha=+1$, it commutes, which corresponds after analytic continuation to a twisted gauge field which transforms under parity reversal as an ordinary 1-form. 

Both possibilities are acceptable, and they both lead to symmetry protected topological phases, but the mechanism in each case is different:\begin{itemize}
\item If $\alpha=+1$, then parity forbids not only the mass term \eq{mterm33}, but also the only additional possibility 
\begin{equation}\lambda_1^T\epsilon \lambda_2.\label{qwe}\end{equation}
Thus, these fermions are protected by virtue of $\mathcal{S}_\alpha$-symmetry alone; the fact that they are also charged under a $U(1)$ is irrelevant to the question of existence of protected massless modes. The system is actually the $\nu=2$ topological superconductor \cite{Witten:2015aba}; from this discussion we have only learnt that it can be consistently coupled to a twisted gauge field. Since gauge transformations commute with inversions, this is a $\Pin^c$ structure.
\item If $\alpha=-1$, then $\mathcal{S}_\alpha$-symmetry would allow for a mass term \eq{qwe}, so it is not enough to protect the existence of massless modes. However, this mass term is in turn forbidden by the $U(1)$ symmetry, so that the massless fermions are indeed protected: this is the standard topological insulator \cite{Witten:2015aba}.
\end{itemize}
This state of affairs is summarized in table \ref{t55}.

\begin{table}[!hbt]
\begin{center}

\begin{tabular}{|c|c|c|}\hline\textbf{Symmetry}& $\lambda_i^T\epsilon \lambda_i$& $\lambda_i^T\epsilon \lambda_j$
\\\hline $\mathcal{S}_{-1}$ & \xmark&\xmark\\\hline
$\mathcal{S}_{+1}$ & \xmark&\cmark\\\hline
$Q$ & \cmark&\xmark\\\hline
 \end{tabular}
\end{center}
\caption{Different symmetry generators and the mass terms they allow. $\mathcal{S}_{+1}$ is enough to forbid all mass terms - this is the symmetry of the topological superconductor. The combination of $Q$ and $\mathcal{S}_{-1}$ are also enough to forbid all mass terms - this is the symmetry of the topological insulator. Neither $Q$ or $\mathcal{S}_{+1}$ on their own are able to ensure the existence of a massless fermion.}

\label{t55}
\end{table}

\section{Alternate generators for  \texorpdfstring{$\Omega_5^{\Spin}(B\mathbb{Z}_n)$}{Omega5SpinBZn}}
\label{sec:modular-inverse}
Here, we present an alternate set of generators for $\Omega_5^{\Spin}(B\mathbb{Z}_n)$, different to the one used in section \ref{sec:spinzn}. To do this, we have to generalize the notion of a lens space. Pick a vector $\vec{q}=(q_1,q_2,\ldots,q_l)$, where all the entries are coprime. Then we define the generalized lens space $L(n;\vec{q})$ as the quotient of the unit sphere $S^{2l-1}\subset \mathbb{C}^l$ by the equivalence relation
\begin{equation}(z_1,\ldots,z_l)\,\equiv \, (z_1e^{\frac{2\pi i q_1}{n}},\ldots,z_le^{\frac{2\pi i q_l}{n}}).\end{equation}
With this notation, we have $L^l(n)=L(n;1,1\ldots)$.  There is a general expression \cite{GilkeyBook} for the $\eta$ invariant
\begin{equation}\eta(L^l(n; \vec{q}),s)=-\frac{d}{n}\text{Td}_l(n,q_1,\ldots q_l; s-l(q_1+\ldots +q_l),\end{equation}
where $\text{Td}_l$ is a specific linear combination of Todd
polynomials (we refer the reader to \cite{GilkeyBook} for details),
and $d$ is an integer that must satisfy
$d q_1\ldots q_l\equiv 0\,\text{mod}\, 24 n$.

Reference \cite{GilkeyBook} also shows that the bordism group $\Omega_5^{\Spin}(B\mathbb{Z}_n)$ is generated by $L(n;1,1,1)$ and $L(n;1,1,2)$. The $L(n;1,1,1)$ case is straightforward and worked out in the main text. 

The $L(n;1,1,2)$ case is more involved. This is because
$\gcd(2,24n)\neq 1$, so we cannot straightforwardly apply theorem
4.5.4 of \cite{GilkeyBook}. Nevertheless, it is clear from the
definitions above that $L(n;1,1,2)=L(n;1,1,2+3n)$, and it is easy show
that $\gcd(2+3n,24n)=1$ for $n$ odd.\footnote{This is most easily done
  in terms of $k=(n+1)/2$. Then the equality becomes
  $\gcd(6k-1, 48k-24)=1$. The second term is divisible by 8, while the
  first is not, so $\gcd(6k-1, 48k-24)=\gcd(6k-1, 6k-3)$. Since
  $\gcd(a,a+2)$ is either one or 2, and $a$ is odd here, the result
  follows.} So the conditions of the theorem apply to this
presentation of the space. A somewhat tricky point now comes from $d$,
which is defined to be the inverse of $(2+3n)$ modulo $24n$. We would like
to find a polynomial expression for $d$ such that
\begin{equation}
  d(2+3n)\equiv 1 \mod 24n\, .
\end{equation}
From Euler's theorem:
\begin{equation}
  d\equiv(2+3n)^{-1} \equiv (2+3n)^{\phi(24n)-1}\mod 24n
\end{equation}
where $\phi(x)$ is Euler's totient function (counting the number of
positive integers smaller or equal to $x$ that are relatively prime to
$x$). Expanding, we have
\begin{equation}
  (2+3n)^{\phi(24n)-1} = \sum_{p=0}^{\phi(24n)-1} \begin{pmatrix}
    \phi(24n)-1 \\ p \end{pmatrix} 2^p (3n)^{\phi(24n)-1-p}\, .
\end{equation}
Since we work modulo $24n=2^3\cdot 3n$ we can drop the terms in the
sum with $\phi(24n)-2\geq p \geq 3$, and we find
\begin{equation}
  \begin{split}
    (2+3n)^{\phi(24n)-1} & \equiv (3n)^{\phi(24n)-1} +
    (\phi(24n)-1)2(3n)^{\phi(24n)-2} \\ &+
    \frac{1}{2}(\phi(24n)-1)(\phi(24n)-2)2^2(3n)^{\phi(24n)-3} +
    2^{(\phi(24n)-1)}\mod 24n\, .
  \end{split}
\end{equation}
Using $\phi(24n)=\phi(8)\phi(3n)=4\phi(3n)$, this simplifies to:
\begin{equation}
  (2+3n)^{4\phi(3n)-1} \equiv (3n)^{4\phi(3n)-1} - 2(3n)^{4\phi(3n)-2} +
  4(3n)^{4\phi(3n)-3} + 2^{(4\phi(3n)-1)}\mod 24n\, .
\end{equation}

We can simplify this further using that:
\begin{subequations}
  \begin{align}
    (3n)^{4\phi(3n)-1} & \equiv 3n \mod 24n\, , \label{eq:3n-relation-24n}\\
    (3n)^{4\phi(3n)-2} & \equiv 9n^2 \mod 12n \, , \label{eq:3n-relation-12n}\\
    (3n)^{4\phi(3n)-3} & \equiv 3n \mod 6n\, . \label{eq:3n-relation-6n}
  \end{align}
\end{subequations}
These relations can be proven as follows. Consider for
instance~\eqref{eq:3n-relation-24n}. Since $4\phi(3n)-1>0$ for the cases
of interest, both sides include a common factor of $3n$. So
\eqref{eq:3n-relation-24n} is equivalent to
\begin{equation}
  (3n)^{4\phi(3n)-2}  \equiv 1 \mod 8\, .
\end{equation}
We have $\gcd(8,3n)=1$ and $\phi(8)=4$, so $(3n)^4=1$ mod 8, which
implies
\begin{equation}
    (3n)^{4\phi(3n)-2}  \equiv (3n)^2 \mod 8\, .
\end{equation}
Subtracting both equations, we get
\begin{equation}
  (3n)^2 - 1 \equiv (3n+1)(3n-1) \equiv 0 \mod 8\, .
\end{equation}
This follows since we are multiplying two consecutive even numbers,
which necessarily gives a multiple of 8. The two other relations can
be proven similarly:~\eqref{eq:3n-relation-6n} follows from
$3n^k\equiv 1$ mod 2 for all $k>0$ (since $3n$ is odd),
while~\eqref{eq:3n-relation-12n} follows from
\begin{equation}
  (3n)^{4\phi(3n)-3} \equiv 3n \mod 4\, .
\end{equation}

Using these relations, we find that
\begin{equation}
  (2+3n)^{4\phi(3n)-1} \equiv 3n(5-6n) + 2^{(4\phi(3n)-1)}\mod 24n\, .
\end{equation}
We can in fact do better. From Euler's theorem we have that
\begin{equation}
  (2+3n)^{4\phi(3n)-1}(2+3n) \equiv \bigl[3n(5-6n) +
  2^{(4\phi(3n)-1)}\bigr] (2+3n) \equiv 1 \mod 24n\, .
\end{equation}
Expanding, this leads to
\begin{equation}
  2\cdot 2^{(4\phi(3n)-1)} \equiv 1 - (3n)^2 \mod 24n\, .
\end{equation}
As explained above, $(3n)^2-1$ is a multiple of 8, so we can try
dividing both sides by 2 to get a stronger result. Since
$\gcd(2,24n)\neq 1$ we should not expect that dividing by two gives a
correct result. And indeed, after some trial and error we obtain an
ansatz (which we will prove to be correct momentarily) with a
correction term:
\begin{equation}
  \begin{split}
    \label{eq:modular-guess}
    2^{(4\phi(3n)-1)} & \equiv \frac{1}{2} (1-3n^2) + \frac{1}{2} (3n) (3 n^{2} - 8 n + 13)\\
    & \equiv \frac{1}{2} \cdot (9 n^{3} - 33 n^{2} + 39 n + 1) \mod 24n
  \end{split}
\end{equation}
for all odd $n$. Our final result is then that
\begin{equation}
  d \equiv(2+3n)^{-1} \equiv \frac{1}{2} (9 n^{3} - 69 n^{2} + 69 n + 1) \mod 24n \, .
\end{equation}
It is easy to check that $d(2+3n)\equiv 1$ mod $24n$ for $n$ odd
holds, as required.

Using this expression we obtain (again after some simplifications)
\begin{equation}
  \label{eq:eta-L112}
  \eta(L(n;1,1,2), s) \equiv \frac{1}{24n}\left((6n^2-2)s^3 - (7n^2-3)s\right) \mod 1 \, .
\end{equation}

Summarizing, so far we find that a $\bZ_n$ symmetry, with $n$ odd, is
anomaly-free if and only if both
 \begin{equation}\sum_i \left[4s_i^3 - (n^2+3)s_i\right]  \equiv 0\mod 24n\end{equation}

 and \eqref{eq:eta-L112} vanish modulo integers, when summed over all
fermions:
\begin{equation}
  \sum_{i} \left[4s_i^3 - (n^2+3)s_i\right] \equiv \sum_i \left[(6n^2-2)s_i^3 - (7n^2-3)s_i\right] \equiv 0 \mod 24n\, .
\end{equation}

These equations can be simplified: expressing them in terms of
$k=(n+1)/2$, and removing an overall factor, they become:
\begin{subequations}
  \label{eq:Zn-odd-anomaly-mixed}
  \begin{align}
    \label{eq:Zn-odd-anomaly-mixed-a}
    \sum_i \left[s_i^3 - (k^2-k+1)s_i\right] & \equiv 0\mod 6(2k-1)\\
    \label{eq:Zn-odd-anomaly-mixed-b}
    \sum_i \left[(6k^2-6k+1)s_i^3 - (7k^2 - 7k +1)s_i\right] & \equiv 0 \mod 6(2k-1)\, .
  \end{align}
\end{subequations}
Subtracting $(6k^2-6k+1)$ times the first equation from the second we
are led to:
\begin{equation}
  k^2(k-1)^2\sum_i s_i \equiv 0 \mod (2k-1)\, .
\end{equation}
Since $\gcd(k^2(k-1)^2, 2k-1)=1$,\footnote{Clearly $k^2$ and $(k-1)^2$
  do not share any factors, so it suffices to show $\gcd(k,2k-1)=1$
  and $\gcd(k-1,2k-1)=1$ separately. To prove the first relation,
  assume $k=pu$, $2k-1=pv$, for $p>1$ a prime and $u,v\in\bZ$. We have
  $2(pu)-1=pv$ or equivalently $p(2u-v)=1$. But $p$ has no inverse
  over $\bZ$. For the second relation we proceed similarly: $k-1=pu$,
  $2k-1=pv$. Subtracting both equations we learn $k=p(v-u)$, which is
  incompatible with $k=pu+1$ unless $p=1$.} we can invert the
coefficient, and we obtain the equivalent equation
\begin{equation}
  \sum_i s_i \equiv 0 \mod (2k-1)\, .
\end{equation}
So we have simplified~\eqref{eq:Zn-odd-anomaly-mixed} to
\begin{subequations}
  \begin{align}
    \sum_i \left[s_i^3 - (k^2-k+1)s_i\right] & \equiv 0\mod 6(2k-1)\\
    \sum_i s_i & \equiv 0 \mod (2k-1)\, .
  \end{align}
\end{subequations}
or equivalently in terms of $n$
\begin{subequations}
  \begin{align}
    \sum_i \left[s_i^3 - \frac{1}{4}(n^2+3)s_i\right] & \equiv 0\mod 6n\\
    \sum_i s_i & \equiv 0 \mod n\, .
  \end{align}
\end{subequations}
which are precisely \eq{eeeeee}.

\bibliographystyle{JHEP}
\bibliography{refs}

\providecommand{\href}[2]{#2}\begingroup\raggedright\begin{thebibliography}{100}

\bibitem{tHooft:1980xss}
G.~'t~Hooft, C.~Itzykson, A.~Jaffe, H.~Lehmann, P.~K. Mitter, I.~M. Singer
  et~al., \emph{{Recent Developments in Gauge Theories. Proceedings, Nato
  Advanced Study Institute, Cargese, France, August 26 - September 8, 1979}},
  \href{http://dx.doi.org/10.1007/978-1-4684-7571-5}{\emph{NATO Sci. Ser. B}
  {\bf 59} (1980) pp.1--438}.

\bibitem{Witten:1982fp}
E.~Witten, \emph{{An SU(2) Anomaly}},
  \href{http://dx.doi.org/10.1016/0370-2693(82)90728-6}{\emph{Phys. Lett.} {\bf
  117B} (1982) 324--328}.

\bibitem{Kapustin:2014dxa}
A.~Kapustin, R.~Thorngren, A.~Turzillo and Z.~Wang, \emph{{Fermionic Symmetry
  Protected Topological Phases and Cobordisms}},
  \href{http://dx.doi.org/10.1007/JHEP12(2015)052}{\emph{JHEP} {\bf 12} (2015)
  052}, [\href{http://arxiv.org/abs/1406.7329}{{\tt 1406.7329}}].

\bibitem{Hsieh:2015xaa}
C.-T. Hsieh, G.~Y. Cho and S.~Ryu, \emph{{Global anomalies on the surface of
  fermionic symmetry-protected topological phases in (3+1) dimensions}},
  \href{http://dx.doi.org/10.1103/PhysRevB.93.075135}{\emph{Phys. Rev.} {\bf
  B93} (2016) 075135}, [\href{http://arxiv.org/abs/1503.01411}{{\tt
  1503.01411}}].

\bibitem{Witten:2015aba}
E.~Witten, \emph{{Fermion Path Integrals And Topological Phases}},
  \href{http://dx.doi.org/10.1103/RevModPhys.88.035001,
  10.1103/RevModPhys.88.35001}{\emph{Rev. Mod. Phys.} {\bf 88} (2016) 035001},
  [\href{http://arxiv.org/abs/1508.04715}{{\tt 1508.04715}}].

\bibitem{Freed:2006mx}
D.~S. Freed, \emph{{Pions and Generalized Cohomology}}, {\emph{J. Diff. Geom.}
  {\bf 80} (2008) 45--77}, [\href{http://arxiv.org/abs/hep-th/0607134}{{\tt
  hep-th/0607134}}].

\bibitem{Hsieh:2018ifc}
C.-T. Hsieh, \emph{{Discrete gauge anomalies revisited}},
  \href{http://arxiv.org/abs/1808.02881}{{\tt 1808.02881}}.

\bibitem{Witten:1985xe}
E.~Witten, \emph{{GLOBAL GRAVITATIONAL ANOMALIES}},
  \href{http://dx.doi.org/10.1007/BF01212448}{\emph{Commun. Math. Phys.} {\bf
  100} (1985) 197}.

\bibitem{AlvarezGaume:1984nf}
L.~Alvarez-Gaume, S.~Della~Pietra and G.~W. Moore, \emph{{Anomalies and Odd
  Dimensions}},
  \href{http://dx.doi.org/10.1016/0003-4916(85)90383-5}{\emph{Annals Phys.}
  {\bf 163} (1985) 288}.

\bibitem{Bilal:2008qx}
A.~Bilal, \emph{{Lectures on Anomalies}},
  \href{http://arxiv.org/abs/0802.0634}{{\tt 0802.0634}}.

\bibitem{Hatcher:478079}
A.~Hatcher, \emph{{Algebraic topology}}.
\newblock Cambridge Univ. Press, Cambridge, 2000.

\bibitem{Freed:1986zx}
D.~S. Freed, \emph{{Determinants, Torsion, and Strings}},
  \href{http://dx.doi.org/10.1007/BF01221001}{\emph{Commun. Math. Phys.} {\bf
  107} (1986) 483--513}.

\bibitem{Garcia-Etxebarria:2017crf}
I.~García-Etxebarria, H.~Hayashi, K.~Ohmori, Y.~Tachikawa and K.~Yonekura,
  \emph{{8d gauge anomalies and the topological Green-Schwarz mechanism}},
  \href{http://dx.doi.org/10.1007/JHEP11(2017)177}{\emph{JHEP} {\bf 11} (2017)
  177}, [\href{http://arxiv.org/abs/1710.04218}{{\tt 1710.04218}}].

\bibitem{AlvarezGaume:1983cs}
L.~Alvarez-Gaume and P.~H. Ginsparg, \emph{{The Topological Meaning of
  Nonabelian Anomalies}},
  \href{http://dx.doi.org/10.1016/0550-3213(84)90487-5}{\emph{Nucl. Phys.} {\bf
  B243} (1984) 449--474}.

\bibitem{Yonekura:2016wuc}
K.~Yonekura, \emph{{Dai-Freed theorem and topological phases of matter}},
  \href{http://dx.doi.org/10.1007/JHEP09(2016)022}{\emph{JHEP} {\bf 09} (2016)
  022}, [\href{http://arxiv.org/abs/1607.01873}{{\tt 1607.01873}}].

\bibitem{Atiyah:1968mp}
M.~F. Atiyah and I.~M. Singer, \emph{{The Index of elliptic operators. 1}},
  \href{http://dx.doi.org/10.2307/1970715}{\emph{Annals Math.} {\bf 87} (1968)
  484--530}.

\bibitem{Dai:1994kq}
X.-z. Dai and D.~S. Freed, \emph{{eta invariants and determinant lines}},
  \href{http://dx.doi.org/10.1063/1.530747}{\emph{J. Math. Phys.} {\bf 35}
  (1994) 5155--5194}, [\href{http://arxiv.org/abs/hep-th/9405012}{{\tt
  hep-th/9405012}}].

\bibitem{Atiyah:1975jf}
M.~F. Atiyah, V.~K. Patodi and I.~M. Singer, \emph{{Spectral asymmetry and
  Riemannian Geometry 1}},
  \href{http://dx.doi.org/10.1017/S0305004100049410}{\emph{Math. Proc.
  Cambridge Phil. Soc.} {\bf 77} (1975) 43}.

\bibitem{Atiyah:1976jg}
M.~F. Atiyah, V.~K. Patodi and I.~M. Singer, \emph{{Spectral asymmetry and
  Riemannian geometry II}},
  \href{http://dx.doi.org/10.1017/S0305004100051872}{\emph{Math. Proc.
  Cambridge Phil. Soc.} {\bf 78} (1976) 405}.

\bibitem{APS-III}
M.~F. Atiyah, V.~K. Patodi and I.~M. Singer, \emph{{Spectral asymmetry and
  Riemannian geometry. III}},
  \href{http://dx.doi.org/10.1017/S0305004100052105}{\emph{Mathematical
  Proceedings of the Cambridge Philosophical Society} {\bf 79} (1976) 71–99}.

\bibitem{Monnier:2014txa}
S.~Monnier, \emph{{The global anomalies of (2,0) superconformal field theories
  in six dimensions}},
  \href{http://dx.doi.org/10.1007/JHEP09(2014)088}{\emph{JHEP} {\bf 09} (2014)
  088}, [\href{http://arxiv.org/abs/1406.4540}{{\tt 1406.4540}}].

\bibitem{Giddings:1987cg}
S.~B. Giddings and A.~Strominger, \emph{{Axion Induced Topology Change in
  Quantum Gravity and String Theory}},
  \href{http://dx.doi.org/10.1016/0550-3213(88)90446-4}{\emph{Nucl. Phys.} {\bf
  B306} (1988) 890--907}.

\bibitem{Witten:1998zw}
E.~Witten, \emph{{Anti-de Sitter space, thermal phase transition, and
  confinement in gauge theories}},
  \href{http://dx.doi.org/10.4310/ATMP.1998.v2.n3.a3}{\emph{Adv. Theor. Math.
  Phys.} {\bf 2} (1998) 505--532},
  [\href{http://arxiv.org/abs/hep-th/9803131}{{\tt hep-th/9803131}}].

\bibitem{Freed:2014iua}
D.~S. Freed, \emph{{Anomalies and Invertible Field Theories}},
  \href{http://dx.doi.org/10.1090/pspum/088/01462}{\emph{Proc. Symp. Pure
  Math.} {\bf 88} (2014) 25--46}, [\href{http://arxiv.org/abs/1404.7224}{{\tt
  1404.7224}}].

\bibitem{Monnier:2019ytc}
S.~Monnier, \emph{{A Modern Point of View on Anomalies}},  in \emph{{Durham
  Symposium, Higher Structures in M-Theory Durham, UK, August 12-18, 2018}},
  2019.
\newblock \href{http://arxiv.org/abs/1903.02828}{{\tt 1903.02828}}.

\bibitem{Benini:2018reh}
F.~Benini, C.~Córdova and P.-S. Hsin, \emph{{On 2-Group Global Symmetries and
  their Anomalies}},  \href{http://arxiv.org/abs/1803.09336}{{\tt 1803.09336}}.

\bibitem{nicolaescu2000notes}
L.~Nicolaescu, \emph{Notes on Seiberg-Witten Theory}.
\newblock Graduate studies in mathematics. American Mathematical Society, 2000.

\bibitem{Guo:2017xex}
M.~Guo, P.~Putrov and J.~Wang, \emph{{Time reversal, SU(N) Yang–Mills and
  cobordisms: Interacting topological superconductors/insulators and quantum
  spin liquids in 3+1D}},
  \href{http://dx.doi.org/10.1016/j.aop.2018.04.025}{\emph{Annals Phys.} {\bf
  394} (2018) 244--293}, [\href{http://arxiv.org/abs/1711.11587}{{\tt
  1711.11587}}].

\bibitem{Wang:2018edf}
J.~Wang, K.~Ohmori, P.~Putrov, Y.~Zheng, Z.~Wan, M.~Guo et~al.,
  \emph{{Tunneling Topological Vacua via Extended Operators: (Spin-)TQFT
  Spectra and Boundary Deconfinement in Various Dimensions}},
  \href{http://dx.doi.org/10.1093/ptep/pty051}{\emph{PTEP} {\bf 2018} (2018)
  053A01}, [\href{http://arxiv.org/abs/1801.05416}{{\tt 1801.05416}}].

\bibitem{McCleary}
J.~McCleary, \emph{A User's Guide to Spectral Sequences}.
\newblock Cambridge Studies in Advanced Mathematics. Cambridge University
  Press, 2~ed., 2000.
\newblock 10.1017/CBO9780511626289.

\bibitem{DavisKirk}
J.~Davis and P.~Kirk, \emph{Lecture Notes in Algebraic Topology}.
\newblock Graduate studies in mathematics. American Mathematical Society, 2001.

\bibitem{ABP}
D.~W. Anderson, E.~H. Brown and F.~P. Peterson, \emph{Spin cobordism},
  {\emph{Bull. Amer. Math. Soc.} {\bf 72} (03, 1966) 256--260}.

\bibitem{Stong-11}
{R. Stong}, \emph{{Calculation of $\Omega_{11}^{spin}(K(Z,4))$}},  in
  \emph{Unified String Theories} (M.~Green and D.~Gross, eds.), World
  Scientific, 1986.

\bibitem{TeichnerPhD}
P.~Teichner, \emph{Topological four-manifolds with finite fundamental group}.
\newblock PhD thesis, Johannes-Gutenberg Universität in Mainz, 1992.

\bibitem{Teichner}
P.~Teichner, \emph{{On the signature of four-manifolds with universal covering
  spin}}, {\emph{Mathematische Annalen} (1993) 745--759}.

\bibitem{adams1995stable}
J.~Adams and J.~Adams, \emph{Stable Homotopy and Generalised Homology}.
\newblock Chicago Lectures in Mathematics. University of Chicago Press, 1995.

\bibitem{WuSteenrod}
W.-T. Wu, \emph{{Les i-carrés dans une variété grassmannienne}}, pp.~19--21.
\newblock World Scientific, 2012.

\bibitem{Borel1953}
A.~Borel, \emph{La cohomologie mod 2 de certains espaces homogènes.},
  {\emph{Commentarii mathematici Helvetici} {\bf 27} (1953) 165--197}.

\bibitem{milnor1974characteristic}
E.~Milnor, J.~Milnor, J.~Stasheff, J.~Mather and P.~Griffiths,
  \emph{Characteristic Classes}.
\newblock Annals of mathematics studies. Princeton University Press, 1974.

\bibitem{Fung}
J.~H. Fung, ``{The Cohomology of Lie groups}.''
  {\url{http://math.uchicago.edu/~may/REU2012/REUPapers/Fung.pdf}}.

\bibitem{Marathe:2010ncz}
K.~Marathe, \emph{{Topics in Physical Mathematics: Geometric Topology and Field
  Theory}}.
\newblock Springer, 2010.
\newblock 10.1007/978-1-84882-939-8.

\bibitem{Switzer}
R.~Switzer, \emph{Algebraic Topology: Homotopy and Homology}.
\newblock Classics in Mathematics. Springer, 2002.

\bibitem{Intriligator:1997pq}
K.~A. Intriligator, D.~R. Morrison and N.~Seiberg, \emph{{Five-dimensional
  supersymmetric gauge theories and degenerations of Calabi-Yau spaces}},
  \href{http://dx.doi.org/10.1016/S0550-3213(97)00279-4}{\emph{Nucl. Phys.}
  {\bf B497} (1997) 56--100}, [\href{http://arxiv.org/abs/hep-th/9702198}{{\tt
  hep-th/9702198}}].

\bibitem{Redlich:1983kn}
A.~N. Redlich, \emph{{Gauge Noninvariance and Parity Violation of
  Three-Dimensional Fermions}},
  \href{http://dx.doi.org/10.1103/PhysRevLett.52.18}{\emph{Phys. Rev. Lett.}
  {\bf 52} (1984) 18}.

\bibitem{Redlich:1983dv}
A.~N. Redlich, \emph{{Parity Violation and Gauge Noninvariance of the Effective
  Gauge Field Action in Three-Dimensions}},
  \href{http://dx.doi.org/10.1103/PhysRevD.29.2366}{\emph{Phys. Rev.} {\bf D29}
  (1984) 2366--2374}.

\bibitem{husemoller}
D.~Husem{\"o}ller, M.~Joachim, B.~Jurco and M.~Schottenloher, \emph{Basic
  Bundle Theory and K-Cohomology Invariants}.
\newblock Lecture Notes in Physics. Springer Berlin Heidelberg, 2007.

\bibitem{Diaconescu:2000wy}
D.-E. Diaconescu, G.~W. Moore and E.~Witten, \emph{{E(8) gauge theory, and a
  derivation of K theory from M theory}},
  \href{http://dx.doi.org/10.4310/ATMP.2002.v6.n6.a2}{\emph{Adv. Theor. Math.
  Phys.} {\bf 6} (2003) 1031--1134},
  [\href{http://arxiv.org/abs/hep-th/0005090}{{\tt hep-th/0005090}}].

\bibitem{bahri1987}
A.~Bahri and P.~Gilkey, \emph{The eta invariant, ${\rm pin}^c$ bordism, and
  equivariant ${\rm spin}^c$ bordism for cyclic $2$-groups.}, {\emph{Pacific J.
  Math.} {\bf 128} (1987) 1--24}.

\bibitem{Witten:1985bt}
E.~Witten, \emph{{Topological Tools in Ten-dimensional Physics}},
  \href{http://dx.doi.org/10.1142/S0217751X86000034}{\emph{Int. J. Mod. Phys.}
  {\bf A1} (1986) 39}.

\bibitem{FuchsViro}
D.~Fuchs and O.~Viro, \emph{Topology II: Homotopy and Homology. Classical
  Manifolds}.
\newblock Encyclopaedia of Mathematical Sciences. Springer Berlin Heidelberg,
  2003.

\bibitem{borel1955}
A.~Borel, \emph{Topology of lie groups and characteristic classes},
  {\emph{Bull. Amer. Math. Soc.} {\bf 61} (09, 1955) 397--432}.

\bibitem{10.2307/2372495}
A.~Borel and J.-P. Serre, \emph{Groupes de lie et puissances reduites de
  steenrod}, {\emph{American Journal of Mathematics} {\bf 75} (1953) 409--448}.

\bibitem{Tong:2017oea}
D.~Tong, \emph{{Line Operators in the Standard Model}},
  \href{http://dx.doi.org/10.1007/JHEP07(2017)104}{\emph{JHEP} {\bf 07} (2017)
  104}, [\href{http://arxiv.org/abs/1705.01853}{{\tt 1705.01853}}].

\bibitem{Monnier:2017oqd}
S.~Monnier, G.~W. Moore and D.~S. Park, \emph{{Quantization of anomaly
  coefficients in 6D $\mathcal{N}=(1,0)$ supergravity}},
  \href{http://dx.doi.org/10.1007/JHEP02(2018)020}{\emph{JHEP} {\bf 02} (2018)
  020}, [\href{http://arxiv.org/abs/1711.04777}{{\tt 1711.04777}}].

\bibitem{Choi:1992xp}
K.-w. Choi, D.~B. Kaplan and A.~E. Nelson, \emph{{Is CP a gauge symmetry?}},
  \href{http://dx.doi.org/10.1016/0550-3213(93)90082-Z}{\emph{Nucl. Phys.} {\bf
  B391} (1993) 515--530}, [\href{http://arxiv.org/abs/hep-ph/9205202}{{\tt
  hep-ph/9205202}}].

\bibitem{GilkeyBook}
P.~Gilkey, \emph{The Geometry of Spherical Space Form Groups}.
\newblock Series in pure mathematics. World Scientific, 1989.

\bibitem{2001RvMaP..13..953B}
M.~{Berg}, C.~{De Witt-Morette}, S.~{Gwo} and E.~{Kramer}, \emph{{The Pin
  Groups in Physics}},
  \href{http://dx.doi.org/10.1142/S0129055X01000922}{\emph{Reviews in
  Mathematical Physics} {\bf 13} (2001) 953--1034},
  [\href{http://arxiv.org/abs/math-ph/0012006}{{\tt math-ph/0012006}}].

\bibitem{Tachikawa:2018njr}
Y.~Tachikawa and K.~Yonekura, \emph{{Why are fractional charges of orientifolds
  compatible with Dirac quantization?}},
  \href{http://arxiv.org/abs/1805.02772}{{\tt 1805.02772}}.

\bibitem{2016arXiv161200506G}
X.~{Gu}, \emph{{On the Cohomology of the Classifying Spaces of Projective
  Unitary Groups}}, {\emph{ArXiv e-prints} (Dec., 2016) },
  [\href{http://arxiv.org/abs/1612.00506}{{\tt 1612.00506}}].

\bibitem{rotman1998introduction}
J.~Rotman, \emph{An Introduction to Algebraic Topology}.
\newblock Graduate Texts in Mathematics. Springer New York, 1998.

\bibitem{2013arXiv1312.5676B}
L.~{Breen}, R.~{Mikhailov} and A.~{Touz{\'e}}, \emph{{Derived functors of the
  divided power functors}}, {\emph{ArXiv e-prints} (Dec., 2013) },
  [\href{http://arxiv.org/abs/1312.5676}{{\tt 1312.5676}}].

\bibitem{MR0087935}
H.~Cartan, \emph{S\'eminaire {H}enri {C}artan de l'{E}cole {N}ormale
  {S}up\'erieure, 1954/1955. {A}lg\`ebres d'{E}ilenberg-{M}ac{L}ane et
  homotopie}, .

\bibitem{POINTETTISCHLER19971113}
N.~Pointet-Tischler, \emph{La suspension cohomologique des espaces
  d'eilenberg-maclane},
  \href{http://dx.doi.org/https://doi.org/10.1016/S0764-4442(97)88715-0}{\emph{Comptes
  Rendus de l'Académie des Sciences - Series I - Mathematics} {\bf 325} (1997)
  1113 -- 1116}.

\bibitem{TischlerPhD}
N.~Tischler, \emph{Invariants de Postnikov des espaces de lacets}.
\newblock PhD thesis, Universit\'{e} de Lausanne, 1996.

\bibitem{ClementPhD}
A.~Cl\'{e}ment, \emph{Integral Cohomology of Finite Postnikov Towers}.
\newblock PhD thesis, Universit\'{e} de Lausanne, 2002.

\bibitem{Feshbach}
M.~Feshbach, \emph{The integral cohomology rings of the classifying spaces of
  o(n) and so(n)}, {\emph{Indiana University Mathematics Journal} {\bf 32}
  (1983) 511--516}.

\bibitem{10.2307/2044298}
E.~H. Brown, \emph{{The Cohomology of $BSO_n$ and $BO_n$ with Integer
  Coefficients}}, {\emph{{Proceedings of the American Mathematical Society}}
  {\bf 85} (1982) 283--288}.

\bibitem{10.2307/24893350}
M.~FESHBACH, \emph{The integral cohomology rings of the classifying spaces of
  o(n) and so(n)}, {\emph{Indiana University Mathematics Journal} {\bf 32}
  (1983) 511--516}.

\bibitem{QUILLEN1971}
D.~QUILLEN, \emph{The mod 2 cohomology rings of extra-special 2-groups and the
  spinor groups.}, {\emph{Mathematische Annalen} {\bf 194} (1971) 197--212}.

\bibitem{kono1986}
A.~Kono, \emph{On the integral cohomology of $bspin(n)$},
  \href{http://dx.doi.org/10.1215/kjm/1250520870}{\emph{J. Math. Kyoto Univ.}
  {\bf 26} (1986) 333--337}.

\bibitem{2009arXiv0904.0800K}
M.~{Kameko} and M.~{Mimura}, \emph{{On the Rothenberg-Steenrod spectral
  sequence for the mod 2 cohomology of classifying spaces of spinor groups}},
  {\emph{ArXiv e-prints} (Apr., 2009) },
  [\href{http://arxiv.org/abs/0904.0800}{{\tt 0904.0800}}].

\bibitem{Edwards}
S.~R. Edwards, \emph{On the spin bordism of $b(e_{8} \times e_{8})$},
  {\emph{Illinois J. Math.} {\bf 35} (12, 1991) 683--689}.

\bibitem{Ibanez:1991hv}
L.~E. Ibanez and G.~G. Ross, \emph{{Discrete gauge symmetry anomalies}},
  \href{http://dx.doi.org/10.1016/0370-2693(91)91614-2}{\emph{Phys. Lett.} {\bf
  B260} (1991) 291--295}.

\bibitem{Ibanez:1992ji}
L.~E. Ibanez, \emph{{More about discrete gauge anomalies}},
  \href{http://dx.doi.org/10.1016/0550-3213(93)90111-2}{\emph{Nucl. Phys.} {\bf
  B398} (1993) 301--318}, [\href{http://arxiv.org/abs/hep-ph/9210211}{{\tt
  hep-ph/9210211}}].

\bibitem{Dreiner:2005rd}
H.~K. Dreiner, C.~Luhn and M.~Thormeier, \emph{{What is the discrete gauge
  symmetry of the MSSM?}},
  \href{http://dx.doi.org/10.1103/PhysRevD.73.075007}{\emph{Phys. Rev.} {\bf
  D73} (2006) 075007}, [\href{http://arxiv.org/abs/hep-ph/0512163}{{\tt
  hep-ph/0512163}}].

\bibitem{Mohapatra:2007vd}
R.~N. Mohapatra and M.~Ratz, \emph{{Gauged Discrete Symmetries and Proton
  Stability}}, \href{http://dx.doi.org/10.1103/PhysRevD.76.095003}{\emph{Phys.
  Rev.} {\bf D76} (2007) 095003}, [\href{http://arxiv.org/abs/0707.4070}{{\tt
  0707.4070}}].

\bibitem{Lee:2011dya}
H.~M. Lee, S.~Raby, M.~Ratz, G.~G. Ross, R.~Schieren, K.~Schmidt-Hoberg et~al.,
  \emph{{Discrete R symmetries for the MSSM and its singlet extensions}},
  \href{http://dx.doi.org/10.1016/j.nuclphysb.2011.04.009}{\emph{Nucl. Phys.}
  {\bf B850} (2011) 1--30}, [\href{http://arxiv.org/abs/1102.3595}{{\tt
  1102.3595}}].

\bibitem{Nilles:2012cy}
H.~P. Nilles, M.~Ratz and P.~K.~S. Vaudrevange, \emph{{Origin of Family
  Symmetries}}, \href{http://dx.doi.org/10.1002/prop.201200120}{\emph{Fortsch.
  Phys.} {\bf 61} (2013) 493--506}, [\href{http://arxiv.org/abs/1204.2206}{{\tt
  1204.2206}}].

\bibitem{Chen:2015aba}
M.-C. Chen, M.~Fallbacher, M.~Ratz, A.~Trautner and P.~K.~S. Vaudrevange,
  \emph{{Anomaly-safe discrete groups}},
  \href{http://dx.doi.org/10.1016/j.physletb.2015.05.047}{\emph{Phys. Lett.}
  {\bf B747} (2015) 22--26}, [\href{http://arxiv.org/abs/1504.03470}{{\tt
  1504.03470}}].

\bibitem{Gilkey1996}
P.~B. Gilkey and B.~Botvinnik, \emph{The eta invariant and the equivariant spin
  bordism of spherical space form 2 groups}, pp.~213--223.
\newblock Springer Netherlands, Dordrecht, 1996.

\bibitem{Gilkey1998}
P.~B. Gilkey, \emph{The eta invariant of pin manifolds with cyclic fundamental
  groups}, \href{http://dx.doi.org/10.1023/A:1004629725347}{\emph{Periodica
  Mathematica Hungarica} {\bf 36} (Apr, 1998) 139--170}.

\bibitem{Ibanez:1991pr}
L.~E. Ibanez and G.~G. Ross, \emph{{Discrete gauge symmetries and the origin of
  baryon and lepton number conservation in supersymmetric versions of the
  standard model}},
  \href{http://dx.doi.org/10.1016/0550-3213(92)90195-H}{\emph{Nucl. Phys.} {\bf
  B368} (1992) 3--37}.

\bibitem{Gaiotto:2017yup}
D.~Gaiotto, A.~Kapustin, Z.~Komargodski and N.~Seiberg, \emph{{Theta, Time
  Reversal, and Temperature}},
  \href{http://dx.doi.org/10.1007/JHEP05(2017)091}{\emph{JHEP} {\bf 05} (2017)
  091}, [\href{http://arxiv.org/abs/1703.00501}{{\tt 1703.00501}}].

\bibitem{Gaiotto:2017tne}
D.~Gaiotto, Z.~Komargodski and N.~Seiberg, \emph{{Time-reversal breaking in
  QCD$_{4}$, walls, and dualities in 2 + 1 dimensions}},
  \href{http://dx.doi.org/10.1007/JHEP01(2018)110}{\emph{JHEP} {\bf 01} (2018)
  110}, [\href{http://arxiv.org/abs/1708.06806}{{\tt 1708.06806}}].

\bibitem{vanderBij:2007fe}
J.~J. van~der Bij, \emph{{A Cosmotopological relation for a unified field
  theory}}, \href{http://dx.doi.org/10.1103/PhysRevD.76.121702}{\emph{Phys.
  Rev.} {\bf D76} (2007) 121702}, [\href{http://arxiv.org/abs/0708.4179}{{\tt
  0708.4179}}].

\bibitem{Volovik:2016mre}
G.~E. Volovik and M.~A. Zubkov, \emph{{Standard Model as the topological
  material}}, \href{http://dx.doi.org/10.1088/1367-2630/aa573d}{\emph{New J.
  Phys.} {\bf 19} (2017) 015009}, [\href{http://arxiv.org/abs/1608.07777}{{\tt
  1608.07777}}].

\bibitem{Cheng:1985bj}
T.~P. Cheng and L.~F. Li, \emph{{GAUGE THEORY OF ELEMENTARY PARTICLE PHYSICS}}.
\newblock Oxford, Uk: Clarendon ( 1984) 536 P. ( Oxford Science Publications),
  1984.

\bibitem{Patrignani:2016xqp}
{\scshape Particle Data Group} collaboration, C.~Patrignani et~al.,
  \emph{{Review of Particle Physics}},
  \href{http://dx.doi.org/10.1088/1674-1137/40/10/100001}{\emph{Chin. Phys.}
  {\bf C40} (2016) 100001}.

\bibitem{Chang:1994sv}
L.~N. Chang and C.~Soo, \emph{{The Standard model with gravity couplings}},
  \href{http://dx.doi.org/10.1103/PhysRevD.53.5682}{\emph{Phys. Rev.} {\bf D53}
  (1996) 5682--5691}, [\href{http://arxiv.org/abs/hep-th/9406188}{{\tt
  hep-th/9406188}}].

\bibitem{Csaki:1996ks}
C.~Csaki, \emph{{The Minimal supersymmetric standard model (MSSM)}},
  \href{http://dx.doi.org/10.1142/S021773239600062X}{\emph{Mod. Phys. Lett.}
  {\bf A11} (1996) 599}, [\href{http://arxiv.org/abs/hep-ph/9606414}{{\tt
  hep-ph/9606414}}].

\bibitem{dillen1999handbook}
F.~Dillen and L.~Verstraelen, \emph{Handbook of Differential Geometry, Vol.1}.
\newblock Elsevier Science, 1999.

\bibitem{Banks:1991xj}
T.~Banks and M.~Dine, \emph{{Note on discrete gauge anomalies}},
  \href{http://dx.doi.org/10.1103/PhysRevD.45.1424}{\emph{Phys. Rev.} {\bf D45}
  (1992) 1424--1427}, [\href{http://arxiv.org/abs/hep-th/9109045}{{\tt
  hep-th/9109045}}].

\bibitem{Berasaluce-Gonzalez:2013sna}
M.~Berasaluce-Gonz\'alez, M.~Montero, A.~Retolaza and A.~M. Uranga,
  \emph{{Discrete gauge symmetries from (closed string) tachyon condensation}},
  \href{http://dx.doi.org/10.1007/JHEP11(2013)144}{\emph{JHEP} {\bf 1311}
  (2013) 144}, [\href{http://arxiv.org/abs/1305.6788}{{\tt 1305.6788}}].

\bibitem{Garcia-Etxebarria:2015ota}
I.~Garc\'{i}a-Etxebarria, M.~Montero and A.~M. Uranga, \emph{{Closed tachyon
  solitons in type II string theory}},
  \href{http://arxiv.org/abs/1505.05510}{{\tt 1505.05510}}.

\bibitem{Hellerman:2004zm}
S.~Hellerman, \emph{{On the landscape of superstring theory in $D > 10$}},
  \href{http://arxiv.org/abs/hep-th/0405041}{{\tt hep-th/0405041}}.

\bibitem{Shiozaki:2016zjg}
K.~Shiozaki, H.~Shapourian and S.~Ryu, \emph{{Many-body topological invariants
  in fermionic symmetry-protected topological phases}},
  \href{http://dx.doi.org/10.1103/PhysRevB.95.205139}{\emph{Phys. Rev.} {\bf
  B95} (2017) 205139}, [\href{http://arxiv.org/abs/1609.05970}{{\tt
  1609.05970}}].

\bibitem{doi:10.1142/9789812772107_0002}
R.~Thom, \emph{Some “global” properties of differentiable manifolds},
  pp.~131--209.
\newblock WORLD SCIENTIFIC, 2012.

\bibitem{Chen:2013dpa}
M.-C. Chen, M.~Ratz and A.~Trautner, \emph{{Non-Abelian discrete R
  symmetries}}, \href{http://dx.doi.org/10.1007/JHEP09(2013)096}{\emph{JHEP}
  {\bf 09} (2013) 096}, [\href{http://arxiv.org/abs/1306.5112}{{\tt
  1306.5112}}].

\bibitem{Banks:2010zn}
T.~Banks and N.~Seiberg, \emph{{Symmetries and Strings in Field Theory and
  Gravity}}, \href{http://dx.doi.org/10.1103/PhysRevD.83.084019}{\emph{Phys.
  Rev.} {\bf D83} (2011) 084019}, [\href{http://arxiv.org/abs/1011.5120}{{\tt
  1011.5120}}].

\bibitem{BerasaluceGonzalez:2012vb}
M.~Berasaluce-Gonzalez, P.~G. Camara, F.~Marchesano, D.~Regalado and A.~M.
  Uranga, \emph{{Non-Abelian discrete gauge symmetries in 4d string models}},
  \href{http://dx.doi.org/10.1007/JHEP09(2012)059}{\emph{JHEP} {\bf 09} (2012)
  059}, [\href{http://arxiv.org/abs/1206.2383}{{\tt 1206.2383}}].

\bibitem{Polchinski:2003bq}
J.~Polchinski, \emph{{Monopoles, duality, and string theory}},
  \href{http://dx.doi.org/10.1142/S0217751X0401866X}{\emph{Int. J. Mod. Phys.}
  {\bf A19S1} (2004) 145--156},
  [\href{http://arxiv.org/abs/hep-th/0304042}{{\tt hep-th/0304042}}].

\bibitem{Gaiotto:2014kfa}
D.~Gaiotto, A.~Kapustin, N.~Seiberg and B.~Willett, \emph{{Generalized Global
  Symmetries}}, \href{http://dx.doi.org/10.1007/JHEP02(2015)172}{\emph{JHEP}
  {\bf 02} (2015) 172}, [\href{http://arxiv.org/abs/1412.5148}{{\tt
  1412.5148}}].

\bibitem{Tachikawa:2016cha}
Y.~Tachikawa and K.~Yonekura, \emph{{On time-reversal anomaly of 2+1d
  topological phases}},
  \href{http://dx.doi.org/10.1093/ptep/ptx010}{\emph{PTEP} {\bf 2017} (2017)
  033B04}, [\href{http://arxiv.org/abs/1610.07010}{{\tt 1610.07010}}].

\bibitem{Witten:1996hc}
E.~Witten, \emph{{Five-brane effective action in M theory}},
  \href{http://dx.doi.org/10.1016/S0393-0440(97)80160-X}{\emph{J. Geom. Phys.}
  {\bf 22} (1997) 103--133}, [\href{http://arxiv.org/abs/hep-th/9610234}{{\tt
  hep-th/9610234}}].

\bibitem{Freed:2004yc}
D.~S. Freed and G.~W. Moore, \emph{{Setting the quantum integrand of
  M-theory}}, \href{http://dx.doi.org/10.1007/s00220-005-1482-7}{\emph{Commun.
  Math. Phys.} {\bf 263} (2006) 89--132},
  [\href{http://arxiv.org/abs/hep-th/0409135}{{\tt hep-th/0409135}}].

\bibitem{Hattori}
A.~Hattori, \emph{Integral characteristic numbers for weakly almost complex
  manifolds},
  \href{http://dx.doi.org/10.1016/0040-9383(66)90010-3}{\emph{Topology} {\bf 5}
  (sep, 1966) 259--280}.

\bibitem{STONG1965267}
R.~Stong, \emph{{Relations among characteristic numbers—I}},
  \href{http://dx.doi.org/https://doi.org/10.1016/0040-9383(65)90011-X}{\emph{Topology}
  {\bf 4} (1965) 267 -- 281}.

\bibitem{HopkinsHovey}
M.~J. Hopkins and M.~A. Hovey, \emph{{Spin cobordism determines real
  K-theory}}, \href{http://dx.doi.org/10.1007/BF02571790}{\emph{Mathematische
  Zeitschrift} {\bf 210} (Dec, 1992) 181--196}.

\bibitem{Bergman:2013ala}
O.~Bergman, D.~Rodríguez-Gómez and G.~Zafrir, \emph{{Discrete $\theta$ and
  the 5d superconformal index}},
  \href{http://dx.doi.org/10.1007/JHEP01(2014)079}{\emph{JHEP} {\bf 01} (2014)
  079}, [\href{http://arxiv.org/abs/1310.2150}{{\tt 1310.2150}}].

\bibitem{Sethi:2013hra}
S.~Sethi, \emph{{A New String in Ten Dimensions?}},
  \href{http://dx.doi.org/10.1007/JHEP09(2013)149}{\emph{JHEP} {\bf 09} (2013)
  149}, [\href{http://arxiv.org/abs/1304.1551}{{\tt 1304.1551}}].

\bibitem{10.2307/1970030}
A.~Dold and H.~Whitney, \emph{Classification of oriented sphere bundles over a
  4-complex}, {\emph{Annals of Mathematics} {\bf 69} (1959) 667--677}.

\bibitem{woodward1982}
L.~M. Woodward, \emph{The classification of orientable vector bundles over
  cw-complexes of small dimension},
  \href{http://dx.doi.org/10.1017/S0308210500032467}{\emph{Proceedings of the
  Royal Society of Edinburgh: Section A Mathematics} {\bf 92} (1982)
  175–179}.

\bibitem{KirbyTaylor}
R.~C. Kirby and L.~R. Taylor, \emph{{A calculation of $\Pin^+$ bordism
  groups}}, \href{http://dx.doi.org/10.1007/BF02566617}{\emph{Commentarii
  Mathematici Helvetici} {\bf 65} (Dec, 1990) 434--447}.

\bibitem{ABPPin-}
D.~W. Anderson, E.~H. Brown and F.~P. Peterson, \emph{Pin cobordism and related
  topics}, \href{http://dx.doi.org/10.1007/BF02564545}{\emph{Commentarii
  Mathematici Helvetici} {\bf 44} (Dec, 1969) 462--468}.

\bibitem{Novikov}
S.~P. Novikov, \emph{Homotopy properties of Thom complexes}, pp.~211--250.
\newblock World Scientific, 2012.

\bibitem{barth2015compact}
W.~Barth, K.~Hulek, C.~Peters and A.~van~de Ven, \emph{{Compact Complex
  Surfaces}}.
\newblock Ergebnisse der Mathematik und ihrer Grenzgebiete. 3. Folge / A Series
  of Modern Surveys in Mathematics. Springer Berlin Heidelberg, 2015.

\end{thebibliography}\endgroup

\end{document}